\newcommand{\be}{\begin{equation}}
\newcommand{\ee}{\end{equation}}
\newcommand{\bea}{\begin{eqnarray}}
\newcommand{\eea}{\end{eqnarray}}
\documentclass[aps,pra,amsmath,amssymb,amsfonts,onecolumn,nofootinbib]{revtex4}
\usepackage{graphicx}
\usepackage{dcolumn}
\usepackage{bm}
\usepackage{amsmath}
\usepackage{amssymb}
\def\d{d\kern-.8 ex\vrule height 1.3 ex depth-1.24 ex width .7 ex \kern .15 ex}
\def\D{D\kern-1.7 ex\vrule height .87 ex depth-.8 ex width .7 ex \kern .95 ex}

\begin{document}
\title{Quantum criticality in photorefractive optics: Vortices in laser beams and antiferromagnets}

\author{Mihailo~\v{C}ubrovi\'{c}}\email{mcubrovic@gmail.com}\affiliation{Scientific Computing Laboratory, Institute
of Physics, University of Belgrade, Pregrevica 118, 11080 Belgrade,
Serbia}

\author{Milan S. Petrovi\'c}
\affiliation{Institute of Physics, P.O.Box 57, 11001 Belgrade,
Serbia} \affiliation{Texas A\&M University at Qatar, P.O.Box 23874,
Doha, Qatar}

\date{\today}

\begin{abstract}
We study vortex patterns in a prototype nonlinear optical system:
counterpropagating laser beams in a photorefractive crystal, with or
without the background photonic lattice. The vortices are
effectively planar and have two "flavors" because there are two
opposite directions of beam propagation. In a certain parameter
range, the vortices form stable equilibrium configurations which we
study using the methods of statistical field theory and generalize
the Berezinsky-Kosterlitz-Thouless transition of the XY model to the
"two-flavor" case. In addition to the familiar conductor and
insulator phases, we also have the perfect conductor (vortex
proliferation in both beams/"flavors") and the frustrated insulator
(energy costs of vortex proliferation and vortex annihilation
balance each other). In the presence of disorder in the background
lattice, a novel phase appears which shows long-range correlations
and absence of long-range order, thus being analogous to glasses. An
important benefit of this approach is that qualitative behavior of
patterns can be known without intensive numerical work over large
areas of the parameter space. The observed phases are analogous to
those in magnetic systems, and make (classical) photorefractive
optics a fruitful testing ground for (quantum) condensed matter
systems. As an example, we map our system to a doped $O(3)$
antiferromagnet with $\mathbb{Z}_2$ defects, which has the same
structure of the phase diagram.
\end{abstract}

\pacs{05.70.Fh,42.65.Sf,64.60.De,64.70.qd,75.10.Nr}

\maketitle

\section{Introduction}

Nonlinear and pattern-forming systems \cite{patt1,pattbook,korn} have numerous analogies with strongly correlated systems encountered in condensed matter physics \cite{ft1,ft2}, and on the methodological level they are both united through the language of field theory, which has become the standard language to describe strongly correlated electrons \cite{fradkin,tsvelik} as well as nonlinear dynamical systems \cite{cvetbook}. In the field of pattern formation, some connections to condensed matter systems have been observed, see e.g. \cite{ft1}. More recently, extensive field-theoretical studies of laser systems were performed, e.~g. \cite{conti1,conti2,conti3,conti4}, and also compared to experiment \cite{ghofraniha}. However, this topic is far from exhausted and we feel many analogies between quantum many-body systems and pattern-formation dynamics remain unexplored and unexploited. In particular, nonlinear optical systems and photonic lattices are flexible and relatively cheap to build \cite{korn} and they can be used to "simulate" a broad spectrum of phenomena concerning band structure, spin ordering and conduction in strongly correlated electron systems; some of the work in this direction can be found in \cite{primeri1,primeri2}.

Our goal is to broaden the connections between the strongly
correlated systems and nonlinear optics and to put to work the
mighty apparatus of field theory to study the patterns in a
nonlinear optical system from the viewpoint of phase transition
theory: pattern dynamics in certain cases shows critical behavior
which is analogous to phenomena seen in magnetic systems. To that
end, we use the formalism of perturbative field theory and
renormalization group analysis but we also perform numerical
simulations from the first principles, i.e. directly integrating the
equations of motion to provide an \emph{independent} check of our
main conclusions. We also establish a connection to an $O(3)$
antiferromagnetic model which is encountered in the study of
strongly correlated electron systems. The analogy is not just
qualitative: we construct the phase diagrams of both systems and
find they have the same structure. Introducing disorder into the
system further enriches the physics, and it is physically motivated:
in optics, disorder is rooted in the imperfections of the photonic
lattice, and in magnetic systems it comes from the quenched spin
impurities which are regularly found in realistic samples. It turns
out that in both cases a glassy phase arises. This is another
important research topic and it is again appealing to realize
glasses in photonic lattice systems, where the parameters are easy
to tune.

\subsection{On topology and vortices}

The key phenomenon which governs the phenomenology of the systems
studied is the existence of topologically nontrivial solutions or
\emph{topological solitons} \cite{raja}: these are the solutions
which map the physical boundary of the system to the whole
configuration space of the field, so one explores all field
configurations by "going around the system". For example, in a
two-dimensional system (in the $x-y$ plane) with $U(1)$ phase
symmetry, the configuration space is a circle (the phase lies
between $0$ and $2\pi$) and the boundary of the physical space (i.e.
the two-dimensional plane) is again a circle, the "boundary" of the
plane at infinity. The topological soliton is a pattern of the
$U(1)$ field which spans the whole phase circle (its phase goes from
$0$ to $2\pi$), as one moves around the far-away circle in the $x-y$
plane. Of course, this is the vortex -- the most famous and best
studied topological configuration. Similar logic leads to the
classification of topological defects of other, more complicated
symmetry groups. A potential source of confusion is that in
nonlinear dynamics and theory of partial differential equations, the
"integrable" solutions, i.e. linearly (often also nonlinearly)
stable solutions which can be obtained by inverse scattering or
similar methods and which propagate through each other without
interacting, are also called solitons, or more precisely
\emph{dynamical solitons}. In optics, they are often called spatial
solitons. Dynamical solitons in nonlinear optics are a celebrated
and well-studied topic
\cite{rev,kivshar1,kivshar2,kivshar3,newsol1,newsol2}, they show an
amazing variety of patterns and phenomena like localization, Floquet
states \cite{primeri1}, etc. But in general they do not have a
topological charge. In contrast, topological solitons carry a
topological charge (winding number for vortices) and their stability
is rooted in topological protection (conservation of topological
charge).

The phenomenon of vortices is perhaps best known in three spatial dimensions. The phase of the wavefunction can wind, forming a vortex line. These vortices are stable when the phase symmetry is broken by magnetic field. Famously, vortices may coexist with the superconducting order ($U(1)$ symmetry breaking) in type-II superconductors or exist only in the normal phase, upon destroying the superconductivity (type I). The primary example in two spatial dimensions is the vortex unbinding (BKT) phase transition of infinite order found by Berezinsky, Kosterlitz and Thouless for the planar XY model \cite{kosterthoul}. The formal difference between the two- and three-dimensional vortices is that the latter give rise to an emergent gauge field; this does not happen in the XY-like system in two dimensions \cite{kleinert}. While the nonlinear optical system we study is three-dimensional, its geometry and relaxational dynamics make it natural to treat it as a $2+1$-dimensional system (the $x$ and $y$ coordinates are spatial dimensions, the $z$-direction has the formal role of time, and physical time $t$ has the role of a parameter). We therefore have a similar situation to the XY model: point-like vortices in the plane (and no gauge field).

Vortex matter is known to emerge in liquid helium \cite{liqhel}, Bose-Einstein condensates \cite{bec} and magnetic systems \cite{vortcupsach}. The basic mechanisms of vortex dynamics are thus well known. However, novel physics can arise if the system has multiple components and each of them can form vortices which mutually interact. This is precisely our situation -- we have a system of two laser beams propagating in opposite directions, and we will compare it to a two-component antiferromagnet. So far, such situations have been explored in multi-component superconductors \cite{multisc} which have attracted some attention, as they can be realized in magnesium diboride \cite{multiscmg}. But these are again bulk systems, not planar. Vortices in planar multi-component systems have not been very popular, an important exception being the two-component Bose-Einstein condensates of \cite{becmulti}, which were found to exhibit complex vortex dynamics; in these systems, contrary to our case, the two components have an explicit attractive interaction, unlike our case where they interact indirectly, by coupling to the total light intensity (of both components).

\subsection{The object of our study}

In this paper we study phases and critical behavior of topological
configurations (vortices and vortex lattices) in a specific and
experimentally realizable nonlinear optical system: laser beams
counterpropagating (CP) through a photorefractive (PR) crystal. This
means we have an elongated PR crystal (with one longitudinal and two
transverse dimensions) and two laser beams shone onto each end. We
thus effectively have two fields, one forward-propagating and one
backward propagating. The optical response of the crystal depends
nonlinearly on the \emph{total} intensity of both beams, which means
the beams effectively interact with each other. This system has been
thoroughly investigated for phenomena such as dynamical solitons
\cite{prlpet,stabanal,rev}, vortex stability on the photonic lattice
\cite{vortlat2pet,prapet,vortprepet,kivshar1,kivshar2,kivshar3,kivshar4}
and global rotation \cite{gaussrotpet}. We will see that the CP
beams are an analog of the two-component planar antiferromagnet,
which can further be related to some realistic strongly-correlated
materials \cite{sachbook,jurms0,jurms2}. The two beams are now
equivalent to two sublattices which interact through a lattice
deformation or external field. The PR crystal is elongated and the
axial propagation direction has the formal role of time, which has a
finite span, the length of the crystal. For the antiferromagnet, the
third axis is the usual imaginary time compactified to the radius
$1/T$, i.e. inverse temperature. Both systems contain vortices as
topological defects, i.e. solutions with integer topological charge.
In the PR optical system, vortices arise as a consequence of the
$U(1)$ symmetry of the electromagnetic field. In antiferromagnets we
consider, the $O(3)$ symmetry of the antiferromagnet gives rise to
$\mathbb{Z}_2$-charged defects, which exhibit the same interactions
as the vortices. The optical system is not subject to noise, i.e. it
lives at zero temperature, thus the criticality we talk about is
obviously not the same as thermodynamic phase transitions. Phase
transitions happen upon varying the parameters, not temperature, so
they may be described as quantum critical phenomena in the broad
sense taken in \cite{sachbook} -- any critical behavior controlled
not by thermal fluctuations but by parameter dependence.

In the PR counterpropagating beam system, our focus are the vortices but in order to study them we need to do some preparational work. We first recast the system in Lagrangian and then in Hamiltonian form so it can be studied as a field theory, which depends parametrically on the time $t$. Then we consider the time dynamics of the system and show that in a broad parameter range the patterns relax to a static configuration which can be studied within \emph{equilibrium} field theory. Along the way, we also study the stability of topologically trivial (vortex-free) configurations and then consider the phases of the static vortex configurations. The analytical insight we obtain also allows us to avoid overextensive numerics -- analytical construction of the phase diagram tells us which patterns can in principle be expected in different corners of the parameter space. By "blind" numerical approach this result could only be found through many runs of the numerics.

In the antiferromagnetic spin system the non-topological excitations
are simple -- they are spin waves, perturbed away from the
noninteracting solution by the quartic terms in the potential. There
are no dynamical solitons. But we will see that topological
excitations lead to a phase diagram which, after reasonable
approximations, can be \emph{exactly} mapped to the phase diagram of
the photorefractive crystal. The reason is that both can be reduced
to an effective Hamiltonian for a \emph{two-component} vortex
system, i.e., every vortex has two charges, or two "flavors". In the
photorefractive crystal it happens naturally, as there are two
beams, forward- and backward propagating. In the Heisenberg
antiferromagnet it is less obvious, and is a crucial consequence of
the collinearity of the spin pattern. We will focus on common
properties of the two systems and map the phase diagrams onto each
other. In the antiferromagnetic system, different phases are
separated by quantum phase transitions -- phase transitions driven
by the quantum fluctuations instead of temperature.

\subsubsection{On disorder}
It is known that impurities pin the vortices and stabilize them.
This leads to frozen dynamics even though no symmetry is broken, the
phenomenon usually associated with glasses. In simple systems such
as the Ising model with disorder one generically has two phases: the
disordered (paramagnetic) phase remains and the ordered (magnetic)
phase is replaced by a regime with algebraic correlations and no
true order. In many cases, such phases are called glasses. The exact
definition of a glass is lacking; normally, they show (i) long-range
correlations (ii) absence of long-range order, i.e. of a nonzero
macroscopic order parameter (iii) "frozen dynamics", i.e. free
energy landscape with numerous local minima in which the system can
spend a long time \cite{parisi,spinglass}. While the most popular
example are probably spin glasses in Ising-like models such as
Sherington-Kirkpatrick and Edwards-Anderson model, glasses are also
known to appear in the XY model with disorder in two dimensions, the
Cardy-Ostlund model, which postulates both random couplings and a
random magnetic field \cite{xyzoran,xy1,xy2}. Our model is
essentially a two-flavor generalization of the XY model, although in
order to solve it we need to simplify it. According to
\cite{xyzoran,xy1,xy2} the details differ depending on how the
disorder is implemented but the two-phase system (paramagnetic i.e.
disordered, and glass) is ubiquitous. In the two-component version,
the phase diagram becomes richer, and on top of the glassy phase and
the insulator (disordered) phase we find a few other phases. In
nonlinear optics, the topic of random lasers has attracted
considerable attention \cite{conti1,conti2,conti3,conti4,conti5}.
Here one has a complex version of the XY model, with the additional
complication that not only phase but also amplitude is free to vary,
but only with random couplings (no random field). On top of the
glassy and the disordered phase, one or two additional phases
appear.

In the presence of disorder the relation to magnetic systems in condensed matter physics is very inspiring, since a number of complex materials show different ordering mechanisms (spin and charge density waves, superconductivity, etc) in parallel with significant influence of disorder. Just as in the disorder-free case, we are particularly interested in possible spin-glass phenomena in doped insulating $O(3)$ antiferromagnets \cite{jurms0,jurms1,jurms2,jurms3,jurms4} and in the last section we will discuss also the spin-glass phase in such systems.

\subsection{The plan of the paper}

The structure of the paper is as follows. In the next section we
describe the dynamical system which lies at the core of this paper:
counterpropagating laser beams in a photorefractive crystal. We give
the equations of motion and repackage them in the Lagrangian form.
In section III we study the vortex dynamics: we construct the vortex
Hamiltonian and classify the order parameters. Then we study the
renormalization group (RG) flow and obtain the phase diagram.
Finally, we discuss the important question of how to recognize the
various phases in experiment: what do the light intensity patterns
look like and how they depend on the tunable parameters. Section IV
brings the same study for the system with disorder. After describing
the disordered system, we perform the replica trick for the
disordered vortex Hamiltonian and solve the saddle-point equations
to identify the phases and order parameters, again refining the
results with RG calculations. The fifth section takes a look at a
doped collinear antiferromagnet, a model encountered in the
description of many strongly coupled materials, and shows how the
dynamics of topological solitons is again described by a two-flavor
vortex Hamiltonian. We discuss the relation between the phase
diagrams of the two systems and the possibilities of modeling the
condensed matter systems experimentally by the means of
photorefractive optics. The last section sums up the conclusions. In
Appendix A we describe the numerical algorithm we use to check the
analytical results for the phase diagram. In Appendix B we show in
detail that the CP beams are capable of reaching equilibrium (i.e. stop
changing in time)-- if they would not, the application of equilibrium
field theory would not be justified. Appendix C discusses the
stability of nonvortex configurations -- although somewhat
peripheral to the main topic of the paper, it is useful to better
understand the geometry of patterns. In Appendix D we give the
(routine) algebra that yields the vortex interaction Hamiltonian
from the microscopic equations. Appendix E contains an improved
mean-field theory for the clean system, which we do not use much
throughout the paper but we include it for completeness (we prefer
either the simplest single-vortex mean-field reasoning, or the full
RG analysis, which are described in the main text). Appendix F
discusses an important technicality concerning the CP geometry, i.e.
the specific boundary conditions of the CP beam system where the
boundary conditions for one beam are given at the front face and for
the other at the back face of the crystal. Appendix G contains some
details on mean-field and RG calculations of the phase diagram for
the dirty system: the dirty case includes some tedious algebra we
feel appropriate to leave out from the main text.


\section{The model of counterpropagating beams in the photorefractive crystal}

\begin{figure}\centering
\includegraphics[width=70mm]{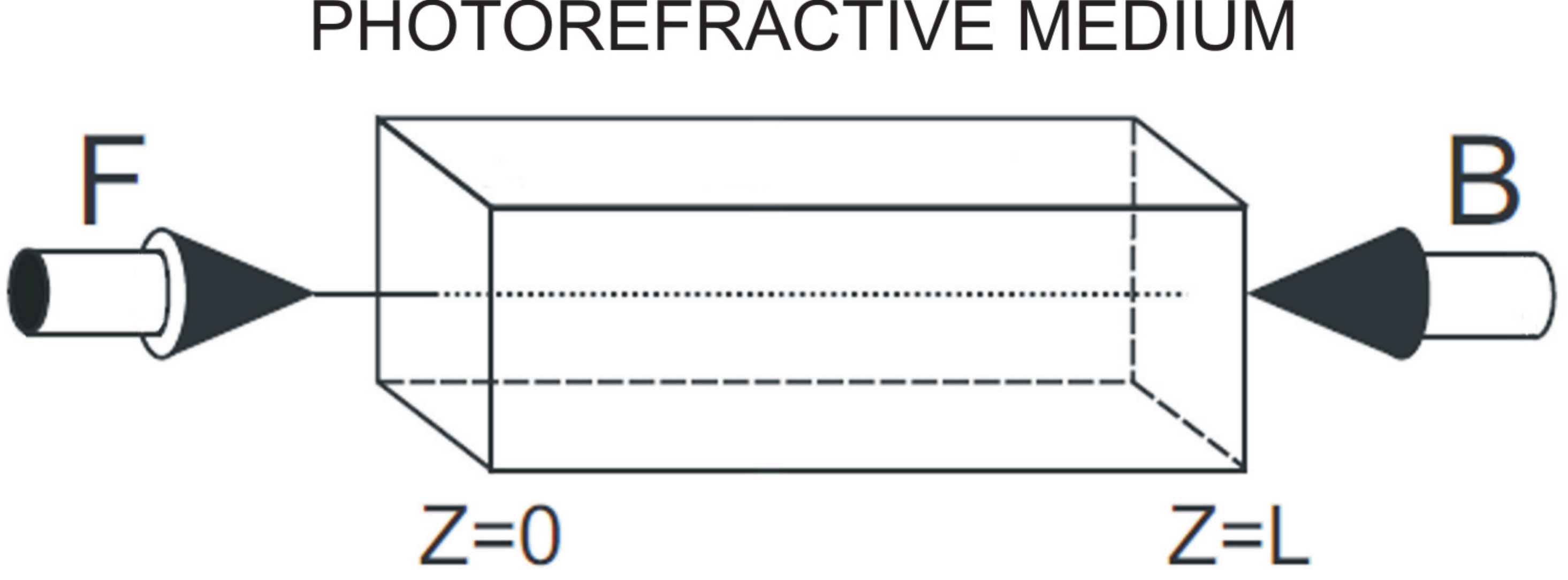}
\caption{\label{figexp} Sketch of the experimental setup for the
study of the CP beams in the PR crystal. The crystal has the shape
of a parallelepiped, and the beams propagate along the longitudinal,
$z$-axis: the forward ($F$)-beam from $z=0$ to $z=L$, and the
backward ($B$)-beam the other way round. The intensity patterns are
observed at the transverse faces of the crystal, at $z=0$ and
$z=L$.}
\end{figure}

We consider a photorefractive crystal of length $L$ irradiated by two laser beams. The beams are paraxial and propagate head-on from the opposite faces of the crystal in the $z$-direction. Photorefractive crystals induce self-focusing of the beams -- the vacuum (linear) wave equation is modified by the addition of a friction-like term, so the diffusion of the light intensity (the broadening of the beam) is balanced out by the convergence of the beam onto an "attractor region". The net result is the balance between the dissipative and scattering effects, allowing for stable patterns to form. The physical ground for this is the redistribution of the charges in the crystal due to the Kerr effect. The nonlinearity, i.e. the response of the crystal to the laser light is contained in the change of the refraction index which is determined by the local charge density. A sketch of the system is given in Fig.~\ref{figexp}. Before entering the crystal, the laser beams can be given any desirable pattern of both intensity and phase. In particular, one can create vortices (winding of the phase) making use of the phase masks \cite{korn} or other, more modern ways.

Assuming the electromagnetic field of the form $\mathbf{E}=e^{\imath\omega t+\imath\mathbf{q}\cdot\mathbf{r}}\left(Fe^{ikz}+Be^{-ikz}\right)$, we can write equations for the so-called envelopes $F$ and $B$ of the forward- and backward propagating beams along the $z$-axis (the frequency, transverse and longitudinal momentum are denoted respectively by $\omega,\mathbf{q},k$). The wave equations for $F$ and $B$ are now:
\be
\label{prpsieq}\pm\imath\partial_z\Psi_\pm(z;x,y;t)+\Delta\Psi_\pm(z;x,y;t)=\Gamma E(z;x,y;t)\Psi_\pm(z;x,y;t),
\ee
where the plus and minus signs on the left-hand side stand for the forward- and backward-propagating component of the beam amplitude doublet $\Psi\equiv (\Psi_+,\Psi_-)\equiv(F,B)$, and $\Gamma$ is the dimensionless PR coupling constant. The two beams (flavors of the field $\Psi$) will from now on be denoted either by $F/B$ or more often by $\Psi_\pm$. We will use $\alpha$ as the general flavor index for summation, e.g. $\Psi_{1\alpha}\Psi_{2\alpha}=\Psi_{1+}\Psi_{2+}+\Psi_{1-}\Psi_{2-}$. The charge field $E$ on the right-hand side of the equation is the electric field sourced by the charges in the crystal (i.e., it does not include the external electric field of the beams). Its evolution is well represented by a relaxation-type equation \cite{rev}:
\be
\label{preeq}\frac{\tau}{1+I(z;x,y;t)}\partial_tE(z;x,y;t)+E(z;x,y;t)=-\frac{I(z;x,y;t)}{1+I(z;x,y;t)}.
\ee
Here, $I\equiv I_\Psi+I_x$ is the total light intensity at a given point, $I_\Psi\equiv\vert F\vert^2+\vert B\vert^2$ is the beam intensity and $I_x$ the intensity of the fixed background. The meaning of $I_x$ is that the crystal is all the time irradiated by some constant light source, independent of the counter-propagating beams with envelopes $F,B$. We will usually take a periodic lattice as the background, allowing also for the defects (missing cells) in the lattice when studying the effects of disorder. The relaxation time is $\tau$. The time derivative $\partial_tE$ is divided by $1+I$, meaning that the polarizability of the crystal depends on the total light intensity: strongly irradiated regions react faster. In the numerical calculations, we solve Eqs.~(\ref{prpsieq}-\ref{preeq}) with no further assumptions, as explained in Appendix \ref{app0}. For analytical results we will need to transform them further assuming a vortex pattern.

The equation for the charge field has no microscopic basis, it is completely phenomenological, but it excellently represents the experimental results \cite{korn}. Notice that the derivative $\partial_tE$ in (\ref{preeq}) is strictly negative (since intensity is non-negative): it thus has the form of a relaxation equation, and one expects that a class of solutions exists where $\partial_tE(t\to\infty)\to 0$, i.e. the system relaxes to a time-independent configuration. We show this in Appendix \ref{appa}; in the main text we will not discuss this issue but will simply take the findings of the Appendix \ref{appa} for granted. Notice that there are also parameter values for which no equilibrium is reached \cite{fisher,fisher2,gaussrotpet}.

For slow time evolution (in the absence of pulses), we can Laplace-transform the equation (\ref{preeq}) in time ($E(t)\mapsto E(u)=\int_0^\infty dte^{-ut}E(t)$) to get the algebraic relation
\be
\label{eueq}E(z;x,y;u)=-\frac{\Psi^\dagger\Psi+I_x-\tau E_0}{1+\tau u+I_x+\Psi^\dagger\Psi}=-1+\frac{1+\tau u+\tau E_0}{1+\tau u+I_x+\Psi^\dagger\Psi}.
\ee
The original system (\ref{prpsieq}) can now be described by the Lagrangian:
\be
\label{psilag}\mathcal{L}=\imath\Psi^\dagger\sigma_3\partial_z\Psi-\vert\nabla\Psi\vert^2+\Gamma\Psi^\dagger\Psi-\Gamma(1+\tau E_0+\tau u)\log (1+\tau u+I_x+\Psi^\dagger\Psi),
\ee
where $\sigma_3$ is the Pauli matrix $\sigma_3=\mathrm{diag}(1,-1)$. One can introduce the effective potential
\be
\label{veff}V_\mathrm{eff}(\Psi^\dagger,\Psi)=-\Gamma\log\frac{e^{\Psi^\dagger\Psi}}{\left(1+\tau u+I_x+\Psi^\dagger\Psi\right)^{1+\tau(E_0+u)}},
\ee
so we can write the Lagrangian as $\mathcal{L}=\imath\Psi^\dagger\sigma_3\partial_z\Psi-\vert\nabla\Psi\vert^2-V_{eff}(\Psi^\dagger,\Psi)$. This is the Lagrangian of a non-relativistic field theory (a nonlinear Schr\"{o}dinger field equation) in $2+1$ dimensions $(x,y;z)$, where the role of time is played by the longitudinal distance $z$, and the physical time $t$ (or $u$ upon the Laplace transform) is a parameter. The span of the $z$ coordinate $0<z<L$ will influence the behavior of the system, while the dimensions of the transverse plane are not important for the effects we consider.

Our main story is now the nature and interactions of the topologically nontrivial excitations in the system (\ref{psilag}). A task which is in a sense more basic, the analysis of the topologically trivial vacua of (\ref{psilag}) and perturbative calculation of their stability, is not of our primary interest now, in part because this was largely accomplished by other methods in \cite{stabanal,prlpet}. We nevertheless give a quick account in Appendix \ref{appb}; first, because some conclusions about the geometry of the patterns can be carried over to vortices, and second, to give another example of applying the field-theoretical formalism whose power we wish to demonstrate and popularize in this paper.

\section{Vortices and mean field theory of vortex interactions}

\subsection{The classification of topological solutions and the vortex Hamiltonian}

Now we discuss the possible topological solitons in our system. Remember once again that they differ from dynamical solitons such as those studied in \cite{rev} and references therein. In order to classify the topologically nontrivial solutions, consider first the symmetries of the Lagrangian (\ref{psilag}). It describes a doublet of 2D complex fields which interact solely through the phase-invariant total intensity $I=\Psi^\dagger\Psi$ (and the spatial derivative term $\vert\nabla\Psi\vert^2$), while in the kinetic term $\Psi^\dagger\sigma_3\partial_z\Psi$ the two components have opposite signs of the "time" derivative, so this term cannot be reduced to a functional of $I$. The intensity $I$ has the symmetry group $SU(2)$ (the isometry group of the three-dimensional sphere in Euclidean space) and the kinetic term has the group $SU(1,1)$ (the transformations which leave the combination $\vert F\vert^2-\vert B\vert^2$ invariant, i.e. the isometry of the hyperboloid). The intersection of these two is the product $U(1)_F\otimes U(1)_B$: the forward- and backward-propagating doublet $(F,B)$ has phases $\theta_{F,B}$ which can be transformed independently, as $\theta_{F,B}\mapsto\theta_{F,B}+\delta\theta_{F,B}$.

The classification of possible topological solitons is straightforward from the above discussion \cite{mermin}. They can be characterized in terms of homotopy groups. To remind, the homotopy group $\pi_n$ of the group $G$ is the group of transformations which map the group manifold of $G$ onto the $n$-dimensional sphere $\mathbb{S}_n$. In $D$-dimensional space the group $\pi_{D-1}$ therefore classifies what a field configuration looks like from far away (from infinity): it classifies the mappings from the manifold of the internal symmetry group of the system to the spherical "boundary shell" in physical space at infinity. Since the beams in our PR crystal effectively see a two-dimensional space (we regard $z$ as time), we need the first homotopy group $\pi_1$ to classify the topological solitons. Since $\pi_1(U(1))=\pi_1(\mathbb{S}_1)=\mathbb{Z}$ and $\pi_1(\mathbb{G}\otimes\mathbb{G})=\pi_1(\mathbb{G})\otimes\pi_1(\mathbb{G})$ for any group $\mathbb{G}$, the topological solutions are flavored vortices, and the topological charge is the pair of integers $\lbrace Q_F,Q_B\rbrace$.

Let us now derive the effective interaction Hamiltonian for the vortices and study the phase diagram. In principle, this story is well known: for a vortex at $\mathbf{r}_0$, in the polar coordinates $(r,\phi)$, we write $\Psi(\mathbf{r})=\psi\exp\left(\imath\theta\left(\mathbf{r}\right)\right)$ for $\vert\mathbf{r}-\mathbf{r}_0\vert/\vert\mathbf{r}_0\vert\ll 1$, and a vortex of charge $Q$ has $\theta(\phi)=Q\phi/2\pi$. In general the phase has a regular and a singular part, $\nabla\Psi=\psi(\nabla\delta\theta+\nabla\times\zeta\mathbf{e}_z)$, where finally $\zeta=Q\log\vert\mathbf{r}-\mathbf{r}_0\vert$. The difference in the CP beam system lies in the existence of two beam fields (flavors) \emph{and} the non-constant amplitude field $\psi_\pm(r)$, so the vortex looks like
\be
\label{psivort}\Psi_{0\pm}(\mathbf{r})=\psi_{0\pm}(r)e^{\imath\delta\theta_\pm(\phi)+\imath\theta_{0\pm}(\phi)}.
\ee
Inserting this solution into the equations of motion (or, equivalently, the Lagrangian) it is just a matter of algebra to obtain the vortex Hamiltonian, analogous to the well-known one but with two components (flavors) and their interaction. We refer the reader to the Appendix \ref{app00} for the full derivation. The outcome is perhaps expected: we get the straightforward generalization of the familiar Coulomb gas picture for the XY model where all interactions of different flavors -- $F$-$F$, $B$-$B$ and $F$-$B$ are allowed. In order to write the Hamiltonian (and further manipulations with it) in a concise way, it is handy to introduce shorthand notation $\vec{Q}\equiv (Q_+,Q_-)$, $\vec{Q}_1\cdot\vec{Q_2}\equiv Q_{1+}Q_{2+}+Q_{1-}Q_{2-}$, and $\vec{Q}_1\times\vec{Q}_2\equiv Q_{1+}Q_{2-}+Q_{1-}Q_{2+}$. For the self-interaction within a vortex $\vec{Q}_1$, we have $\vec{Q}_1\cdot\vec{Q}_1=Q_{1+}^2+Q_{1-}^2$ but $\vec{Q}_1\times\vec{Q}_1\equiv Q_{1+}Q_{1-}$ (i.e. there is a factor of $2$ mismatch with the case of two different vortices). Now for vortices at locations $\mathbf{r}_i,i=1,\ldots, N$ with charges $\lbrace Q_{i+},Q_{i-}\rbrace$ we get:
\be
\label{hamvort}\mathcal{H}_\mathrm{vort}=\sum_{i<j}\left(g\vec{Q}_i\cdot\vec{Q}_j+g'\vec{Q}_i\times\vec{Q}_j\right)\log r_{ij}+\sum_i\left(g_0\vec{Q}_i\cdot\vec{Q}_i+g_1\vec{Q}_i\times\vec{Q}_i\right).
\ee
The meaning of the Hamiltonian (\ref{hamvort}) is obvious. The first term is the Coulomb interaction of vortices; notice that only like-flavored charges interact through this term (because the kinetic term $\vert\nabla\Psi\vert^2$ is homogenous quadratic). The second term is the forward-backward interaction, also with Coulomb-like (logarithmic) radial dependence. This interaction comes from the mixing of the $F$- and $B$-modes in the fourth term in Eq.~(\ref{psivort2}), and it is generated, as we commented in Appendix \ref{app00}, when the amplitude fluctuations $\delta\psi_\alpha(r)$, which couple linearly to the phase fluctuations, are integrated out. In a system without amplitude fluctuations, i.e. classical spin system, this term would not be generated. The third and fourth terms constitute the energy of the vortex core. The self-interaction constants $g_0,g_1$ are of course dependent on the vortex core size and behave roughly as $g\log a/\epsilon,g'\log a/\epsilon$, where $\epsilon$ is the UV cutoff. The final results will not depend on $\epsilon$, as expected, since $g_0,g_1$ can be absorbed in the fugacity $y$ (see the next subsection). Expressions for the coupling constants in terms of original parameters are given in (\ref{gmicro}).

In three space dimensions, vortices necessitate the introduction of a gauge field \cite{kleinert} which, in multi-component systems, also acquires the additional flavor index \cite{multisc,multi15}. In our case, there is no emergent gauge field and the whole calculation is a rather basic exercise at the textbook level but the results are still interesting in the context of nonlinear optics and analogies to magnetic systems: they imply that the \emph{phase} structure (vortex dynamics) can be spotted by looking at the \emph{intensity} patterns (light intensity $I$ or local magnetization $\mathcal{M}$, see the penultimate section).

\subsection{The phase diagram}

\subsubsection{The mean-field theory for vortices}

The phases of the system can be classified at the mean field level, following e.g. \cite{parisi,kleinert}. In order to do that, one should construct the partition function, assuming that well-defined time-independent configuration space exists. We have already mentioned the question of equilibration, and address it in detail in Appendix \ref{appa}. Knowing that the system reaches equilibrium (in some part of the parameter space), we can count the ways in which a system of vortices can be placed in the crystal -- this is by definition the partition function $\mathcal{Z}$. First, the number of vortices $N$ can be anything from $0$ to infinity, second, the vortex charges can be arbitrary, and finally the number of ways to place each vortex in the crystal is simply the total surface section of the crystal divided by the size of the vortex. Then, each vortex carries a Gibbs weight proportional to the energy, i.e. the vortex Hamiltonian (\ref{hamvort}) for a single vortex.\footnote{Again, this is not generally true for out-of-equilibrium configurations but if the system reaches equilibrium, i.e., stable fixed point, this follows by usual statistical mechanics reasoning.} Let us focus first on a single vortex. If the vortex core has linear dimension $a$ and the crystal cross section linear dimension $\Lambda$, the vortex can be placed in any of the $(\Lambda/a)^2$ cells (and in the mean-field approach we suppose the vortex survives all the way along the crystal, from $z=0$ to $z=L$, so there is no additional freedom of placing it along some subinterval of $z$). This gives
\be
\label{zdef}\mathcal{Z}=\sum_{Q_+,Q_-}\left(\frac{\Lambda}{a}\right)^2e^{-L\mathcal{H}_1}=\sum_{Q_+,Q_-}e^{2\log\frac{\Lambda}{a}-L\left(g\vec{Q}\cdot\vec{Q}+g'\vec{Q}\times\vec{Q}\right)\log\frac{\Lambda}{a}}
\ee
Remember that $\mathcal{H}$ is energy density along the $z$ axis, so it appears multiplied by $L$. The factor $\log(\Lambda/a)$ in the second term of the exponent comes from the Coulomb potential of a single vortex (in a plane of size $\Lambda$). The exponent can be written as $-L\mathcal{F}^{(1)}$, with $\mathcal{F}^{(1)}=\mathcal{H}_1-(1/L)S_1$, recovering the relation between the free energy $\mathcal{F}^{(1)}$ and entropy $S^{(1)}$ of a single vortex. The entropy comes from the number of ways to place a vortex of core size $a$ in the plane of size $\Lambda\gg a$: $S\sim\log(\Lambda/a)^2$. Suppose for now that elementary excitations have $\vert Q_\pm\vert\leq 1$, as higher values increase the energy but not the entropy, so they are unlikely (when only a single vortex is present). Now we can consider the case of single-charge vortices with possible charges $(1,0),(-1,0),(0,1),(0,-1)$, and the case of two-charge vortices where $F$- and $B$-charge may be of the same sign or of the opposite sign, $(1,1),(-1,-1),(1,-1),(-1,1)$:
\bea
\label{vort0}\mathcal{F}_0^{(1)}=\left(g-\frac{2}{L}\right)\log\frac{\Lambda}{a},~~~\vec{Q}=(\pm 1,0)~\textrm{or}~\vec{Q}=(0, \pm 1)\\
\label{vort1}\mathcal{F}_1^{(1)}=\left(2g-g'-\frac{2}{L}\right)\log\frac{\Lambda}{a},~~~(Q_+,Q_-)=(\pm 1,\mp 1)\\
\label{vort2}\mathcal{F}_2^{(1)}=\left(2g+g'-\frac{2}{L}\right)\log\frac{\Lambda}{a},~~~(Q_+,Q_-)=(\pm 1,\pm 1).
\eea
Now we identify four regimes, assuming that $g,g'>0$:\footnote{One specificity of multi-component vortices is that the coupling constants may be negative, as can be seen from (\ref{gmicro}). In that case the ordering of the four regimes (how they follow each other upon dialing $L$) changes but the overall structure remains.}
\begin{enumerate}
\item{For $L>2/g$ a vortex always has positive free energy so vortices are unstable like in the low-temperature phase of the textbook BKT system. This is the vortex-free phase where the phase $U(1)_F\otimes U(1)_B$ does not wind. This phase we logically call \emph{vortex insulator} in analogy with the single-flavor case.}
\item{For $2/g>L>1/(g-g'/2)$ a double-flavor vortex always has positive free energy but single-flavor vortices are stable; in other words, there is proliferation of vortices of the form $\vec{Q}=(Q_+,0)$ or $\vec{Q}=(0,Q_-)$. This phase is like the conductor phase in a single-component XY model, and the topological excitations exist for the reduced symmetry group, i.e. for a single $U(1)$. We thus call it \emph{vortex conductor}; it is populated mainly by single-flavor vortices $(Q,0)$, $(0,Q)$.}
\item{For $1/(g-g'/2)>L>1/(g+g'/2)$ double-vortex formation is only optimal if the vortex has $Q_++Q_-=0$ which corresponds to the topological excitations of the diagonal $U(1)_d$ symmetry subgroup, the reduction of the total phase symmetry to the special case $(\theta_F,\theta_B)\mapsto (\theta_F+\delta\theta,\theta_B-\delta\theta)$. In other words, vortices of the form $(Q_+,-Q_+)$ proliferate. Here higher charge vortices may be more energetically favorable than unit-charge ones, contrary to the initial simplistic assumption, the reason being that the vortex core energy proportional to $gQ_+^2$ may be more than balanced out by the intra-vortex interaction proportional to $-g'Q_+^2$ (depending on the ratio of $g$ and $g'$). This further means that there may be multiple ground states of equal energy (frustration). We thus call this case \emph{frustrated vortex insulator} (FI); it is populated primarily with vortices of charge $(Q,-Q)$.}
\item{For $1/(g+g'/2)>L$ vortex formation always reduces the free energy, no matter what the relation between $Q_+$ and $Q_-$ is, and each phase can wind separately: $(\theta_F,\theta_B)\mapsto (\theta_F+\delta\theta_F,\theta_B+\delta\theta_B)$. Vortices of both flavors proliferate freely at no energy cost and for that reason we call this phase \emph{vortex perfect conductor} (PC). We deliberately avoid the term superconductor to avoid the (wrong) association of this phase with the vortex lines and type I/type II superconductors familiar from the 3D vortex systems: remember there is no emergent gauge field for the vortices in two spatial dimensions, and we only have perfect conductivity in the sense of zero resistance for transporting the (topological) charge, but no superconductivity in the sense of breaking a gauge symmetry.}
\end{enumerate}
A more systematic mean-field calculation will give the phase diagram also for an arbitrary number of vortices. This is not so interesting as it already does not require much less work than the RG analysis, which is more rigorous and more accurate for this problem. For completeness, we give the multi-vortex mean-field calculation in Appendix \ref{appmf}.

One might worry that the our whole approach approach misses the CP geometry of the problem, i.e. the fact that the $\Psi_+$ field has a source at $z=0$ and the $\Psi_-$ field at $z=L$. In Appendix \ref{appc} we show that nothing is missed at the level of approximations taken in this paper, i.e. mean-field theory in this subsubsection and the lowest-order perturbative RG in the next one. Roughly speaking, it is because the sources are irrelevant in the RG sense -- the bulk configuration dominates over the boundary terms. The Appendix states this in much more precise language.

\subsubsection{RG analysis}

We have classified the symmetries and thus the phases of our system at the mean-field level. To describe quantitatively the borders between the phases and the phase diagram, we will perform the renormalization group (RG) analysis. Here we follow closely the calculation for conventional vortex systems \cite{kleinert}. We consider the fluctuation of the partition function $\delta\mathcal{Z}$ upon the formation of a virtual vortex pair at positions $\mathbf{r}_1,\mathbf{r}_2$ with charges $\vec{q},-\vec{q}$, (with $\mathbf{r}_1+\mathbf{r}_2=2\mathbf{r}$ and $\mathbf{r}_1-\mathbf{r}_2=\mathbf{r}_{12}$), in the background of a vortex pair at positions $\mathbf{R}_1,\mathbf{R}_2$ (with $\mathbf{R}_1+\mathbf{R}_2=2\mathbf{R}$ and $\mathbf{R}_1-\mathbf{R}_2=\mathbf{R}_{12}$) with charges $\vec{Q}_1,\vec{Q}_2$. This is a straightforward but lengthy calculation and we state just the main steps. First, it is easy to show that the creation of single-charge vortices is irrelevant for the RG flow so we disregard it. Also, we can replace the core self-interaction constants $g_{0,1}$ with the fugacity parameter defined as $y\equiv\exp\left[-\beta\left(g_0+g_1\right)\log\epsilon\right]$. Here we introduce the notation $\beta\equiv L$ in analogy with the inverse temperature $\beta$ in standard statistical mechanics, in order to facilitate the comparison with the literature on vortices in spin systems, and also with antiferromagnetic systems in Section V.\footnote{Of course, the physical meaning of $\beta$ in our system is very different: we have no thermodynamic temperature or thermal noise, and the third law of thermodynamics is not satisfied for the "temperature" $1/\beta=1/L$. We merely use the $\beta$-notation to emphasize the similarity between free energies of different systems, not as a complete physical analogy.}

\begin{figure}\centering
\includegraphics[width=150mm]{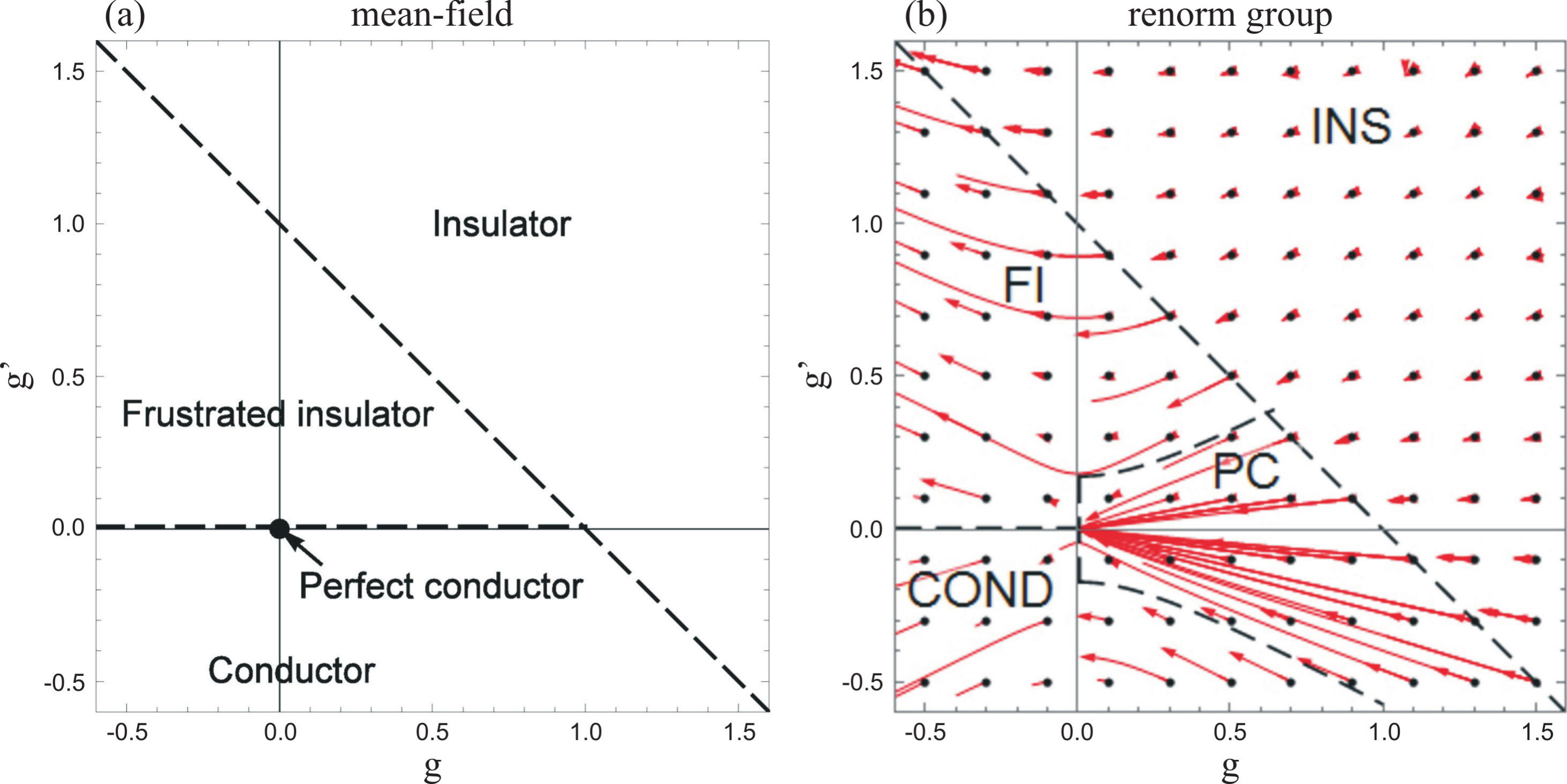}
\caption{\label{figphdiag} Phase diagram for the clean system in the
$g$-$g'$ plane, at the mean-field level (a) and with RG flows (b).
We show the flows for a grid of initial points, denoted by black
dots; red lines are the flows. Four phases exist, whose boundaries
are delineated by black dashed lines. In the mean-field calculation
(a) all phase boundaries are analytical. In the RG calculation, the
straight line $g+g'=1$ is obtained analytically whereas the other
phase boundaries can only be found by numerical integration of the
flow equations (\ref{rgfloweqs}). The flows going to infinity are
the artifacts of the perturbative RG; they probably correspond to
finite values which are beyond the scope of our analytical approach.
Notice how the flows in the $g+g'>1$ phase all terminate at
different values.}
\end{figure}

Now from the vortex Hamiltonian $\mathcal{H}_\mathrm{vort}$ the fluctuation equals (at the quadratic order in $y$ and $r$):
\be
\label{zfluc}\frac{\delta\mathcal{Z}}{\mathcal{Z}}=1+\frac{y^4}{4}\sum_{q_\pm}\int dr_{12}r_{12}^3e^{g\vec{q}\cdot\vec{q}+g'\vec{q}\times\vec{q}}\left[\int drr^2\left(g\vec{Q}_1\cdot\vec{q}+g'\vec{Q}_1\times\vec{q}\right)\nabla\log\vert\mathbf{R}_1-\mathbf{r}\vert+\left(g\vec{Q}_2\cdot\vec{q}+g'\vec{Q}_2\times\vec{q}\right)\nabla\log\vert\mathbf{R}_2-\mathbf{r}\vert\right]^2.
\ee
Notice that $\nabla$ is taken with respect to $\mathbf{r}$. The above result is obtained by expanding the Coulomb potential in $r_{12}$ (the separation between the virtual vortices being small because of their mutual interaction) and then expanding the whole partition function (i.e. the exponent in it) in $y$ around the equilibrium value $\mathcal{Z}$. The term depending on the separation $r_{12}$ is the mutual interaction energy of the virtual charges, and the subsequent term proportional to $r^2$ is the interaction of the virtual vortices with the external ones (the term linear in $r$ cancels out due to isotropy). Then by partial integration and summation over $q_\pm\in\lbrace 1,-1\rbrace$ we find
\bea
\nonumber\frac{\delta\mathcal{Z}}{\mathcal{Z}}=1+y^4\left(8\pi g^2\vec{Q}_1\cdot\vec{Q}_2+8\pi\left(g'\right)^2\vec{Q}_1\cdot\vec{Q}_2+16\pi gg'\vec{Q}_1\times\vec{Q}_2\right)I_3\log R_{12}+\\
\label{zfluc2}+y^4\left(4\pi g\left(g+g'\right)\left(\vec{Q}_1\times\vec{Q}_1+\vec{Q}_2\times\vec{Q}_2\right)I_1+8\left(g'\right)^2I_1\right)\log\epsilon,
\eea
with $I_n=\int_{\epsilon a}^{\Lambda a}drr^{n+g+g'}$. Now, taking into account the definition of the fugacity $y$, rescaling $\Lambda\mapsto\Lambda (1+\ell)$, performing the spatial integrals, and expanding over $\ell$, we can equate the bare quantities $g,g',y$ in (\ref{hamvort}) with their corrected values in $\mathcal{Z}+\delta\mathcal{Z}$ to obtain the RG flow equations:
\bea
\nonumber\frac{\partial g}{\partial\ell}=-16\pi (g^2+g'^2)y^4\\
\nonumber\frac{\partial g'}{\partial\ell}=-2\pi gg'y^4\\
\label{rgfloweqs}\frac{\partial y}{\partial\ell}=2\pi(1-g-g')y.
\eea
Now let us consider the fixed points of the flow equations. If one puts $g'=0$, they look very much like the textbook XY model RG flow, except that the fugacity enters as $y^4$ instead of $y^2$ (simply because every vortex contributes two charges). They yield the same phases as the mean-field approach as it has to be, but now we can numerically integrate the flow equations to find exact phase borders. The fugacity $y$ can flow to zero (meaning that the vortex creation is suppressed and the vortices tend to bind) or to infinity, meaning that vortices can exist at finite density. At $y=0$ there is a fixed line $g+g'=1$. This line is attracting for the half-plane $g+g'>1$; otherwise, it is repelling. There are three more attraction regions when $g+g'<1$. First, there is the point $y\to\infty,g=g'=0$ which has no analog in single-component vortex systems. Then, there are two regions when $g\to\infty$ and $g'\to\pm\infty$ (and again $y\to\infty$). Of course, the large $g,g'$ regime is strongly interacting and the perturbation theory eventually breaks down, so in reality the coupling constants grow to some finite values $g_*,g'_*$ and $g_{**},g'_{**}$ rather than to infinities. The situation is now the following:
\begin{enumerate}
\item{The attraction region of the fixed line is the vortex insulator phase: the creation rate of the vortices is suppressed to zero.}
\item{The zero-coupling fixed point attracts the trajectories in the vortex perfect conductor phase: only the fugacity controls the vortices and arbitrary charge configurations can form. Numerical integration shows that this point also has a finite extent in the parameter space.}
\item{In the attraction region of the fixed point with $g_*<0$ and $g'_*>0$ (formally they flow to $-\infty$ and $+\infty$, respectively), same-sign $F$- and $B$-charges attract each other and those with the opposite sign which repel each other. This is the frustrated insulator.}
\item{The fixed point with $g_{**},g'_{**}<0$ (formally both flow to $-\infty$) corresponds to the conductor phase.}
\end{enumerate}
The RG flows in the $g-g'$ plane are given in Fig.~\ref{figphdiag}. Full RG calculation is given in panel (b); for comparison, we include also the mean-field phase diagram (following from the previous subsubsection and Appendix \ref{appmf}) in the panel (a). In the half-plane $g+g'>1$ every point evolves toward a different, finite point $(g,g')$ in the same half-plane. In the other half-plane we see the regions of points moving toward the origin or toward one of the two directions at infinity. The PC phase (the attraction region of the point $(0,0)$) could not be obtained from the mean field calculation (i.e. it corresponds to the single point at the origin at the mean field level).

It may be surprising that the coupling constants can be negative, with like charges repelling and opposite charges attracting each other. However, this is perfectly allowed in our system. In the usual XY model, the stiffness is proportional to the kinetic energy coefficient and thus has to be positive. Here, the coupling between the fluctuations of $F$- and $B$-beams introduces other contributions to $g,g'$ and the resulting expressions (\ref{gmicro}) give bare values of $g,g'$ that can be negative, and the stability analysis of the RG flow clearly shows that for nonzero $g'$, the flow can go toward negative values even if starting from a positive value in some parameter range. If we fix $g'=0$, the flow equations reproduce the ones from the single-component XY model, and the phase diagram is reduced to just the $g'=0$ line. If we additionally suppose that the bare value of $g$ is non-negative, than we are on the positive $g'=0$ semiaxis in the phase diagram -- here we see only two phases, insulator (no vortices, $g\to\mathrm{const.}$) and perfect conductor ($g\to 0$). However, for $g'$ fixed to zero (that is, with a single flavor only), the perfect conductor reduces to the usual conductor phase of the single-component XY model -- in other words, we reproduce the expected behavior.

Physically, it is preferable to give the phase diagram in terms of the quantities $\Gamma,\tau,I,I_x,L$ that appear in the initial equations of motion (\ref{prpsieq}-\ref{preeq}): the light intensities can be directly measured and controlled, whereas the relaxation time and the coupling cannot, but at least they have a clear physical interpretation. The relations between these and the effective Hamiltonian quantities $y,g,g'$ are found upon integrating out the intensity fluctuations to obtain (\ref{hamvort}) and the explicit relations are stated in (\ref{gmicro}). Making use of these we can easily plot the phase diagram in terms of the physical quantities for comparison with experiment. However, for the qualitative understanding we want to develop here it is much more convenient to use $g,g'$ as the phase structure is much simpler.

\begin{figure}\centering
\includegraphics[width=70mm]{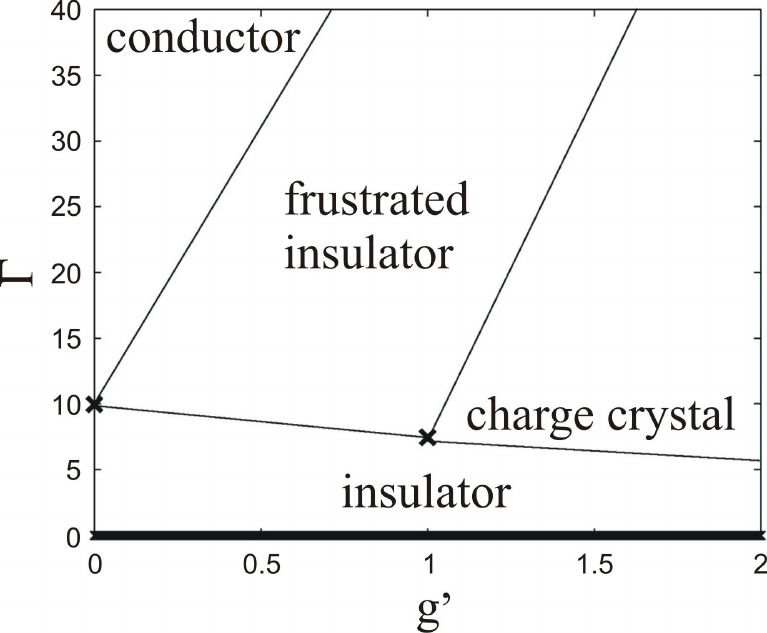}
\caption{\label{figphdiag1} Typical phase diagram for the system
without disorder, in the $\Gamma-g'$ plane. There are two discrete
fixed points and the critical line at $\Gamma=0$, which corresponds
to the critical line $g+g'=1$ in the previous figure. We also see
two discrete fixed points, corresponding to $g_{*,**},g'_{*,**}$.
The advantage of physical parameters is that the location of these
fixed points in the $\Gamma-I$ plane can be calculated directly from
the numerics (or measured from the experiment).}
\end{figure}

As an example, we plot the $\Gamma$-$g'$ diagram in Fig.~\ref{figphdiag1} (we have kept $g'$ to keep the picture more informative; the $\Gamma$-$L$ and $\Gamma$-$I$ diagrams contain multiple disconnected regions for each phase). The non-interacting fixed point $g=g'=0$ is now mapped to $\Gamma=0$. The tricritical point where the PC, the FI and the conductor phases meet is at $R=1$. Therefore, the rule of thumb is that low couplings $\Gamma$ produce stable vortices with conserved charges -- the perfect vortex conductor. Increasing the coupling pumps the instability up, and the kind of instability (and the resulting phase) is determined by the relative strength of the photonic lattice compared to the propagating beams. Obviously, such considerations are only a rule of thumb and detailed structure of the diagram is more complex. This is one of the main motives of this study -- blind numerical search for patterns without the theoretical approach adopted here would require many runs of the numerics for a good understanding of different phases.

\subsection{Geometry of patterns}

Now we discuss what the intensity pattern $I(\mathbf{r})$ looks like in various phases, for various boundary conditions. This is very important as this is the only thing which can be easily measured in experiment -- phases $\theta_\alpha$ are not directly observable, while the intensity distribution is the direct outcome of the imaging of the crystal \cite{prlpet}. We shall consider three situations. The first is a single Gaussian beam on zero background ($I_x=0$), with Gaussian initial intensity profile $\vert F(z=0,\mathbf{r})\vert^2=\vert B(z=L,\mathbf{r})\vert^2=\mathcal{N}\exp(-r^2/2s^2)$ and possibly non-zero vortex charges: $\mathrm{arg}\Psi_\pm(\mathbf{r})\sim\exp\left(Q_{F,B}\phi\right)$, with $\mathbf{r}=(r\cos\phi,r\sin\phi)$. The second case is a quadratic vortex lattice of $F$- and $B$-beams, so the initial beam intensity is $I_0=\sum_{i,j}\exp\left(-\left(x-x_i\right)^2/2s_0^2-\left(y-y_i\right)/2s_0^2\right)$), with $x_{i+1}-x_i=y_{i+1}-y_i\equiv b=\mathrm{const.}$, the situation particularly relevant for analogies with condensed matter systems. In the third case we have again a quadratic vortex lattice but now on top of the background photonic square lattice, which is either coincident or off-phase (shifted for half a lattice spacing) with the beam lattice. The background intensity is thus of the form $I_x=\sum_{i,j}\exp\left(-\left(x-x_i\right)^2/2s^2-\left(y-y_i\right)/2s^2\right)$.

First of all, it is important to notice that there are two kinds of instabilities that can arise in a vortex beam:\footnote{They are distinct from the bifurcations which happen also in topologically trivial beam patterns and lead to the instability which eventually destroys optical (non-topological) solitons. These instabilities have been analyzed in the Appendix \ref{appb} and in more detail in \cite{stabanal}, where the authors have found them to start from the edge of the beam and result in the classical "walk through the dictionary of patterns".}
\begin{itemize}
\item{There is an instability which originates in the disbalance between the diffusion and self-focusing (crystal response) in favor of diffusion in \emph{high-gradient regions}: if a pattern $I(x,y)$ has a large gradient $\nabla I$, the kinetic term in the Lagrangian (\ref{psilag}), i.e. the diffusion term in (\ref{prpsieq}) is large and the crystal charge response is not fast enough to balance it as we travel along the $z$-axis, so the intensity rapidly dissipates and the pattern changes. Obviously, the vortex core is a high-gradient region so we expect it to be vulnerable to this kind of instability. This is indeed the case: in the center of the vortex the intensity diminishes, a dark region forms and the intensity moves toward the edges. We dub this the core or central instability (CI), and in the effective theory it can be understood as the decay of states with low fugacity $y$, i.e. high self-interaction constants $g_0,g_1$. This instability prevents the formation of vortices in the insulator phase, or limits it in the frustrated insulator and conductor phases.}
\item{There is an instability stemming from the dominance of diffusion over self-focusing in \emph{low-intensity regions of sufficient size and/or convenient geometry}. At low intensity, the charge response is nearly proportional to $I$ (from Eq.~\ref{preeq}), so if $I$ is small diffusion wins and the intensity dissipates. If there is sufficient inflow of intensity from more strongly illuminated regions, it may eventually balance the diffusion; but if the pattern has a long "boundary", i.e. outer region of low intensity, it will not happen and the pattern will dissipate out, or reshape itself to reduce the low-intensity region. We call this case the edge instability (EI). For a vortex, it happens when the positive and negative vortex charges tend to redistribute due to Coulomb attraction and repulsion. In our field theory Hamiltonian (\ref{hamvort}), this instability dominates in the conductor and perfect conductor phases.}
\end{itemize}

\begin{figure}\centering
\includegraphics[width=70mm]{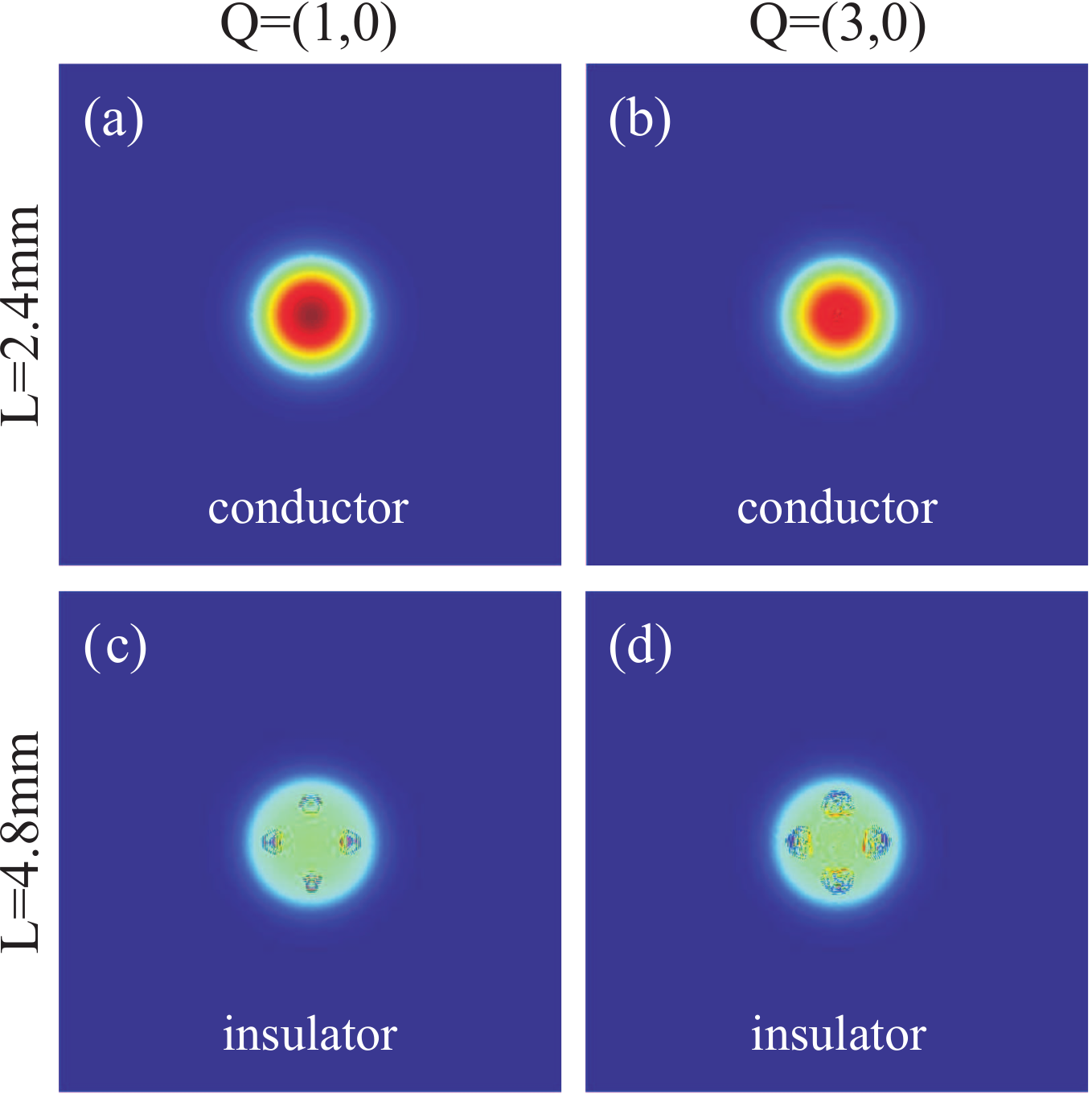}
\caption{\label{figcleangauss1} Transverse profiles for a single
Gaussian beam for two different propagation distances,
$L=2.4\mathrm{mm}$ (top) and $L=4.8\mathrm{mm}$ (bottom), with
vortex charges $(1,0)$ (a, c) and $(3,0)$ (b, d), at the back face
of the crystal ($z=L$). The regime on top (a, b) corresponds to the
conductor phase, which has a single conserved vortex charge $Q_F$.
This vortex charge conservation prevents significant instabilities;
nevertheless, the multi-quantum vortex $(3,0)$ shows the onset of CI
- notice the reduced intensity and incoherent distribution of the
beam in the central region in the top right panel (the CI is
expected to grow roughly as $Q_+^2+Q_-^2$). The insulator phase only
preserves the $F-B$ invariance but not the vortex charge and in the
absence of topological protection the vortices can annihilate into
the vacuum -- here we see the EI taking over for both charges - four
unstable regions appear near the boundary, violating the circular
symmetry and dissipating away the intensity of the vortex. Parameter
values: FWHM $40\mu\mathrm{m}$, $\Gamma I_0=41$, $t=10\tau$.}
\end{figure}

\begin{figure}\centering
\includegraphics[width=70mm]{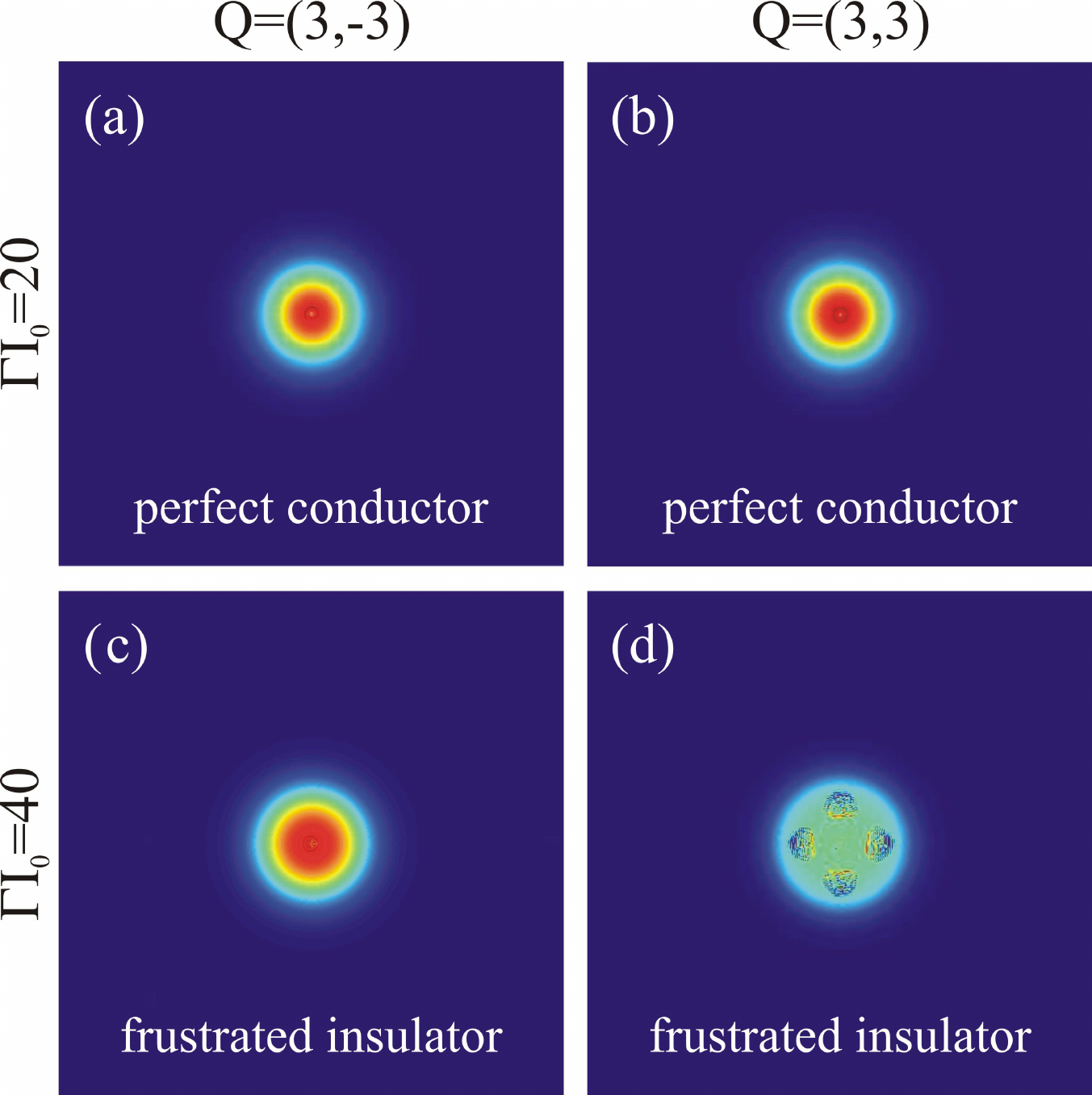}
\caption{\label{figcleangauss2} Transverse profiles for a single
Gaussian beam for two different coupling strengths, $\Gamma I_0=20$
(top) and $\Gamma I_0=40$ (botom), with vortex charges $(3,-3)$ (a,
c) and $(3,3)$ (b, d) at the back face of the crystal ($z=L$). The
regime on top corresponds to the perfect conductor phase, where the
vortices of all charges freely proliferate -- both vortices are
reasonably stable. The bottom case is in the frustrated insulator
phase -- the forward-backward coupling makes the $(3,3)$ vortex
unstable from EI while the $(3,-3)$ vortex survives. Parameter
values: FWHM $40\mu\mathrm{m}$, $L=2\mathrm{mm}$, $t=10\tau$.}
\end{figure}

Let us first show how the CI and EI work for a single beam with nonzero vortex charge. In Fig.~\ref{figcleangauss1} we show the intensity patterns for a single vortex with charges $(1,0)$ and $(3,0)$ as the $x-y$ cross sections (transverse profiles) in the middle of the crystal, i.e. for $z=L/2$. The parameters chosen ($\Gamma,I_0,R,L$) correspond to the conductor phase (top) and the insulator phase (bottom). In top panels, for $Q_+^2+Q_-^2=1$ the core energy is not so large and CI is almost invisible. For $Q_+^2+Q_-^2=9$ we see the incoherence and the dissipation in the core region, signifying the CI. The conductor phase allows the proliferation of vortices but only those with $\vert Q_\pm\vert\le 1$ are stable. In the bottom panels, both vortices have almost dissipated away due to EI, which starts from discrete poles near the boundary.\footnote{As a rule, it follows the sequence (\ref{symmseq}) found in the Appendix \ref{appb} from the pole structure of the propagator, though some of the steps can be absent, e.g. for a single Gaussian vortex there is no $\mathbb{C}_2$ stage.} Indeed, the insulator phase has no free vortices, no matter what the charge. In Fig.~\ref{figcleangauss2} we see no instability even for a high-charge vortex in the perfect conductor phase (top), whereas the frustrated insulator phase (bottom) shows strong EI for the like-charged vortex $(3,3)$ since this fixed point has $g'_*>0$, but the $(3,-3)$ vortex is stable. Notice that we could not expect CI for this case since the sum $Q_+^2+Q_-^2=9$ is the same in both cases -- if for $Q_-=-Q_+$ the vortex has no CI, then for $Q_-=Q_+$ it cannot have it either (since the value $Q_+^2+Q_-^2$ is the same).

We have thus seen what patterns to expect from CI and EI, and also what kind of stable vortices to expect in different phases: \emph{the perfect conductor phase allows free proliferation of vortices of any charge, the conductor phase allows only single-quantum vortices (or vortices with sufficiently low $Q_+^2+Q_-^2$) while others dissipate from CI, the frustrated insulator supports the vortices with favorable charges (or favorable charge distribution in multiple-vortex systems) while others disintegrate from EI, and the insulator phase supports no vortices -- they all dissipate from CI or EI, whichever settles first (depending on the vortex charges)}.

\begin{figure}\centering
\includegraphics[width=150mm]{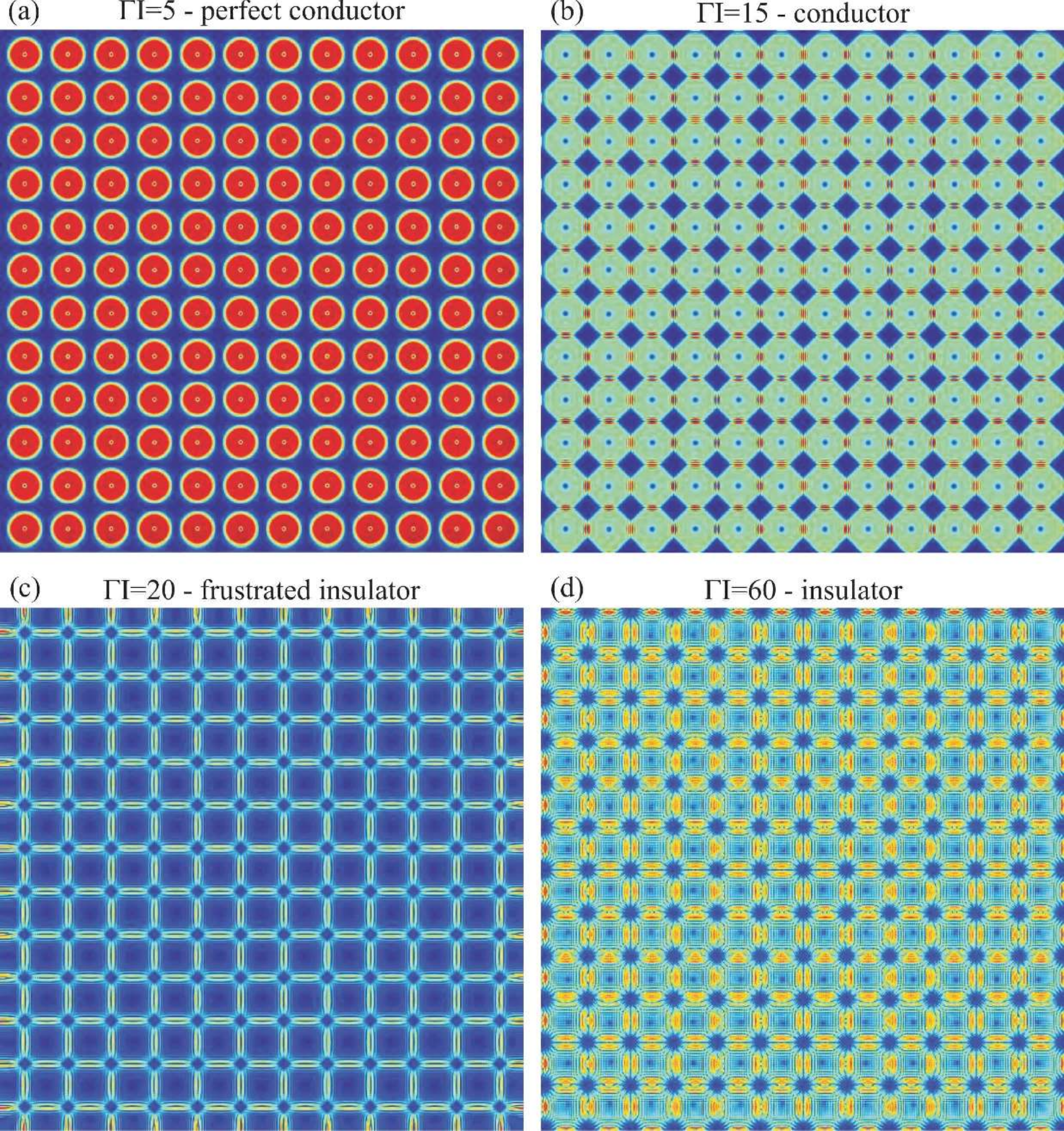}
\caption{\label{figcleanlatt} Vortex lattice with Gaussian profile
for $\Gamma I=5$ (PC, panel a), $\Gamma I=15$ (conductor, panel b),
$\Gamma I=20$ (FI, panel c), and $\Gamma I=60$ (insulator, panel d).
The perfect conductor phase has a coherent vortex lattice and no
instabilities. Conductor exhibits a deformation of the vortex
lattice and the reduction of the full $O(2)$ symmetry, starting from
the \emph{center}, whereas the FI exhibits the reduction of symmetry
and the inversion of the lattice due to \emph{edge} effects. Notice
how both phases have reduced symmetry compared to PC but retain
coherence. Only the insulator phase looses not only symmetry but
also coherence, i.e. the intensity diffuses and the pattern is
smeared out. Transverse size of the lattice is $512\times 512$ in
computational space; same lattice size, FWHM and lattice spacing are
used for all subsequent figures unless specified otherwise.
Parameter values: $L=4.8\mathrm{mm}$, $t=10\tau$, FWHM
$10\mu\mathrm{m}$ and lattice spacing equal to FWHM.}
\end{figure}

\begin{figure}\centering
\includegraphics[width=150mm]{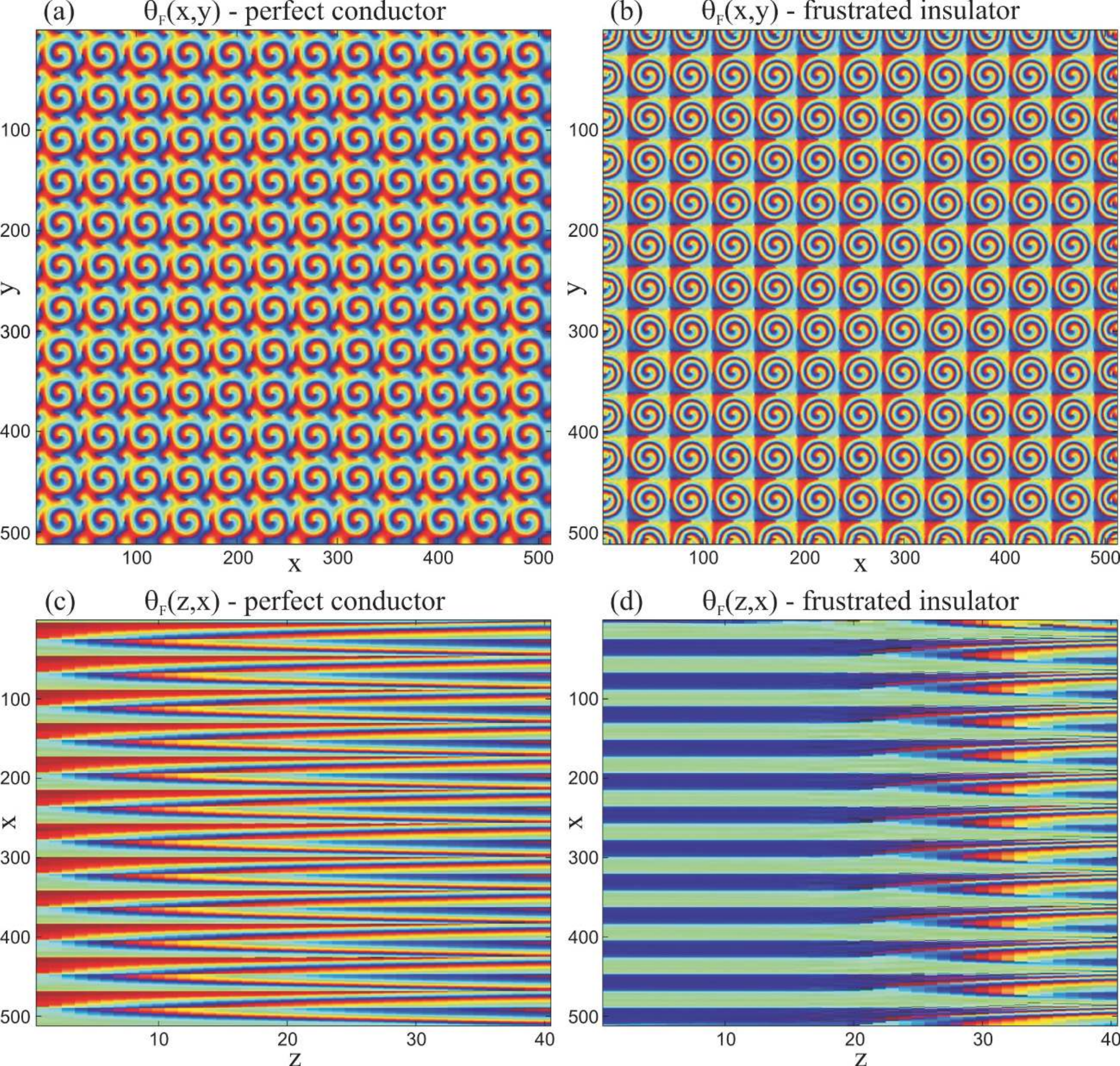}
\caption{\label{figcleanlattph} Same system as in (a) and (c) panels
from the previous figure (PC and FI phases) but now we plot the
phase $\theta_F$, as the transverse cross-section
$\theta_F(x,y;z=L/2)$ (a, b) and as the longitudinal section along
the PR crystal $\theta(x,z;y=320\mu\mathrm{m})$ (c, d). The perfect
conductor phase has well-defined vortices in contact which allows
the transport of the vortex charge through the lattice, and shows as
the periodical modulation of the phase along the $z$-axis (vortex
lines). The frustrated insulator keeps well-defined vorticity even
though the intensity map undergoes inversion
(Fig.~\ref{figcleanlatt}(c)) with frozen phase along the $z$-axis,
so there is no vorticity transport until some $z\approx
L/2=2.4\mathrm{mm}$, when the phase stripes develop into vortex
lines. The unit on the $x$- and $y$-axis is $1\mu\textrm{m}$ ($1$
in computational space) and on the $z$-axis $0.12\mathrm{mm}$ ($120$ in
computational space).}
\end{figure}

The case rich with analogies with condensed matter systems is the square vortex lattice on the background photonic square lattice, Fig.~\ref{figcleanlatt}. Here we can also appreciate the transport processes. The photonic lattice is coincident with the beam lattice and equal in intensity, so $\Gamma (I_0+I_x)=2\Gamma I_0$. In the perfect conductor phase (a) the vortices are stable and coherent and keep the uniform lattice structure. In the conductor phase (b) the CI is visible but the lattice structure survives. The bottom panels show the non-conducting phases, frustrated insulator (c) and insulator (d). Insulator looses both lattice periodicity and the Gaussian profile of the vortices but the frustrated insulator keeps regular structure: from EI the intensity is \emph{inverted} and the resulting lattice is \emph{dual} to the original one (compare (c) to (a)). The phase patterns $\theta_F(x,y;z=L/2)$ and $\theta_F(x,z;y=320\mu\mathrm{m})$ for the perfect conductor (top) and the frustrated insulator phase (bottom) are shown in the Fig.~\ref{figcleanlattph}. Here we see the vortex charge transport mechanism in a PC: the vortices are connected in the sense that the phase $\theta_F$ is coherently traveling from one vortex to the next. In the FI phase, the phase is initially frozen along the $z$-axis, until the transport starts at some $z\approx L/2$.

\begin{figure}\centering
\includegraphics[width=150mm]{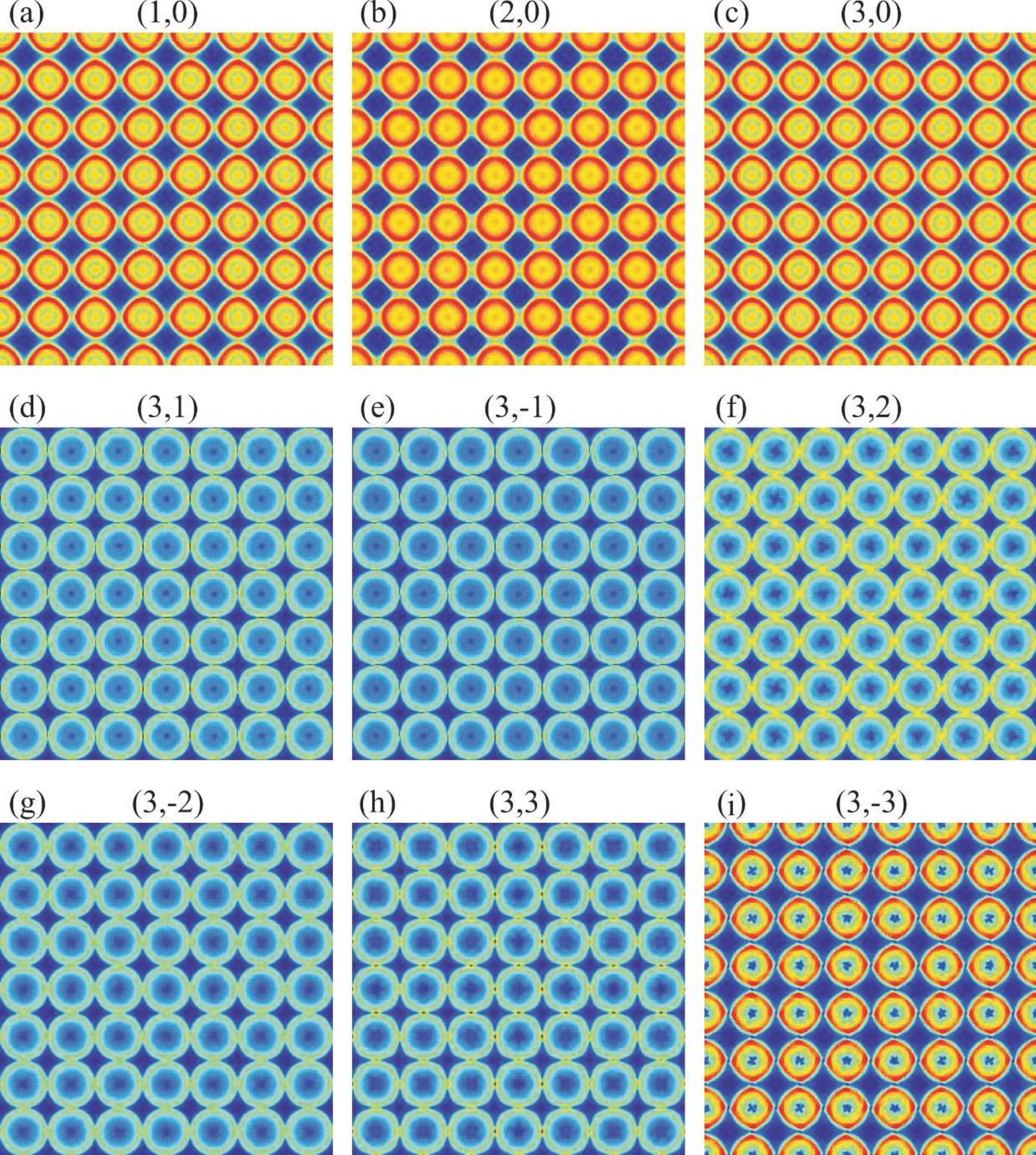}
\caption{\label{figcleanlatt3} Transverse profiles for vortex
lattices with different charges in the FI phase. In the first row
(a-c) we see how the CI gets stronger and stronger as the total
vortex core energy grow (with the square of the total charge). The
second and third row show the growth of CI from $(3,0)$ to $(3,\pm
3)$ (notice the increasingly reduced intensity in the center and the
strong ring-like structure of the beams) but also the
forward-backward interaction which favors the configurations
$(3,-3),(3,-2),(3,-1)$ over $(3,3),(3,2),(3,1)$. In particular, the
$(3,-3)$ lattice is the optimal configuration of all $(3,Q_-)$
configurations even though it has greater CI than say $(3,0)$
(notice the small dark regions in the center), because the
$\sum_{ij}gg'Q_{i+}Q_{i-}\log r_{ij}$ term minimizes the EI --
notice there is no "spilling" of intensity from one vortex to the
next. The parameters are $\Gamma I=20,L=2.5\mathrm{mm}$.}
\end{figure}

It may be instructive to take a closer look at the lattice dynamics of the most interesting phase: the frustrated insulator. In Fig.~\ref{figcleanlatt3} we inspect square lattices on the photonic lattice background for several charges of the form $(Q_+=3,Q_-)$. The first row shows how the vortices loose stability and develop CI as the total square of the charge grows (from (a) to (c)). The panels ((d)-(i)) show how the $g'$ coupling favors the opposite sign of $Q_+$ and $Q_-$ and how the optimal configuration is found for $Q_-=-3$. This is easily seen by minimizing the free energy over $Q_-$: it leads to the conclusion that the forward-backward coupling favors the "antiferromagnetic" ordering in the sense that $Q_++Q_-=0$.

\begin{figure}\centering
\includegraphics[width=150mm]{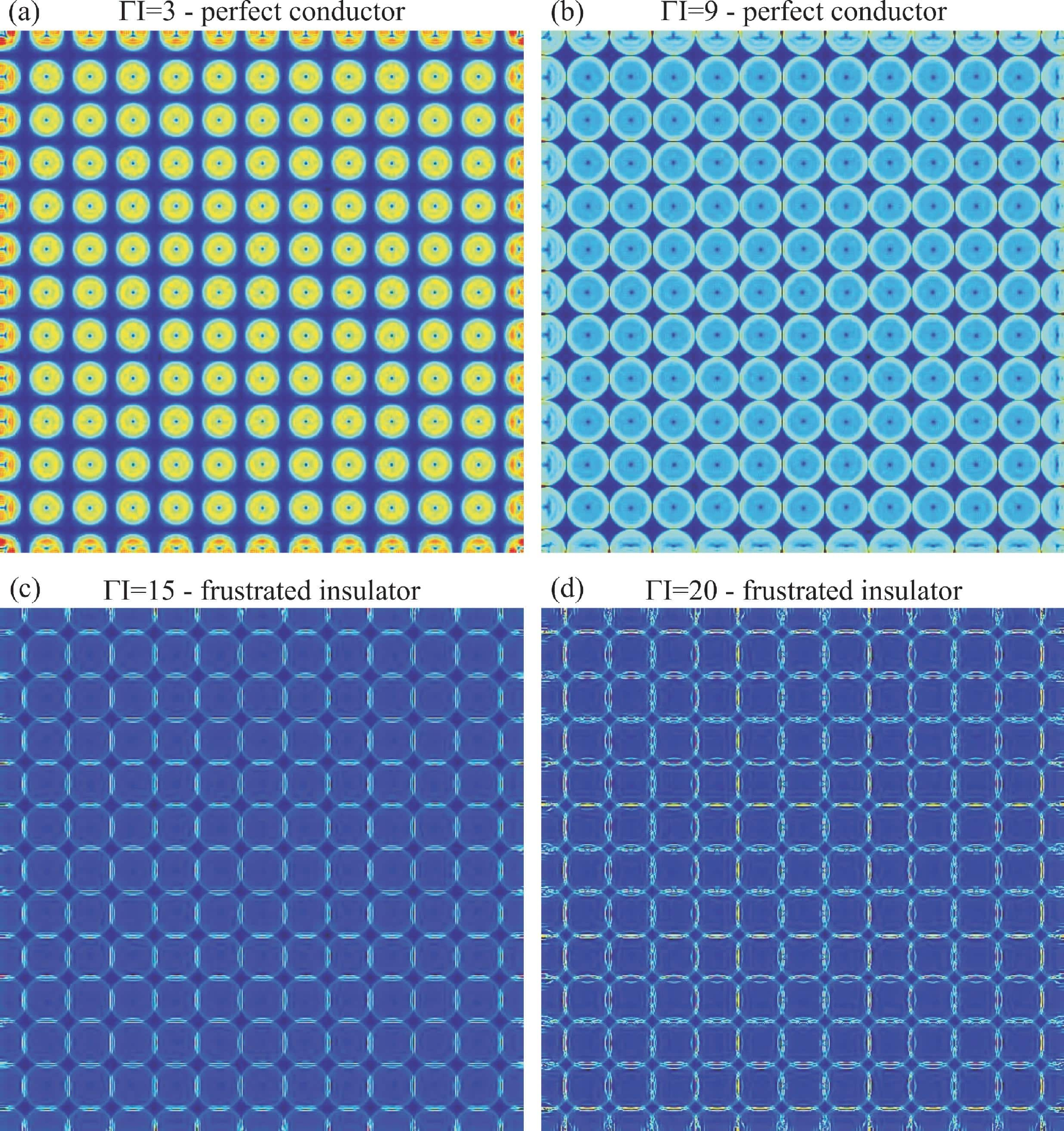}
\caption{\label{figcleangausslatt1} Intensity maps for the quadratic
vortex lattice with charges $(1,1)$, for increasing values of
$\Gamma I=\Gamma (I_0+I_x)$. The transition from the PC phase (a, b)
into the FI phase (c, d) happens at about $\Gamma I\approx 12$. The
edge instability sets in progressively, in accordance with what we
saw in the previous figure, leading eventually to an inverse square
lattice. Propagation length $L=5\mathrm{mm}$.}
\end{figure}

\begin{figure}\centering
\includegraphics[width=150mm]{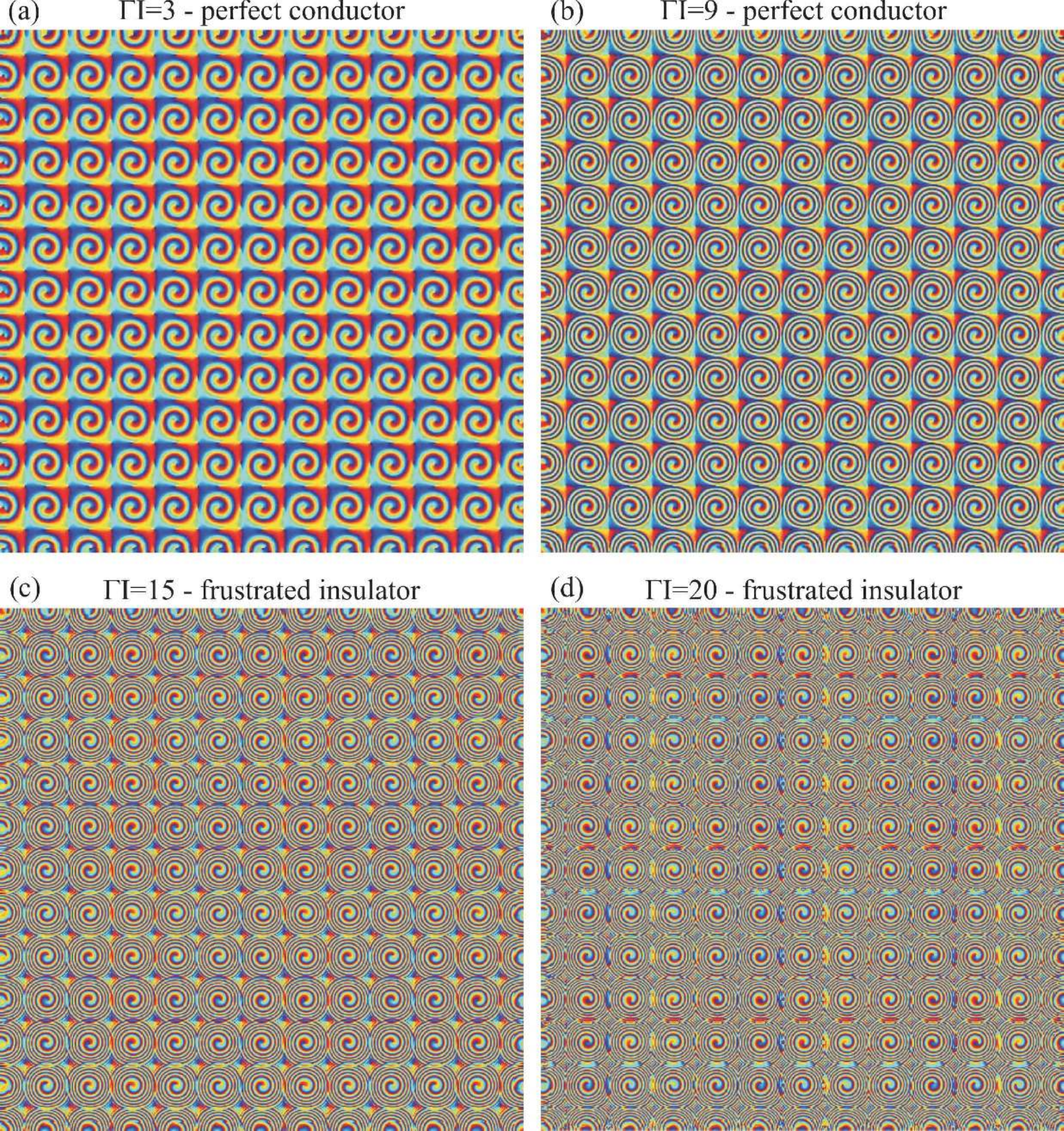}
\caption{\label{figcleangausslatt1ph} Transverse phase maps for the
$F$-beam for the same cases as in Fig.~\ref{figcleangausslatt1}. As
the coupling strength $\Gamma I$ grows toward very large values (d),
the violation of translation symmetry becomes obvious: notice the
vertical and horizontal phase stripes. This instability gives rise
to the charge density wave ordering in the presence of disorder.}
\end{figure}

Finally, it is interesting to see how the FI phase at high
intensities and coupling strengths contains a seed of translation
symmetry breaking which will become important in the presence of
disorder. In
Figs.~\ref{figcleangausslatt1}-\ref{figcleangausslatt1ph} we give
intensity and phase transverse profiles across the PC-FI transition
and deep into the FI phase at large couplings. The intensity maps
show the familiar inverse square lattice but the phase maps show
stripe-like ordering, i.e. translation symmetry breaking along one
direction in Figs.~\ref{figcleangausslatt1ph}(c) and
\ref{figcleangausslatt1ph}(d) -- horizontal and vertical lines with
a repeating constant value of the phase $\theta_F$ on all lattice
cells along the line. This is a new instability, distinct from CI
and EI. We cannot easily derive this instability from the
perturbation theory in the Appendix \ref{appb} as it is a collective
phenomenon and cannot be understood from a single beam.

\section{The system with disorder}

Consider now the same system in the presence of quenched disorder.
This is a physically realistic situation: the disorder corresponds
to the holes in the photonic lattice which are caused by the defects
in the material. The defects are in fixed positions, i.e. they are
quenched, whereas the beam is dynamical and can fluctuate. Now
$I_x(\mathbf{r})\to I_x(\mathbf{r})+I_h(\mathbf{r})$, i.~e., the
quenched random part $I_h(\mathbf{r})$ is superimposed to the
regular background (whose intensity is $I_x$). The disorder is given
by some probability distribution, assuming no correlations between
defects at different places. As in the disorder-free case, the
lattice is static and "hard", i.~e., does not backreact due to the
presence of the beams. One should however bear in mind that the
backreaction on the background lattice can sometimes be important as
disregarding it violates the conservation of the angular momentum
\cite{gaussrotpet}. Disregarding the backreaction becomes exact when
$I_x+I_h\gg\vert\Psi\vert^2$, i.e. when the background irradiation
is much stronger than the propagating beams.

To treat the disorder we use the well-known replica formalism
\cite{cardy}. For vortex-free configurations, typical experimental
values of the parameters suggest that the influence of disorder is
small \cite{prlpet, prapet, vortlat2pet}. However, the influence of
disorder becomes dramatic when vortices are present. This is
expected, since holes in the lattice can change the topology of the
phase field $\theta_\pm$ (the phase now must wind around the holes).
Our equations of motion are still given by the Lagrangian
(\ref{psilag}), but with $I_x\mapsto I_x+I_h$. In our analytical
calculations, we assume that a defect in the photonic lattice
changes the lattice intensity from $I_x$ to $I_x+I_h$, with Gaussian
distribution of "holes" in $I_h$, which translates to the
approximately Gaussian distribution of the couplings $g,g',g_0,g_1$.
In the numerics however, we do a further simplification and model
the defects in a discrete way, i.e. at a given spot either there is
a lattice cell of intensity $I_1$ (with probability $h$), or there
isn't (the intensity is zero, with probability $1-h$). This
corresponds to $I_x=I_1/2,I_h=\pm I_1/2$ so the disorder is
discrete. Due to the central limit theorem, we expect that the
Gaussian analytics should be applicable to our numerics.

\subsection{The replica formalism at the mean-field level}

To study the system with quenched disorder in the photonic lattice,
we need to perform the replica calculation of the free energy of the
vortex Hamiltonian (\ref{hamvort}). We refer the reader to the
literature \cite{parisi,spinglass} for an in-depth explanation of
the replica trick. In short, one needs to average over the various
realizations of the disorder \emph{prior} to calculating the
partition function, i.e. prior to averaging over the dynamical
degrees of freedom (vortices in our case). This means that we need
to perform the disorder-average of the free energy, i.e. the
logarithm of the original partition function $-\log\mathcal{Z}$, and
not the partition function $\mathcal{Z}$ itself. The final twist is
the identity $\log\mathcal{Z}=\lim_{n\to
0}\left(\mathcal{Z}^n-1\right)/n$: we study the Hamiltonian
consisting of $n$ copies (replicas) of the original system and then
\emph{carefully} take the $n\to 0$ limit.\footnote{Care is needed as
the $n\to 0$ limit does not in general commute with the
thermodynamic limit.} The partition function of the replicated
Hamiltonian reads
\be \mathcal{Z}=\lim_{n\to
0}\mathrm{Tr}\exp\left(-\sum_{\mu=1}^n\mathcal{H}_\mathrm{vort}\left(Q^{(\mu)}\right)\right),
\ee
where $Q^{(\mu)}$ are the vortex charges in the $\mu$-th replica
of the system. In the original Hamiltonian (\ref{hamvort}), the
disorder turns the interaction constants into quenched random
quantities $g_{ij},g'_{ij},g_{0;ij},g_{1;ij}$, so we can compactly
write our interaction term as
\be
\label{intterm}\mathcal{H}_\mathrm{vort}=\sum_{ij}\sum_{\alpha\beta}Q_{i\alpha}J_{ij}^{\alpha\beta}Q_{j\beta}
\ee
with $J_{ij}^{++}=J_{ij}^{--}=g_{ij}(1-\delta_{ij})\log
r_{ij}+g_0\delta_{ij}$,
$J_{ij}^{+-}=J_{ij}^{-+}=g'_{ij}(1-\delta_{ij})\log
r_{ij}+g_1\delta_{ij}$. Now we again make the mean-field
approximation for the long-ranged logarithmic interaction. Similar
to the clean case, for $i\neq j$ we approximate $g\log r_{ij}\sim
g'\log r_{ij}\sim\log\Lambda$, knowing that $g,g'\sim 1$ and
assuming that average inter-vortex distance is of the same order of
magnitude as the system size $\Lambda$, and for the core energy we
likewise get $g_0,g_1\sim\log a/\epsilon\sim
-\log\epsilon\sim\log\Lambda$. The result is that all terms in
$J_{ij}^{\alpha\beta}$, both for $i\neq j$ and $i=j$, are on average
of the order $\log\Lambda\gg 1$, and the mean-field approach is
justified. We will sometimes denote the $2\times 2$ matrices in the
flavor space by hats (e.g.~$\hat{J}=J^{\alpha\beta}$).

The final Hamiltonian (\ref{intterm}) has the form of the
random-coupling and random-field Ising-like model: random couplings
stem from the stochasticity of $J_{ij}$ values and random field from
the fact that $\langle J_{ij}\rangle\neq 0$ introduces terms linear
in $Q_{i\alpha}$, i.e. an effective external field coupling to the
"spins". We have arrived at this model through three steps of
simplification: our microscopic model is a type of the XY-glass
model (Cardy-Ostlund model \cite{cardyostlund}), a well-known toy
model for disorder. At this stage our model is similar to the work
of \cite{conti1,conti2}, only with two components instead of one.
Then we have written the effective vortex Hamiltonian with
Coulomb-like interaction, disregarding the topologically trivial
configurations. This is a rather extreme approximation but a
necessary one as it is very complicated to consider the full model
with vortices. Finally, we have approximated the logarithmic
potential with a constant all-to-all vortex coupling. Such an
approximation (essentially the infinite dimension limit) is
frequently taken and lies at the heart of the solvable
Sherington-Kirkpatrick Ising random coupling model \cite{parisi}.
Our case differs from the Sherington-Kirkpatrick model as it (i) has
also a random field (ii) has two flavors (iii) has the Ising spins
taking arbitrary integer values. From the random XY-model it differs
by (i) and (ii) above, and also by considering only vortices and no
non-topological spin configurations. The additional phases we get in
comparison to \cite{conti1,conti2} and its generalization in
\cite{conti3,conti4,conti5} come from the interactions between the
forward and backward flavor. But bearing in mind the drastic
approximations we take we stress that we cannot aspire to solve
neither the XY-model nor the resulting Ising-like model in any
rigorous way (certainly not at the level of rigor of mathematical
physics). We merely try to obtain a crude understanding.

The Gaussian distribution of defects reads $p(J^{\alpha\beta}_{ij})=\exp\left(-\left(J_{ij}^{\alpha\beta}-J^{\alpha\beta}_0\right)\left(\hat{\sigma}^{-2}\right)_{\alpha\beta}\left(J^{\alpha\beta}_{ij}-J^{\alpha\beta}_0\right)\right)$, where the second moments are contained in the matrix $\sigma_{\alpha\beta}$, with $\sigma_{+-}=\sigma_{-+}$. In this case we get the replicated partition function
\be
\label{rephamvort}\bar{\mathcal{Z}}^n=\int\mathcal{D}\left[Q^{(\mu)}_{i\alpha}\right]\int\mathcal{D}\left[J^{\alpha\beta}_{ij}\right]\exp\left[-\frac{1}{2}\sum_{i,j=1}^N\sum_{\alpha,\beta}\left(J_{ij}^{\alpha\beta}-J_0^{\alpha\beta}\right)\sigma_{\alpha\beta}^{-2}\left(J^{\alpha\beta}_{ij}-J_0^{\alpha\beta}\right)-\sum_{\mu=1}^n\sum_{i,j=1}^N\sum_{\alpha,\beta}\beta J_{ij}^{\alpha\beta}Q^{(\mu)}_{i\alpha}Q^{(\mu)}_{j\beta}\right].
\ee
We can now integrate out the couplings $J^{\alpha\beta}_{ij}$ in (\ref{rephamvort}) and get
\be
\label{rephamvort2}\bar{\mathcal{Z}}^n=\mathrm{const.}\times\int\mathcal{D}\left[Q^{(\mu)}_{i\alpha}\right]\exp\left[\frac{1}{2}\beta^2\sum_{\mu,\nu=1}^n\sum_{i,j=1}^N\sum_{\alpha,\beta}Q^{(\mu)}_{i\alpha}Q^{(\nu)}_{i\beta}\left(\hat{\sigma}^2\right)_{\alpha\beta}Q^{(\mu)}_{j\alpha}Q^{(\nu)}_{j\beta}-\beta\sum_{\mu=1}^n \sum_{i,j=1}^N\sum_{\alpha,\beta}J_0^{\alpha\beta}Q^{(\mu)}_{i\alpha}Q^{(\mu)}_{j\beta}\right].
\ee
Integrating out the disorder has generated the non-local quartic term proportional to the elements of $\sigma_{\alpha\beta}^2$. The additional scale given by the average disorder concentration means we cannot scale out $\beta=L$ anymore, and it becomes an additional independent parameter. The partition function can be rewritten in the following way, usual in the spin-glass literature \cite{cardy,spinglass}. We can introduce the non-local order parameter fields
\be
\label{glassop}p_\alpha^{(\mu)}=\frac{1}{N}\sum_{i=1}^NQ_{i\alpha}^{(\mu)},~~q_{\alpha\beta}^{(\mu\nu)}=\frac{1}{N}\sum_{i,j=1}^NQ_{i\alpha}^{(\mu)}Q_{j\beta}^{(\nu)},
\ee
which have the meaning of overlap between different metastable states. The rest is just algebra, although rather tedious: one rewrites the Hamiltonian in terms of new order parameters, and then one can solve the saddle point equations for $p_\alpha$ and $q_{\alpha\beta}$, or do an RG analysis. The calculation is found in Appendix \ref{appd}.

The mean-field analysis yields six phases:
\begin{enumerate}
\item{One phase violates both the replica symmetry and the flavor symmetry, breaking it down to identity. We dub this phase \emph{vortex charge density wave} (CDW), as it implies spatial modulation of the vortex charge, leading to nonzero net charge density $\sum_iQ_{i\alpha}^{(\mu)}$ in some parts of the system even if the boundary conditions are electrically neutral (the \emph{total} net charge density must still be zero due to charge conservation). Vortices take their charges from $\mathbb{Z}\otimes\mathbb{Z}$.}
\item{The second phase violates the replica symmetry in both flavors and reduces the flavor symmetry but does not break it down to identity. Instead, it reduces it to the diagonal subgroup $U(1)_F\otimes U(1)_B\to U(1)_d$, so it has nonzero density of the vortex charge in a given replica $\sum_iQ_{i+}^{(\mu)}=-\sum_iQ_{i-}^{(\mu)}$. Again, the charge density is locally nonzero but now with an additional constraint resulting in frustration (multiple equivalent free energy minima!). This is thus the dirty equivalent of the frustrated insulator phase and we dub it \emph{vortex glass}, as it has long-range correlations (because of the logarithmic interactions between charged areas), does not break spatial symmetry and exhibits frustration; its charges are from $\pi_1\left(U\left(1\right)_d\right)=\mathbb{Z}$.}
\item{The remaining phases have no nonzero vortex charge density fluctuation, and are similar to the phases in the clean system. Vortex perfect conductor violates the replica symmetry of all three fields $q^{++},q^{--},q^{+-}$ and allows free proliferation of vortices with charges $(Q_+,Q_-)\in\mathbb{Z}\otimes\mathbb{Z}$.}
\item{Frustrated vortex insulator preserves the replica symmetry of $q^{\pm\pm}$ but has non-zero value, with broken replica symmetry, of the mixed $q^{+-}$ field, which gives $U(1)_d$ vortices, with charges $Q_+=-Q_-\in\mathbb{Z}$.}
\item{Vortex conductor preserves the replica symmetry of the mixed $q^{+-}$ order parameter but violates it in $q^{\pm\pm}$, resulting in the proliferation of single-flavor vortices with $\mathbb{Z}$ charge.}
\item{Vortex insulator fully preserves the replica symmetry, all order parameters are zero and vortices cannot proliferate. RG analysis will show that insulator surivives only at zero disorder, otherwise it generically becomes CDW.}
\end{enumerate}

The phase diagram (given in Fig.~\ref{figphdiagdirty} in the next subsection) now contains six phases (only five are visible for the parameters chosen in the figure): CDW, insulator, FI, conductor, PC and the glassy phase. The insulator phase is now of measure zero in the $(g,g',\sigma^2)$ plane, existing only for the points at $\sigma^2=0$; for generic nonzero values we have a CDW. For simplicity, we have plotted the phase diagram for $\sigma_{++}^2=\sigma_{--}^2=\sigma_{+-}^2\equiv\sigma^2$.

\subsection{RG analysis and the phase diagram}

To study the RG flow, we can start from the replicated partition function (\ref{rephamvort2}), inserting the definition of the couplings $J_{ij}^{\alpha\beta}$ and keeping the vortex charges $Q^{(\mu)}_{i\alpha}$ as the degrees of freedom (without introducing the quantities $p^\mu_\alpha,q^{(\mu\nu)}_{\alpha\beta}$). The basic idea is the same: we consider the fluctuation $\delta\left(\bar{\mathcal{Z}}^n\right)$ upon the creation of a vortex pair at $\mathbf{r}_{1,2}$ with charges $\vec{q}^{(\mu)}_1,-\vec{q}^{(\mu)}_2$, in the background of the vortices $\vec{Q}^{(\nu)}_{1,2}$ at positions $\mathbf{R}_{1,2}$. Likewise, we introduce the fugacity parameter $y^{(\mu)}$ to account for the vortex core energy. However, this problem is much harder than the clean problem and one has to resort to many approximations to perform the calculation. In its most general form, the problem is still open, in the sense that all known solutions suppose a certain form of replica symmetry breaking or truncate the RG equations \cite{spinglass}. The RG analysis is thus less useful in the disordered case but at least the numerical integration of the flow equations is supposed to give a more precise rendering of the phase diagram compared to the mean field theory. We again describe the calculation in Appendix \ref{appd} and jump to the results.

\begin{figure}\centering
\includegraphics[width=150mm]{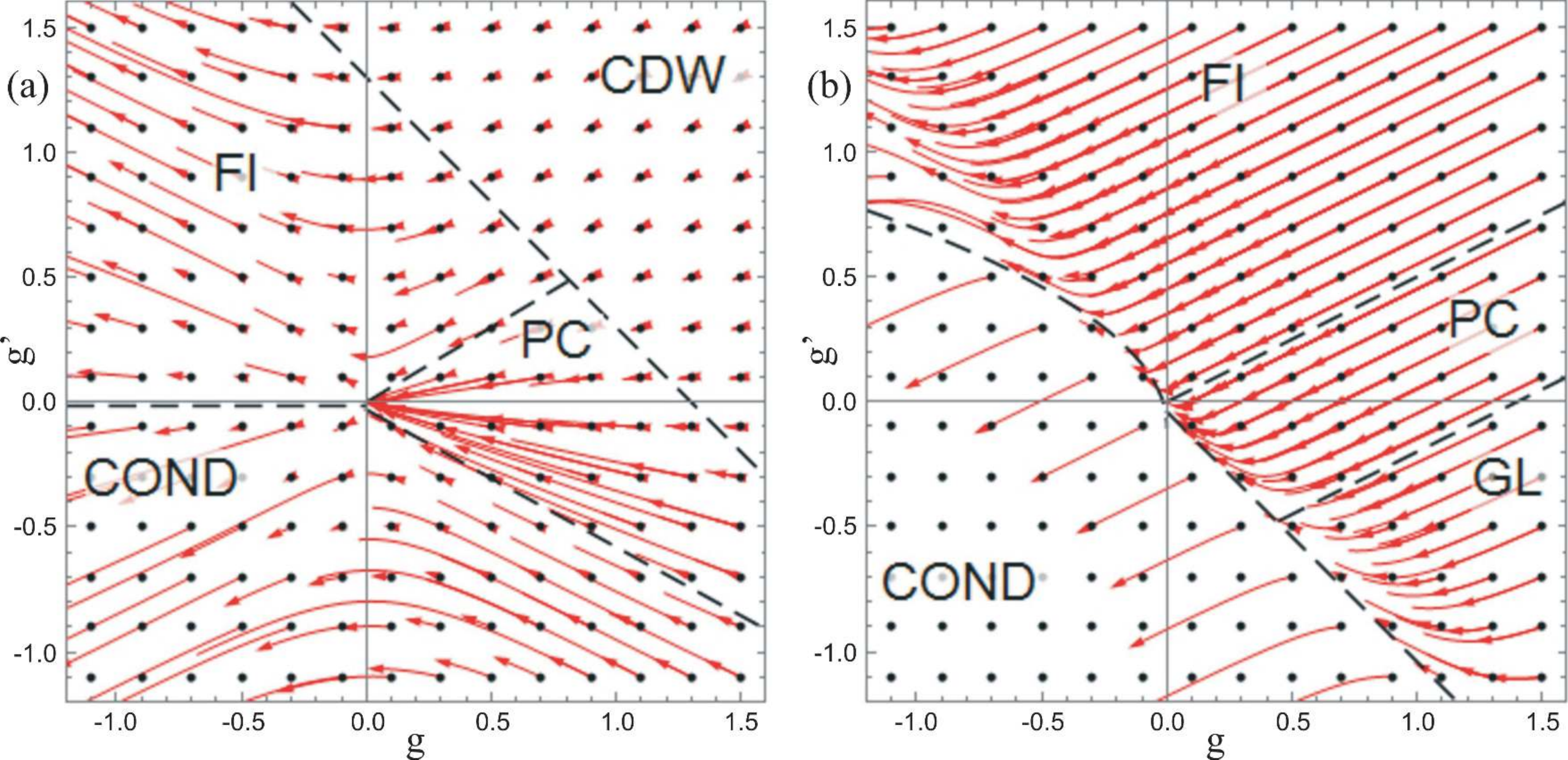}
\caption{\label{figphdiagdirty} Phase diagram for the system with
lattice disorder in the $g$-$g'$ plane together with RG flows, with
red lines denoting the flows starting at the initial conditions
denoted by black points. The dashed black lines are approximate
phase boundaries from mean-field theory, for $\sigma^2=0.4$ (a) and
$\sigma^2=1.2$ (b). In (a), the area where $g+g'+\beta^2\sigma^2>1$
is inhabited by the flows toward nonuniversal values of $(g,g')$
which belong to the CDW phase and the opposite region is divided
between the attraction regions of $(0,0)$,
$(g_*\to\infty,g'_*\to\infty)$ and
$(g_{**}\to\infty,g'_{**}\to-\infty)$ -- the familiar PC, FI and
conductor phases. In the panel (b), for $\sigma^2=1.2$, the disorder
becomes relevant in the glass phase (denoted by "GL"), whose RG
flows end on the half-line of fixed points
$g+g'+\beta^2\sigma^2=1,g'<0$. For our parameter values this line
happens to pass almost through the origin; in general this is not
necessarily the case. The non-disordered phases (flowing to
$\sigma^2=0$) -- FI, conductor and PC have survived. Propagating
length is $L=3.0\mathrm{mm}$.}
\end{figure}

The fixed point of the flow equations lies either at infinite $y$ or at $y=0$ like in the clean case. This is again controlled by the the equation for $\partial y/\partial\ell$ but now depending on the combination $g+g'+\beta^2\sigma^2$ instead of $g+g'$ in the clean case (for simplicity, we consider the case where $\sigma_{\alpha\beta}^2$ are all equal). The following cases appear:
\begin{enumerate}
\item{When the fugacity flows toward infinity, we reproduce the phases and the fixed point values $(g,g',\sigma^2)$ from the clean case: the PC flows toward $(0,0,0)$, the FI toward $(g_*,g'_*,0)$ and the conductor toward $(g_*,g'_{**},0)$ with $g_*\to -\infty,g'_*\to -\infty,g'_{**}\to\infty$. Notice that all these phases flow to $\sigma^2=0$, i.e. disorder is irrelevant.}
\item{When the fixed point lies at $y=0$, one possibility is that all parameters ($g,g',\sigma^2$) flow toward some non-universal nonzero values. The attraction region of this point is the CDW phase: the disorder term stays finite as well as the couplings. In particular, the points on the half-plane $g+g'>0,\sigma^2=0$ stay at $\sigma^2=0$ (with constant coupling values) and this is the insulator phase from the clean case. Notice that $\sigma^2>0$ now, i.e. disorder is relevant. For $\sigma^2<1$, this are the only fixed points when $y=0$.}
\item{However, for sufficiently strong disorder ($\sigma^2>1$), there is a new line of fixed points at $y=0$ with a finite attraction region, corresponding to a new phase. For $\beta>1$, the right-hand side of the second RG equation in (\ref{rgflowdirty}) has a zero at nonzero $g'$ and there are trajectories flowing toward $(y,g,g',\sigma^2)=\left(0,g,g'(g),\sigma^2(g)\right)$ and not toward an arbitrary nonuniversal value of $\sigma^2$. This is precisely the glass phase, where disorder is again relevant. At the lowest order, the relation between $g,g',\sigma^2$ at the fixed point line is given by the relation $g+g'+\beta^2\sigma^2=1$.}
\end{enumerate}
Now we have made contact between the mean-field classification of phases and the fixed points and regions of the RG flow. The flows in the $(g,g')$ plane are given in the Fig.~\ref{figphdiagdirty}. The parameter space is four-dimensional so the phase structure is different at different disorder concentrations $\sigma^2$. In the (A) panel for $\sigma^2=0.4$, the phase structure is similar to the clean case -- we see the same four phases except that insulator (no stable vortices) is replaced by the CDW phase with localized vortices. In the (B) panel, for $\sigma^2=1.2$, the CDW phase is replaced by another disordered phase, the glass-like regime. Importantly, the glass phase does not cross the $g'=0$ axis, meaning that a single-flavor system even with disorder could not support a glass. We thus conjecture that the transition at $\sigma^2=1$ is of first order, as the change is the structure of the $(g,g')$ phase diagram is discontinuous, and we do not see how this could happen if the first derivative $\partial\mathcal{F}/\partial\rho_\pm$ (the derivative of the free energy with respect to vortex charge density) is continuous. However, we have not checked the order of this transition by explicit calculation. The phase structure is further seen in the $\sigma^2-g'$ diagram, where we see the glass phase emerge at some value of the disorder. This is discussed further in the next section, where we study the equivalent antiferromagnetic system (with the same structure of the phase diagram, Fig.~\ref{figphdiagz2}).

\subsection{Geometry of patterns}

The two previously considered mechanisms of instability -- central instability and edge instability -- remain active also in the presence of disorder. However, in the presence of disorder there is a third, inherently collective effect that we dub \emph{domain instability} (DI). It follows from the fact that the self-focusing term $\Gamma E$ grows with intensity $I$: more illuminated regions react faster (Eqs.~\ref{prpsieq}-\ref{preeq}). In the presence of background lattice there will be regions of initially zero beam intensity $I_0$ where the regular lattice cells have some nonzero intensity $I_x$. Approximating $I=I_0+I_x\approx I_x=\mathrm{const.}$, our equations in the vicinity of the defect (hole) in the background lattice becomes the Schr\"odinger equation in a step potential (equal to $I_x$ in the regular parts of the photonic lattice, and equal to zero where a hole is found), so the $z$-dependent part of the solution is of the form $\sum_k e^{\imath\lambda_kz}$ and the eigenenergies along $z$ are gapped by the inverse length: $\lambda_k>1/L$. For small eigenenergies, the transmission coefficient is very low whereas for large energies it approaches unity. Thus for $1/L$ large (i.e., there are few $\lambda_k$'s which are larger than $1/L$) most of the intensity remains confined by the borders of the defect and the intensity does not spill but for small $1/L$ the beam profile is deformed by the "spilling" into the hole regions. For vortices, there is an additional Coulomb interaction in the $x-y$ plane, meaning the effective potential is not piecewise constant anymore (even in the simplest approximation) but the qualitative conclusion remains: large $L$ brings global reshaping of the intensity profile.

\begin{figure}
\includegraphics[width=120mm]{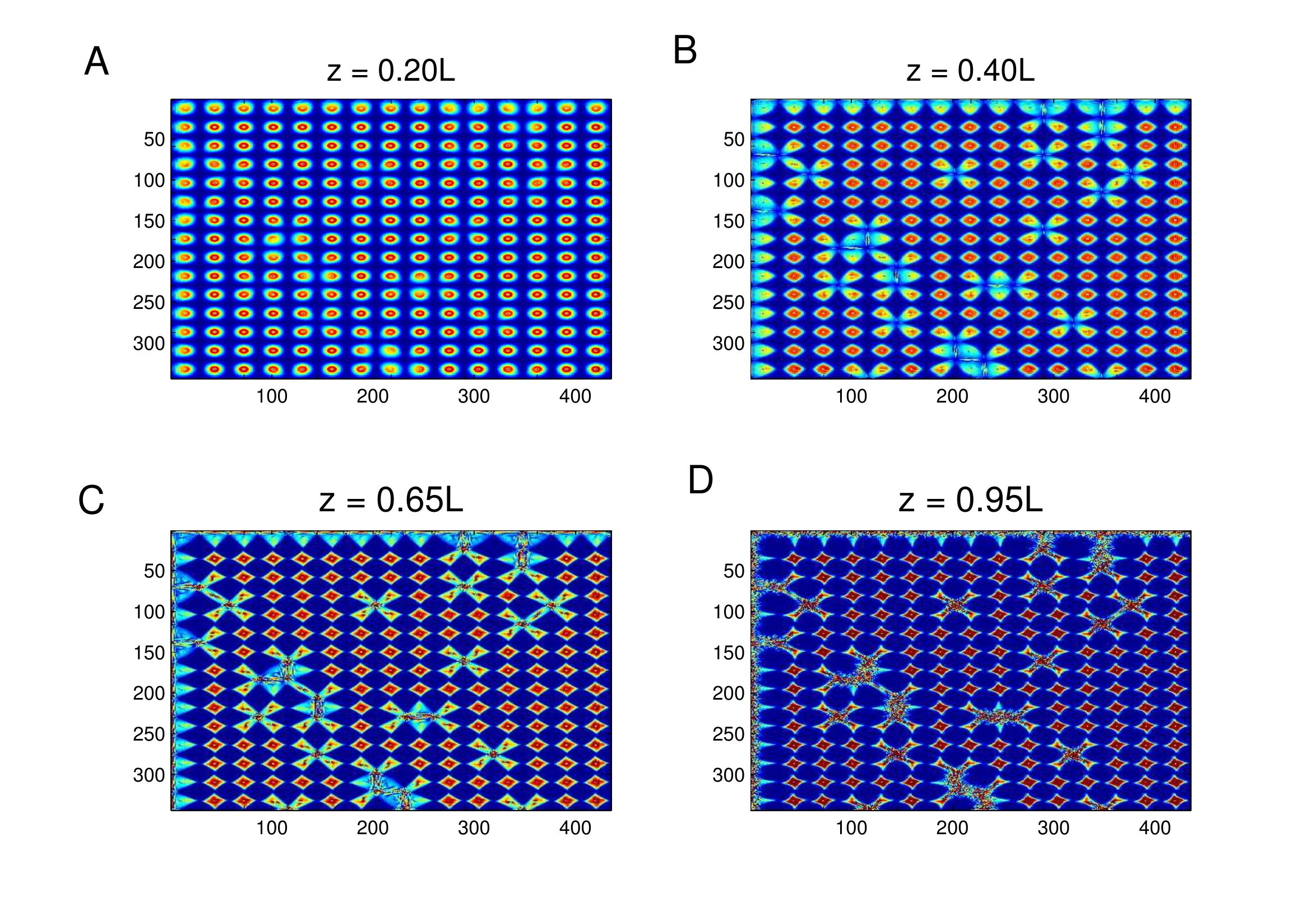}
\caption{\label{figdirtylatt1} Transverse profile for the PC phase
in a Gaussian beam lattice on a background lattice, for four
different propagation distances. The vortex charge is $(1,1)$, which
is sufficiently low that the CI does not destroy the vortices. We
see some CI-induced symmetry reduction from $O(2)$ to $C_4$ but the
overall lattice structure is preserved. Parameter values are:
$\sigma^2=0.1,\Gamma I=20,L=2\mathrm{mm}$, FWHM for the CP beams is
$9\mu\mathrm{m}$ and for the photonic lattice $6\mu\mathrm{m}$.}
\end{figure}

\begin{figure}
\includegraphics[width=120mm]{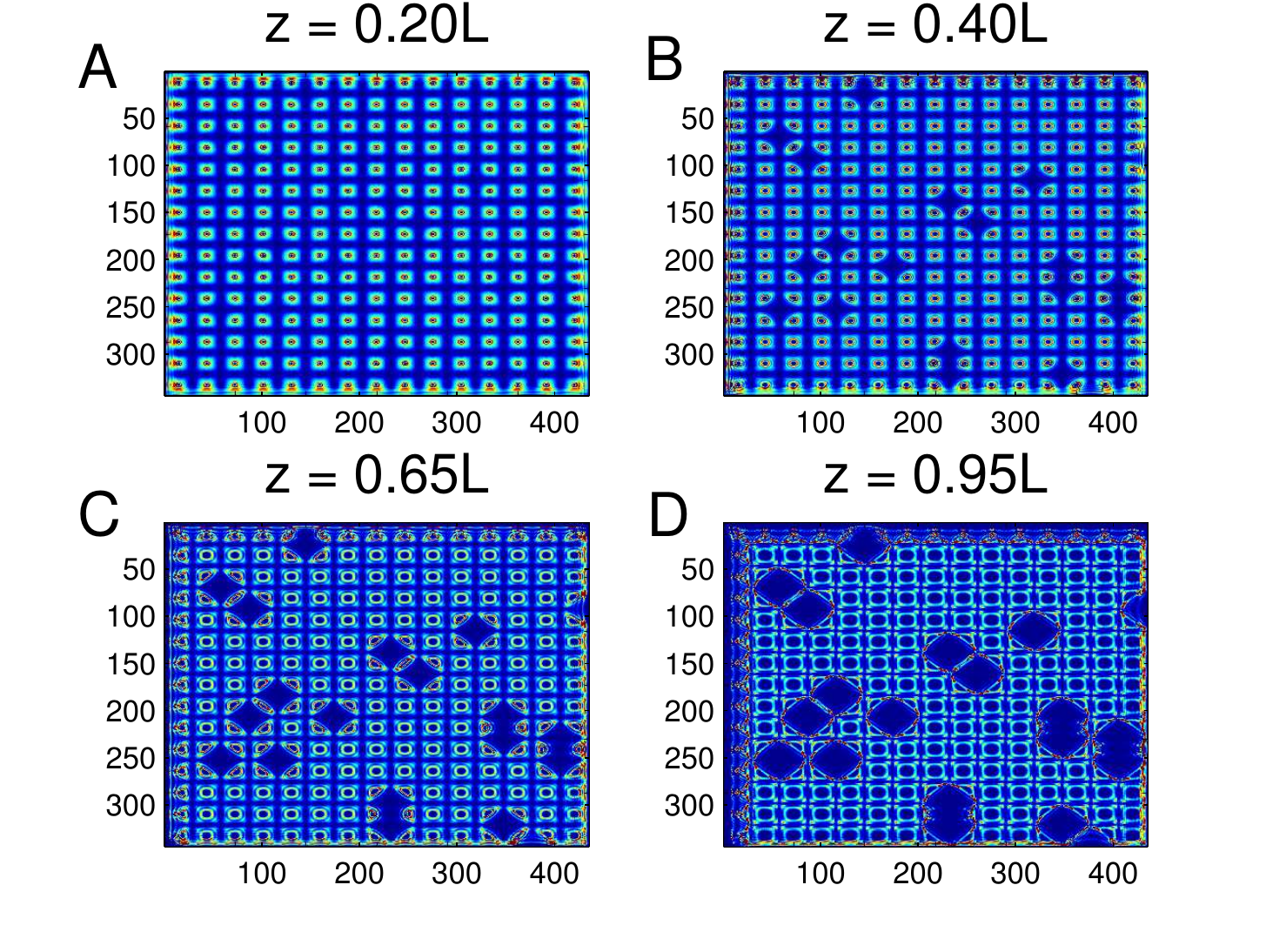}
\caption{\label{figdirtylatt2} Transverse profile for the FI phase,
present in the same system as in Fig.~\ref{figdirtylatt1} but for
$\Gamma I=40$. Now both the CI (low-intensity regions in the beam
center in a, b) and the EI (lattice inversion in c, d) are present.
The net result is the lattice inversion, and the vortex charge
dissipates along the inverse lattice.}
\end{figure}

The other phases are analogous to the ones in the clean case, though with a general trend that the presence of disorder decreases the stability of vortex patterns. The PC and FI phases are shown in Figs.~\ref{figdirtylatt1}-\ref{figdirtylatt2}. In this section we only look at the lattices, as the notion of disorder is inapplicable for a single beam. Consider first the patterns in the PC phase (Fig.~\ref{figdirtylatt1}). Compared to the clean case (Fig.~\ref{figcleanlatt}(a)), the symmetry is much reduced, from $O(2)$ to $C_4$ but the vortices are conserved and the original lattice structure (outside the holes) is clearly visible. The FI (Fig.~\ref{figdirtylatt2}) shows mainly EI (and to a smaller extent CI), which together lead to the lattice inversion. The rule of thumb for differentiating the conductor and PC on one side from the CDW and FI on the other side is precisely the presence of the lattice inversion. The absence of the charge transport is best appreciated in the phase images: the charge pins to the defects and localizes toward the end of the crystal (i.~e., for $z$ near $L$). Only near the edges we see high vorticity, somewhat analogous to topological insulators, which only have nonzero conductivity along the edges of the system.

\begin{figure}
\includegraphics[width=150mm]{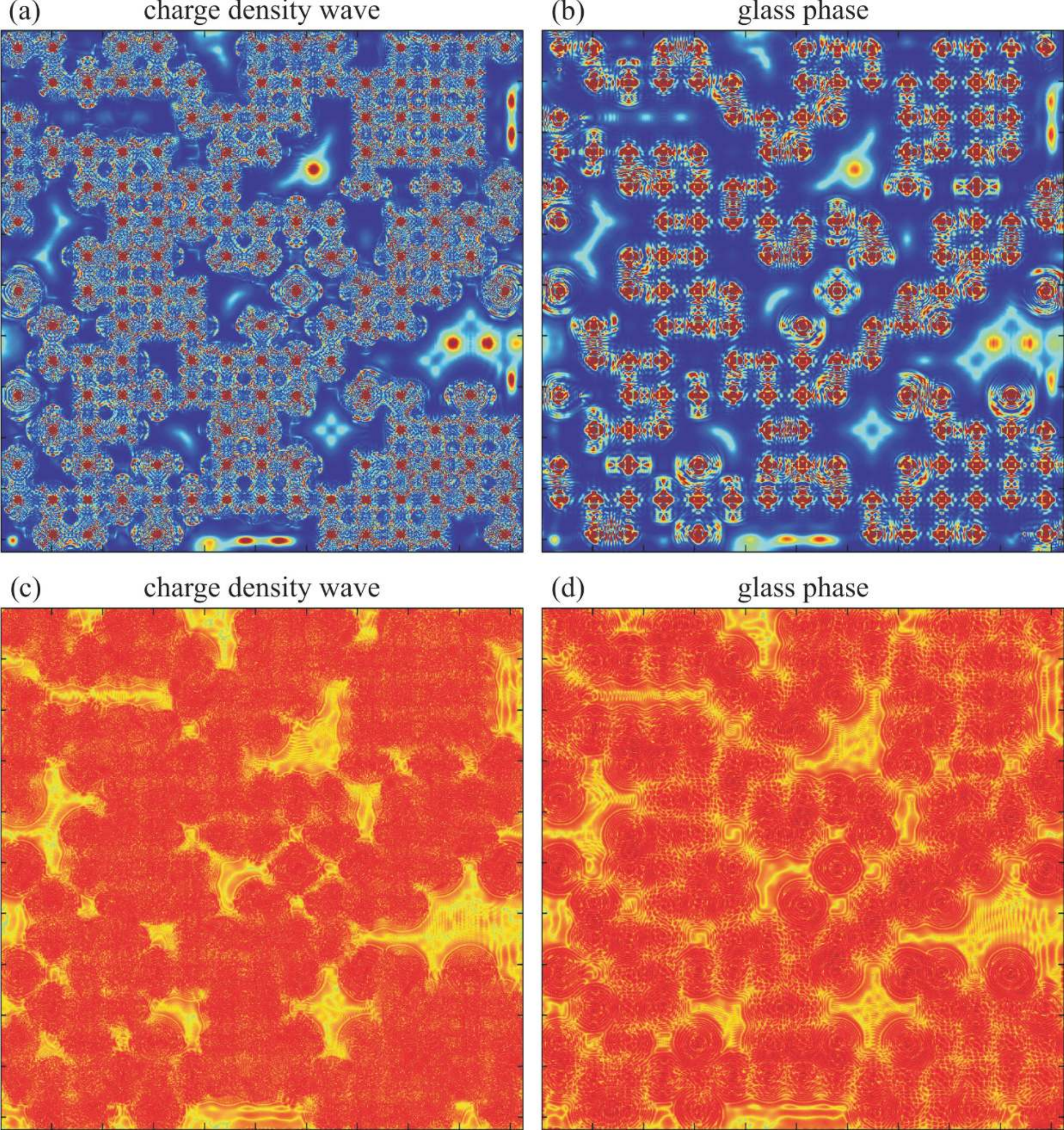}
\caption{\label{figdirtylatt3} Transverse profiles for the charge
density wave (a, c) and the glass phase (b, d): intensity maps (top)
and vortex charge density maps (down). The telltale difference is
that the CDW loses the regular lattice as the intensity "flows"
between the regular and the defect regions and we see the DI at
work. Glass on the other hand consists of domains with coherent
(well-defined) vortices though with reduced symmetry ($C_4$) mostly
due to EI. The charge density forms a connected network in the glass
phase and transport is possible, whereas in a frustrated insulator
the charge is stuck in isolated points.}
\end{figure}

The CDW versus the glass phase is given in Fig.~\ref{figdirtylatt3}. The charge density wave (a, c, $L=240\mu\mathrm{m}$) exhibits the diffusion of intensity due to DI, and the vortex beams are in general asymmetric and not clearly delineated. In (b, d), where $L=120\mu\mathrm{m}$ with all other parameters the same, there is a clear border between defects and the regular parts of the lattice and the intensity is concentrated in the vortex cores. We give also the vortex charge density map in (c, d) in addition to the intensity maps in (a, b) as the charge density shows why the CDW is insulating: even though individual beams diffuse and smear out in \emph{intensity}, the regions of nonzero vortex charge are disjoint and no global conduction can occur. Glass is divided into ordered domains in intensity but the vortex charges form a connected network which supports transport. This is analogous to the percolation transition in a disordered Ising model \cite{parisi1,parisi2} and we may expect that the CDW-glass transition follows the same scaling laws near the critical point. However, we have not checked this explicitly and we leave it for further work.

\section{The condensed matter analogy: collinear doped Heisenberg antiferromagnet}

The two-beam photorefractive system can serve as a good model for quantum magnetic systems. The most obvious connection is to multi-component XY antiferromagnets (i.e., two-dimensional Heisenberg model): planar spins are nothing but complex scalars, and the vortex Hamiltonian remains identical ($\pi_1(SO(2))=\pi_1(U(1))=\mathbb{Z}$). The nonlinearity in the spin system is different and usually much simpler, but that typically does not influence the phase diagram (the symmetry structure remains the same). Such connection is so obvious it does not require further explanations. Our point is that the CP beams in a PR crystal can also describe more general magnetic systems in the presence of topological solutions described by homotopy groups different from $\mathbb{Z}$. In particular, we want to point out to a connection with a two-sublattice antiferromagnetic system which has some time ago enjoyed considerable popularity as a possible description of magnetic ordering in numerous planar strongly-coupled electron systems, including cuprate high-$T_c$ superconductors \cite{ft2,sachbook,modelza}. This is the collinear doped antiferromagnet defined on two sublattices. When coupled to a charge density wave (speaking about the usual, $U(1)$ electromagnetic charge) and a superconducting order parameter, it becomes a toy model of cuprate materials (one variant is given in \cite{modelza}). In the light of what we know today, the ability of this model to realistically describe the cuprate physics is quite questionable; but even so it is an interesting magnetic system on its own, and it was already found in \cite{jurms0,jurms1} to exhibit a spin-glass phase, though in a slightly different variant (in particular, with spiral instead of collinear ordering).

Let us formulate the model. While the material is a lattice on the microscopic level, here we are talking about an effective field theory model. The order parameter is the staggered magnetization
\be
\label{mag}M(\mathbf{r})=\sum_{\alpha=1,2}\mathbf{M}_\alpha(\mathbf{r})\cos(\mathbf{n}\cdot\mathbf{r}),
\ee
where $\alpha\in\lbrace 1,2\rbrace$ is the sublattice "flavor" index (analogous to the $\alpha$-index for the $F$ and $B$ beam in the previous sections)\footnote{Sometimes we will denote the sublattices by $\pm$ instead of $1,2$ for compactness of notation.} and each component $\mathbf{M}_\alpha$ is a three-component spin, describing the internal, i.e. spin degree of freedom (we label the spin axes as X,Y,Z). The total spin is thus the sum of the spins of the two components, and $\mathbf{n}$ is the modulation vector. The modulation gives rows of alternating staggered magnetization in opposite directions as in Fig.~\ref{figpi1}(a). This stands in contrast with the spiral order, where the modulation vectors become $\mathbf{n}_\alpha$, i.e. differ for the two sublattices, and are themselves space-dependent \cite{jurms0}. The ordered phase of the collinear system has the nonzero expectation value of the staggered magnetization along one direction, which can be chosen as the Z-axis ("easy axis"), where the spin fluctuations about the easy axis remain massless, and the symmetry is broken from $O(3)$ to $O(3)/O(2)$. The spiral order, on the other hand, breaks the symmetry down to identity, as the order parameter is a dreibein \cite{jurms0}.

The symmetry conditions (isotropy in absence of external magnetic field) determine the Hamiltonian up to fourth order, as discussed in \cite{modelza}:
\be
\label{afham0}\mathcal{H}_\mathrm{af}=\frac{1}{2g_M}\left[\left(\frac{1}{c_M}\partial_\tau\mathbf{M}_\alpha\right)^2+\vert\nabla\mathbf{M}_\alpha\vert^2+\frac{r}{2}\vert\mathbf{M}_\alpha\vert^2\right]+
\frac{u_0}{2}\vert\mathbf{M}_\alpha\vert^4-v_0\left(\vert\mathbf{M}_1\vert^2+\vert\mathbf{M}_2\vert^2\right)^2.
\ee
The antiferromagnetic coupling is $g_M$, the spin stiffness is $c_M$ and the effective mass of spin wave excitations is $r$. The fourth-order coupling $u_0$ comes from the "soft" implementation of the constraint $\vert\mathbf{M}_\alpha\vert=1$\footnote{One could also enforce the constraint exactly, through the nonlinear sigma model, as was done in \cite{jurms0}. While the leading term of the "vortex" Hamiltonian would remain the same in that case, the amplitude fluctuations have different dynamics which influences some terms of the Hamiltonian and thus its RG flow (though probably not the very existence of the glass phase).} and $v_0$ is the anisotropy between the two sublattices, justified by the microscopic physics \cite{ft2,modelza}. The Hamiltonian can be transformed by rescaling $\tau$ and $x,y$, together with the couplings $u_0\mapsto u$ and $v\mapsto v_0$ to set $g_M=c_M=1$ so that the kinetic term becomes isotropic, giving
\be
\label{afham}\mathcal{H}_\mathrm{af}=\frac{1}{2}\left(\partial_\tau M\right)^2+\frac{1}{2}\vert\nabla M\vert^2+\frac{r}{2}\vert M\vert^2+\frac{u}{2}\left(\vert M\vert^2\right)^2-v\vert\mathbf{M}_1\vert^2\vert\mathbf{M}_2\vert^2,
\ee
where we have also rewritten the quartic terms for convenience. Without anisotropy, the energy of the system is a function of $\vert\mathbf{M}_1\vert^2+\vert\mathbf{M}_2\vert^2$ only and the symmetry group is the full $O(6)$. With $v\neq 0$, the symmetry is reduced to $O(3)_1\otimes O(3)_2$: the internal spin symmetry in each sublattice remains unbroken but the spatial rotation symmetry between the layers is broken down to just the discrete flip. Compare this to the $U(1)\otimes U(1)$ symmetry in the PR system: there, it is the internal phase symmetry that remains unbroken.

\subsection{$\mathbb{Z}_2$ "vortices"}

\begin{figure}\centering
\includegraphics[width=150mm]{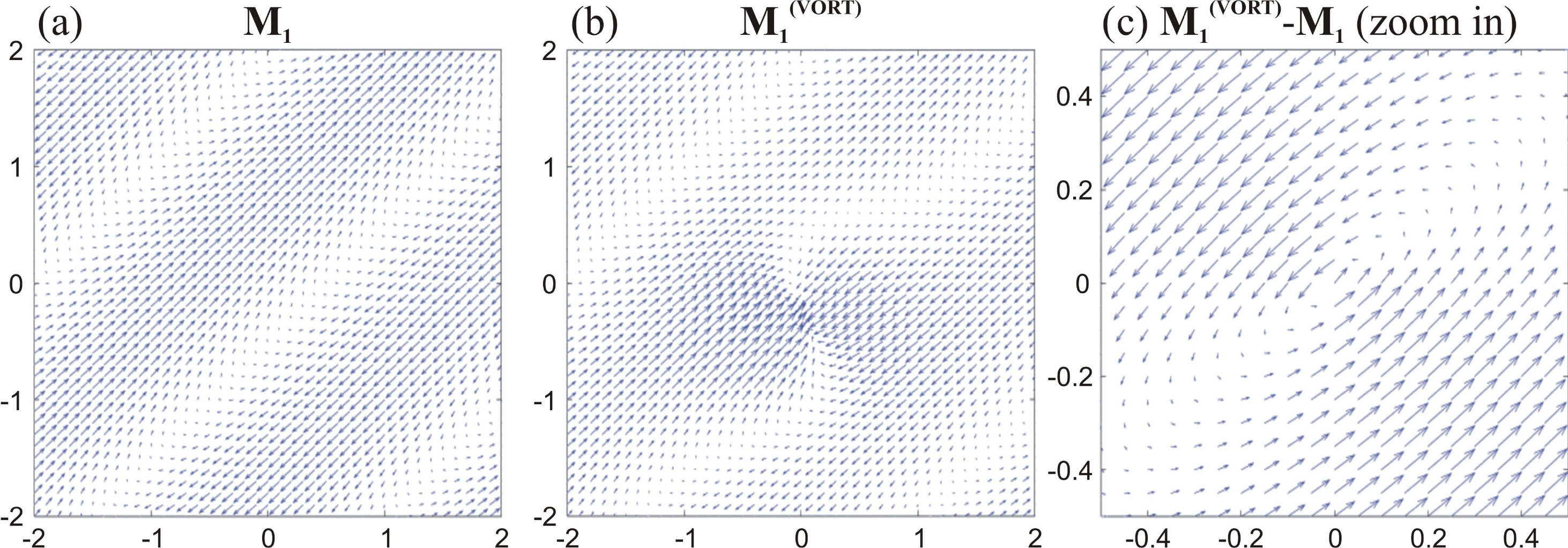}
\caption{\label{figpi1} Numerical realization of the spin pattern
(staggered magnetization $\mathbf{M}_1$) in the collinear $O(3)$
antiferromagnet. Magnetization is three-dimensional and we give the
projection in the XY-plane, $\mathbf{M}_1\cdot\mathbf{n}_{XY}\equiv
M_\perp$. In (a), we show the characteristic collinear spin pattern
in absence of "vortices". In (b) we plot
$\mathbf{M}_1^\mathrm{(vort)}$, a $\mathbb{Z}_2$-charged point
"vortex" defect with $Q=1$. In (c) we give the zoom-in of the "vortex"
from (b) shown as the difference
$\mathbf{M}_1^\mathrm{(vort)}-\mathbf{M}_1$ to show more clearly the
structure of the "vortex" -- now the regular periodic pattern is
absent and we appreciate the point-like structure of the "vortex". The
parameters are $u=r=1$ and $v=0.5$.}
\end{figure}

Remembering that topological solitons are classified by homotopy groups, and that we work in a two-dimensional plane, the relevant group is again the first homotopy group, $\pi_1(O(3))=\mathbb{Z}_2$. For simplicity, we will call these excitations "vortices", bearing in mind that the only possible charges are $Q_\alpha=\pm 1$ and not all integers. A realization of the "vortex" with $Q=1$ is shown in Fig.~\ref{figpi1}(b). Since the spins are three-dimensional (the figure shows the projection in the XY plane), it becomes clear that "vortex" charge is only defined modulo $2$, i.e. it makes no sense to talk about charges $\vert Q\vert>1$. For example, winding around twice in the XY plane can be done along a closed line in the XYZ space which can be contracted to a point. That could not happen for the two-dimensional phase $U(1)$ precisely because there is no extra dimension. In Fig.~\ref{figpi1}(b) the "vortex" is superimposed onto the regular configuration: it is recognizable as a contact point between two lines of alternating staggered magnetization. In. Fig.~\ref{figpi1}(c) we have subtracted the regular part and only the "vortexing" spin pattern is shown: here we see the "vortex" interpolates between two opposite spin orientations in two opposite directions in the plane.

Now let us derive the effective Hamiltonian of the "vortices". For the $\mathbb{Z}_2$ "vortex", a loop in real space is mapped onto a $\pi$-arc in the internal space, so the "vortex" can be represented as
\be
\label{pardef0}\mathbf{M}_\alpha(r,\phi)=\int d\phi'e^{\frac{\imath}{2}\left(\phi'-\phi\right)\hat{\ell}_3}\mathbf{m}_\alpha,
\ee
giving (the matrices $\ell_{1,2,3}$ represent the $so(3)$ algebra):
\be
\label{pardef}\mathbf{M}_\alpha=\left(\begin{matrix}\cos\phi & \mp\sin\phi & 0 \\ \pm\sin\phi & \cos\phi & 0 \\ 0 & 0 & 1\end{matrix}\right)\left(\begin{matrix}m_{1\alpha}\\ m_{2\alpha} \\ m_{3\alpha}\end{matrix}\right),
\ee
where $\mathbf{m}_\alpha$ is the magnetization amplitude, analogous to the beam amplitude $\psi_\alpha$ in the optical system. The leading-order, non-interacting term in (\ref{afham}) gives for the energy of a single "vortex" of charge $\vec{Q}$:
\be
\label{1vorten}E_1=2\pi(\vert\mathbf{m}_X\times\mathbf{e}_Z\vert^2+\vert\mathbf{m}_Y\times\mathbf{e}_Z\vert^2)\log\Lambda=2\pi\vert\mathbf{m}_{\perp\alpha}\vert^2\log\Lambda,
\ee
which is in fact independent of the sign of $\vec{Q}$ (as could be expected, as it is in general proportional to $\vec{Q}\cdot\vec{Q}$ which is a constant for parity "vortices"). The "vortex" singles out an easy axis (Z-axis) around which the staggered magnetization winds ($\phi$ being the winding angle). This allows one to introduce $\mathbf{m}_{\alpha\perp}\equiv\left(m_{X\alpha},m_{Y\alpha},0\right)$. A "vortex" pair with charges $\vec{Q}_i$ and $\vec{Q}_j$ has the binding energy
\be
\label{2vorten}E_2=2\pi\vec{Q}_i\cdot\vec{Q}_j\left(\vert\mathbf{m}_1\times\mathbf{e}_Z\vert^2+\vert\mathbf{m}_2\times\mathbf{e}_Z\vert^2\right)\log r_{ij}=2\pi\vert\mathbf{m}_{\perp\alpha}\vert^2\vec{Q}_i\cdot\vec{Q}_j\log r_{ij}.
\ee
Now we should integrate out the amplitude fluctuations as we did in Appendix \ref{app00} for the CP beams. This again leads to the coupling between different flavors, giving a "vortex" Hamiltonian analogous to (\ref{hamvort}):
\be
\label{hamvortz2}\mathcal{H}_\mathrm{vort}=\sum_{i<j}\left(g\vec{Q}_i\cdot\vec{Q}_j+g'\vec{Q}_i\times\vec{Q}_i\right)\log r_{ij}+\sum_i\vec{\mu}\cdot\vec{Q}_i.
\ee
Two obvious differences with respect to the optical system are (i) the charges are now limited to the values $\pm 1$ (ii) there is a term linear in charge density, which acts as a chemical potential. The latter arises from the coupling of the \emph{three-dimensional} spin waves (i.e., the topologically trivial excitations of the amplitude $\mathbf{m}_\alpha$) to the "vortices". Remember that in the CP system, the amplitude fluctuations also couple to the vortices, but there is no third, $\mathrm{Z}$-axis of the order parameter so no linear term appears. The microscopic expressions for the effective parameters $g,g',\mu_\alpha$ read:
\bea
g=m_\perp^2+\frac{4r+6um_\perp^2}{\left(2v+\frac{3}{2}um_\perp^2+\frac{v}{2}m_\perp^2\right)\left(2r+\frac{3}{2}um_\perp^2-\frac{v}{2}m_\perp^2\right)}\\
g'=-\frac{4vm_\perp^2}{\left(2v+\frac{3}{2}um_\perp^2+\frac{v}{2}m_\perp^2\right)\left(2r+\frac{3}{2}um_\perp^2-\frac{v}{2}m_\perp^2\right)}\\
\mu_\alpha=\frac{1}{2}m_{\perp}m_z,
\eea
assuming $m_{1\perp}=m_{2\perp}\equiv m_\perp$. Now the RG calculation is similar to the optical case but the nonzero chemical potential introduces two differences. First, there is obviously the additional term proportional to the total charge of the virtual pair of "vortices", $\mu_\alpha(q_{1\alpha}+q_{2\alpha})$. Second, there is no charge conservation as the expectation value of the total "vortex" charge is now $\langle\vec{Q}\rangle=\partial\mathcal{F}/\partial\vec{\mu}\neq 0$. Thus we need to take into account not only the fluctuations with zero net charge (virtual "vortex" pairs with charges $\vec{q}_1\equiv\vec{q}$ and $\vec{q}_2\equiv-\vec{q}$) but also the situations with arbitrary pairs $\vec{q}_1,\vec{q}_2$.\footnote{In the CP beam system, the total vortex charge can be nonzero if the boundary conditions at $z=0,L$ have nonzero total vorticity. But there we had no \emph{bulk} chemical potential so the total vorticity in the crystal could not change during the propagation along $z$. Here, we have a \emph{bulk} term in the Hamiltonian which violates charge conservation.} This modifies the variation of the partition function from (\ref{zfluc}-\ref{zfluc2}) to:
\bea
\nonumber&\frac{\delta\mathcal{Z}}{\mathcal{Z}}&=1+\frac{y^4}{4}\sum_{\vec{q}_{1,2}}\int dr_{12}r_{12}^3e^{-g\vec{q}_1\cdot\vec{q}_2-g'\vec{q}_1\times\vec{q}_2-\vec{\mu}\cdot\vec{q}'}\times\\
\nonumber&\times&\left[\int drr^2\left(g\vec{Q}_1\cdot\vec{q}+g'\vec{Q}_1\times\vec{q}\right)\nabla\log\vert\delta\mathbf{R}_1\vert+\left(g\vec{Q}_2\cdot\vec{q}+g'\vec{Q}_2\times\vec{q}\right)\nabla\log\vert\delta\mathbf{R}_2\vert\right]^2+\\
\nonumber&+&\frac{y^4}{4}\sum_{\vec{q}_{1,2}}\int dr_{12}r_{12}^3e^{-g\vec{q}_1\cdot\vec{q}_2-g'\vec{q}_1\times\vec{q}_2-\vec{\mu}\cdot\vec{q}_1}\left[\int drr^2\left(g\vec{Q}_1\cdot\vec{q}_0+g'\vec{Q}_1\times\vec{q}_0\right)\log\vert\delta\mathbf{R}_1\vert+\left(g\vec{Q}_2\cdot\vec{q}_0+g'\vec{Q}_2\times\vec{q}_0\right)\log\vert\delta\mathbf{R}_2\vert\right]^2,
\eea
where we have introduced $2\vec{q}\equiv\vec{q}_1-\vec{q}_2,\vec{q}_0\equiv\vec{q}_1+\vec{q}_2$ and $\delta\mathbf{R}_{1,2}\equiv\mathbf{R}_{1,2}-\mathbf{r}$. The mixed term which includes both $\vec{q}$ and $\vec{q}_0$ vanishes due to isotropy. Matching the terms in the resulting expression with the original Hamiltonian, we find the recursion relations:
\bea
\nonumber\frac{\partial g}{\partial\ell}=-16\pi y^4\left(g^2+g'^2\right)\\
\nonumber\frac{\partial g'}{\partial\ell}=-16\pi y^4gg'\\
\nonumber\frac{\partial\vec{\mu}}{\partial\ell}=0\\
\label{rgflowz2}\frac{\partial y}{\partial\ell}=\left(1-g-g'-\mu_+-\mu_-\right)y.
\eea
Crucially, the chemical potential does not run which could be guessed from dimensional analysis (it couples to dimensionless charge). This is the same system as (\ref{rgfloweqs}) up to the trivial rescaling of the coupling constants and the shift of the critical line $g+g'=1$ in the PR system to the line $g+g'+\mu_++\mu_-=1$. It becomes obvious that the phase diagrams are equivalent and can be mapped onto each other.

\subsection{Influence of disorder}

The disorder in a doped antiferromagnet comes from electrically neutral metallic grains quenched in the bipartite lattice. Being metallic and neutral, they are naturally modeled as magnetic dipoles $\mathbf{X}$ quenched in the bipartite lattice. This picture stems from the microscopic considerations in \cite{jurms3}. We again assume the Gaussian distribution of the disorder as $p(X)\propto\exp(-\vert\mathbf{X}\vert^2/2\sigma_X^2)$. The disorder dipoles are one and the same for both sublattices, so $\mathbf{X}$ has no flavor (sublattice) index. The minimal coupling of the dipoles to the lattice spins $\partial_i\mapsto\partial_i-\imath\hat{\ell}_iX_i$ gives
\be
\label{spindisorder}\mathcal{H}_\mathrm{af}\mapsto\mathcal{H}_\mathrm{dis}=\mathcal{H}_\mathrm{af}+\nabla\mathbf{M}_\alpha\cdot(\mathbf{X}\times\mathbf{M}_\alpha)+M^2X^2.
\ee
Now the replica calculation requires the multiplication of the $M$ field into $n$ copies and performing the Gaussian integral over the disorder. The initial distribution of the disorder $p(X)$ gives rise to two independent Gaussian distributions: for the couplings $J_{ij}^{\alpha\beta}$ with dispersion matrix $\sigma^2_{\alpha\beta}$, and for the chemical potential $\mu_i^\alpha$ with the dispersion vector $\xi^2_\alpha$. The resulting Hamiltonian is
\bea
\nonumber\mathcal{H}_\mathrm{dis}=\sum_{\mu=0}^n\left(\frac{1}{2}\vert\partial_\tau\mathbf{M}_\alpha^{(\mu)}\vert^2+\frac{1}{2}\vert\nabla\mathbf{M}_\alpha^{(\mu)}\vert^2+\frac{u}{2}\vert\mathbf{M}^{(\mu)}_\alpha\vert^2-v\vert\mathbf{M}^{(\mu)}_1\vert^2\vert\mathbf{M}^{(\mu)}_2\vert^2\right)+\\
\label{rephamspin}+\frac{\sigma^2}{4}\sum_{\mu,\nu=0}^n\left(\nabla\mathbf{M}_\alpha^{(\mu)}\times\mathbf{M}_\alpha^{(\mu)}\right)\cdot\left(\nabla\mathbf{M}_\alpha^{(\nu)}\times\mathbf{M}_\alpha^{(\nu)}\right),
\eea
where we have disregarded the subleading logarithmic term ($\sim\log\vert\mathbf{M}_\alpha^{(\mu)}\vert$). Now making use of the representation (\ref{pardef0}) and plugging it in into (\ref{rephamspin}) gives the disordered "vortex" Hamiltonian
\be
\label{rephamvortz2}\beta\mathcal{H}_\mathrm{vort}=\sum_{\mu,\nu=1}^n\sum_{i,j=1}^N\left[\frac{\beta^2}{2}Q_{i\alpha}^{(\mu)}Q_{i\beta}^{(\nu)}Q_{j\alpha}^{(\mu)}Q_{j\beta}^{(\nu)}-\beta Q_{i\alpha}^{(\mu)}J_0^{\alpha\beta}Q_{j\beta}^{(\mu)}+\beta^2Q_{i\alpha}^{(\mu)}\xi^2Q_{i\alpha}^{(\nu)}\right]-\sum_{\mu=1}^n\sum_{i=1}^N\beta\xi^2\mu_0^\alpha Q_{i\alpha}^{(\mu)}.
\ee
Of course, we could have arrived at the same effective action starting from the "vortex" Hamiltonian (\ref{hamvortz2}), taking the infinite range approximation and identifying $J_{ij}^{\alpha\alpha}=g_{ij}\log r_{ij}$ and similarly for other components of $J_{ij}^{\alpha\beta}$ as we demonstrated for the PR system. The final result has to be same at leading order.

The next step is to rewrite the Hamiltonian in terms of the order parameters $p_\alpha^{(\mu)},q_{\alpha\beta}^{(\mu\nu)}$ defined in (\ref{glassop}). Compared to the effective action for the photonic lattice with disorder in Eq.~(\ref{rephamvort3}), there are two extra terms in the resulting action $S_\mathrm{eff}$: one is proportional to the dispersion $\xi^2$ and the other to the mean chemical potential $\vec{\mu}_0$. The former term just introduces the shift $J_0^{\alpha\beta}\mapsto J_0^{\alpha\beta}-\sigma^2/2\beta$ and the latter term, linear in the "vortex" charges and proportional to the chemical potential, introduces solutions with nonzero net "vortex" charge density. Looking back at the results of the saddle-point calculation in Eqs.~(\ref{glassop},\ref{psol}), this tells us that the relation between the phase diagrams is the following. The phases with no net "vortex" charge density -- insulator, conductor, frustrated insulator and perfect conductor -- remain the same as in the PR system, since both the average coupling value $J_0^{\alpha\beta}$ (which gets shifted) and the term proportional to the chemical potential $\mu_\alpha$ couple only to $\vec{p}^{(\mu)}$. For brevity, denote $J_0^{\pm\pm}\equiv J_0^{\pm}$ and notice that $J_0^{-+}=J_0^{+-}$. The structure of phases with nonzero $\vec{p}^{(\mu)}$ depends on the zeros of the saddle point equation
\be
\label{saddlepz2}J_0^\pm p^\pm+\left(\frac{J_0^{+-}}{\beta}-\frac{\beta}{2}\xi^\pm\right)p^\mp+\left(p^\pm\right)^{-1}-\frac{\mu_0^\pm(\sigma^{\pm\pm})^2+\mu_0^\mp\sigma_{+-}^2}{\beta}=0,
\ee
analogous to (\ref{saddlep}), where the 1-step replica symmetry breaking implies $p_{(\mu)}^\pm=\left(p^\pm,\ldots, p^\pm\right)$. Now the equation is cubic and the structure of solutions is different from (\ref{psol}). We could not find the solution in the closed form but it is clear that a pair of cubic equations will have either a single solution $(p^+,p^-)$ or nine combinations $(p^+,p^-)$, not necessarily all different. Numerical analysis of (\ref{saddlepz2}) reveals only two inequivalent solutions, analogous to (\ref{psol}), i.e. one of them has a single free energy minimum, the other one a pair of degenerate minima. Therefore, we again have two disordered solutions, one of which is glassy (frustrated).

\begin{figure}\centering
\includegraphics[width=70mm]{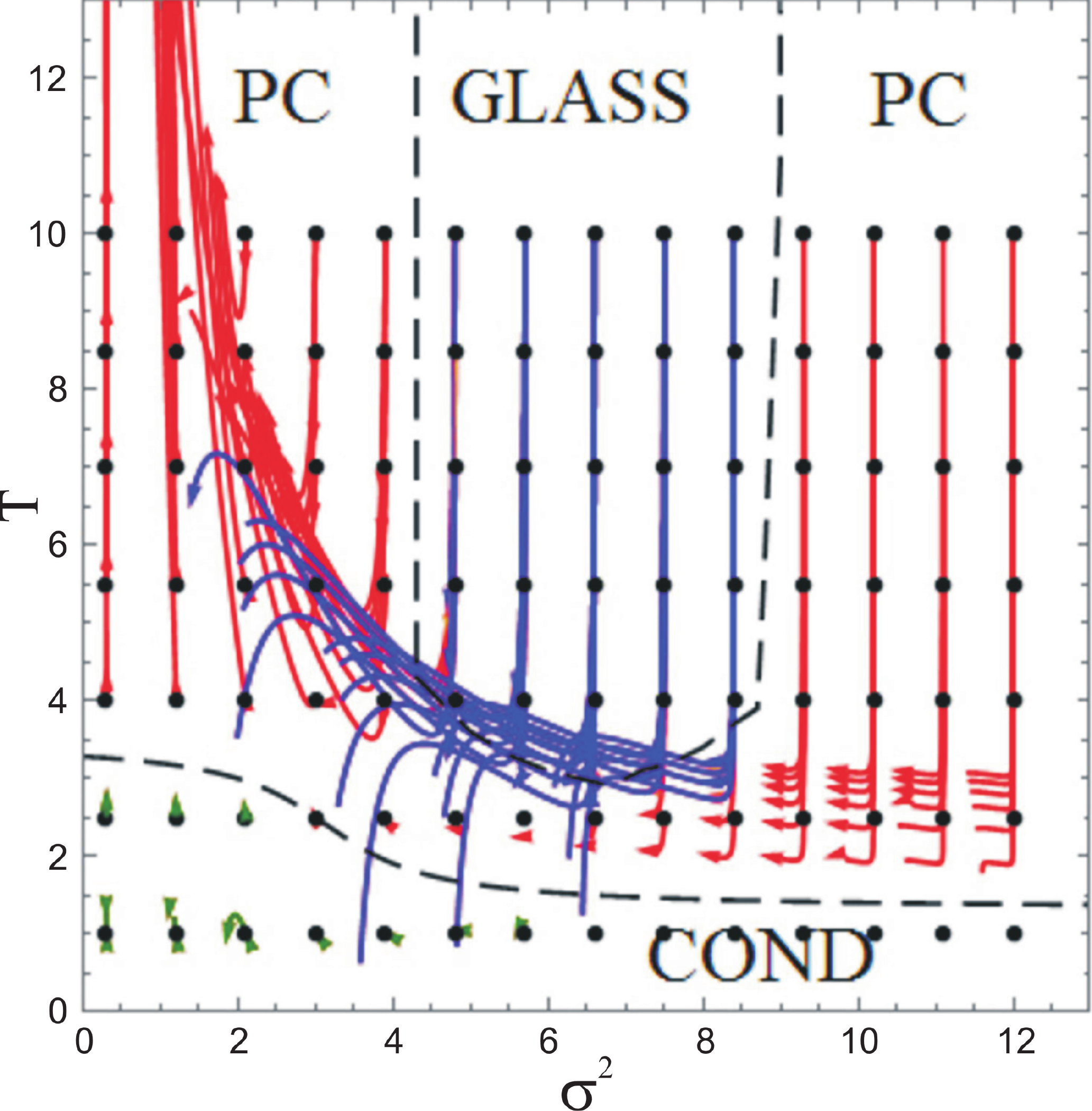}
\caption{\label{figphdiagz2} The phase diagram of the two-sublattice
doped Heisenberg antiferromagnet model in the $\sigma^2$-$T$ plane
(we have rescaled $\sigma^2\mapsto 12\sigma^2$). Since $T\sim 1/g'$,
we can alternatively understand the vertical axis as $1/g'$. Black
dashed curves are approximate phase boundaries. RG flows (starting
from black dots) are colored differently according to the phase they
belong to: spin glass (blue), PC (red), conductor (green). At high
temperatures, the "vortex" conductor becomes either a perfect "vortex"
conductor or a spin glass. Spin glass (blue) is recognized by the
fact that the RG equations flow to nonzero disorder at finite and
large $g'$ (low temperatures). The PC phase (red) flows toward zero
disorder and zero coupling (infinite $T$), collapsing practically to
a single trajectory. The flows for the conductor (green) end up at
$T=\sigma^2=0$ but are not shown to scale in the figure. Parameter
values: $u=r=1$ and varying $v$ so as to have $g=-0.5$ for all
trajectories.}
\end{figure}

Now we can write down also the RG equations for the effective action (\ref{rephamvortz2}). In this calculation, we put $\xi_\alpha^2=\sigma_{\alpha\beta}^2\equiv\sigma^2$ for simplicity. Following the same logic as earlier, the equations are found to be:\footnote{For the most general case of different and non-scalar $\sigma^2_{\alpha\beta}$ and $\xi^2_\alpha$ the flow equations for them complicate significantly and we will not consider them.}
\bea
\nonumber\frac{\partial g}{\partial\ell}=-8\pi\left(g+g'\right)^2y^4\cosh\left(2\beta^2\sigma^2\right)\cosh\left(2\beta^2\sigma^2\right)-8\pi\left(g-g'\right)^2y^4\\
\nonumber\frac{\partial g'}{\partial\ell}=-\pi\left(g+g'\right)^2y^4\cosh\left(2\beta^2\sigma^2\right)\cosh\left(2\beta^2\sigma^2\right)-\pi\left(g-g'\right)^2y^4\\
\nonumber\frac{\partial y}{\partial\ell}=2\pi\left(1-g-g'-\mu_+-\mu_--\beta^2\sigma^2\right)y\\
\nonumber\frac{\partial\mu}{\partial\ell}=-8\pi\mu\\
\label{rgflowdirtyz2}\frac{\partial\sigma^2}{\partial\ell}=-2\pi\beta^4\sigma^4y^4.
\eea
Like in the clean case, the chemical potential is irrelevant and the solutions for fixed point are the same as for the PR beams, including the spin-glass fixed point. We conclude that the phase structure of the optical system is repeated in strongly correlated doped antiferromagnets, which also exhibit the spin-glass phase and have the phase diagram sketched in Fig.~\ref{figphdiagz2}. In this context it is more interesting to plot the phase diagram in the $\sigma^2-1/g'$ plane, mimicking the $x-T$ phase diagram of quantum critical systems \cite{sachbook} (remember that the coupling constants $g,g'$ behave roughly as inverse temperature in XY-like models). Bear in mind that all phases shown are about "vortex" dynamics, i.e. one should not compare Fig.~\ref{figphdiagz2} to the textbook phase diagram of high-temperature superconductors, which accounts also for the charge or stripe order and the superconducting order. All "vortex" phases would be located inside the pseudogap regime of the superconductor, where various exotic orders can coexist (assuming, of course, that our model is an adequate approximation of the magnetic order in a cuprate or similar material, which is a complex question). Crucially, the spin-glass phase (blue curves) flows toward \emph{finite disorder} $\sigma^2$, whereas the remaining two phases end up at zero disorder, either at infinite $1/g'$ (PC, red flows), or at zero $1/g'$ (conductor, green flows). The RG flows in the conductor phase are almost invisible in the figure, as the flows are much slower than in the remaining two phases.

\subsubsection{Discussion}

Early papers which found and explored the spin-glass phase in a very similar model are \cite{jurms0,jurms1,jurms2,jurms4}. The main difference is that the papers cited consider the spiral (non-collinear) spin order. These works are all inspired by the cuprate materials, the most celebrated brand of high-temperature superconductors. While \cite{jurms1,jurms2} explore in detail the transport properties, we have no pretension neither to provide a realistic model of cuprates nor to explore in detail all the properties of the spin-glass phase. We are content to see that the PR system of $\mathbb{Z}$ vortices reproduces the phase structure of a certain kind of dirty Heisenberg antiferromagnets (with $O(3)$ spins and $\mathbb{Z}_2$ "vortices"), besides the more obvious connection to systems which directly reproduce the $\mathbb{Z}$ vortices in multicomponent $U(1)$ systems like multicomponent Bose-Einstein condensates and type-$1.5$ superconductors.

\section{Conclusions}

We have investigated the light intensity patterns in a nonlinear
optical system, a pair of counterpropagating laser beams in a
photorefractive crystal. We have studied this system as a strongly
interacting field theory and have focused mostly on the formation
and dynamics of vortices. The vortices show a remarkable collective
behavior and their patterns are naturally classified in the
framework of statistical field theory: the effective action shows
several different phases with appropriate order parameters, and the
system is essentially an XY model with two flavors, i.e. two kinds
of vortex charge, for the two beams. The interaction between the
flavors is the central reason that the total energy of the Coulombic
interactions between the vortices in general cannot be locally
minimized at every point. In the presence of disorder, a phase with
multiple free energy minima arises, where the absence of long-range
order is complemented by the local islands of ordered vortex
structure, and which resembles spin glasses.

The phase diagram is simple in terms of the effective parameters --
vortex coupling constants -- and quite complex when expressed in
terms of the experimentally controllable quantities -- the intensity
of the laser beams, the intensity of the background photonic lattice
and the properties of the photorefractive crystal (the last is not
controllable but can be estimated reasonably well \cite{korn}). The
lesson is that the approach we adopt can save us from demanding
numerical work if the space of original parameters is "blindly"
explored. Our phase diagrams can serve as a starting point for
"guided" numerical simulations, suggesting what phenomena one should
specifically look for. So far the field-theoretical and statistical
approach was not much used in nonlinear optics (important exceptions
are
\cite{primeri1,primeri2,fisher,fisher2,fisher3,fisher4,conti1,conti2,conti3,conti4}).
We hope to stimulate work in this direction, which is promising also
because of the potential of the photorefractive systems to serve as
models of strongly correlated condensed matter systems. They make an
excellent testing ground for various models because of the
availability and relatively low cost of experiments.

In this work we have focused on the relation of the photorefractive counterpropagating system to the model of an $O(3)$ doped antiferromagnet with two sublattices. The authors of previous works on this model \cite{modelza,jurms1,jurms2,jurms3} were motivated mainly by the ubiquitous problem of understanding the pseudogap phase in cuprate superconductors. The applicability of the model to this particular problem is still an open question -- it may well be that cuprate physics goes far beyond. Nevertheless, it is an important quantum magnetic system in its own right and serves as an illustration on how one can simulate condensed matter systems in photorefractive optics.

Another field where vortices are found as solutions of a non-linear
Schr\"odinger equation are cold atom systems and Bose-Einstein
condensates \cite{bec}. Notice however that Bose-Einstein
condensates in optical traps are usually (but not always, see
\cite{becmulti}) three-dimensional systems with vortex lines (rather
than XY-type systems with point vortices) and our formalism would be
more complicated there: in three spatial dimensions, vortices give
rise to emergent gauge fields. The multi-component systems of this
kind give rise to so-called type $1.5$ superconductors
\cite{multi15} which are a natural goal of further study.

A more complete characterization of the glass-like phase is also left for further work. The reader will notice we have devoted very little attention to the correlation functions in various parameter regimes or the scaling properties of susceptibility, which should further corroborate the glassy character of the system. This is quite difficult in general but very exciting as it offers an opportunity to tune the parameters (e.g. disorder strength) freely in the optical system and study the glass-like phase and its dynamics.

\section*{Acknowledgments}

We are grateful to Mariya Medvedyeva and Vladimir Juri\v{c}i\'{c}
for careful reading of the manuscript. Work at the Institute of
Physics is funded by Ministry of Education, Science and
Technological Development, under grants no. OI171033 and OI171017.
M.~P. is also supported by the NPRP 7-665-1-125 project of the Qatar
National Research Fund (a member of the Qatar Foundation).

\appendix
\section{Numerical algorithm}
\label{app0}

In order to solve numerically the system (\ref{prpsieq}-\ref{preeq}) we employ a variation on the method of \cite{belicalg0,belicalg1}. The method does not make use of any analytical ansatz: it is an \emph{ab initio} numerical procedure which integrates the equations of motion. The system has four independent variables: the transverse coordinates $(x,y)$, the longitudinal coordinate ("formal time") $z$ and the (physical) time $t$. That means we have essentially three nested loops: (i) at every $z$-slice we integrate the transverse Laplacian and the interaction terms (ii) when done with this for the whole $z$-axis we advance the time $t$ (iii) we repeat the whole procedure until reaching some time $t_f$ which certainly should be much longer than the relaxation time $\tau$.

The important point is the very different nature of the initial and boundary conditions for various coordinates. The boundary conditions in the $(x,y)$ plane, i.e. at the crystal edge are not crucial: we have either just one or a few Gaussian beams whose intensity drops exponentially away from the center and is practically zero at the crystal edge, or we have a large lattice consisting of many (of the order of $50-100$) Gaussian beams so the edge effects only affect a small portion of the whole lattice. Therefore, imposing periodic boundary conditions (stemming naturally from the integration in Fourier space, see the next paragraph) are perfectly satisfying. Crucially, however, the CP geometry means that $F(t;z=0;x,y)=F_0(x,y)$ and $B(t;z=L;x,y)=B_0(x,y)$ are given functions, fixed for all times. We thus have a two-point boundary value problem along $z$ and have to iterate the $z$-integration several times until we reach the right solution. Finally, the initial condition for the relaxation equation (\ref{preeq}) is that the crystal is initially at equilibrium, meaning that $E(t=0)=-I_x/(1+I_x)$; specifically, for zero background lattice $E(t=0)=0$.

The algorithm now has the following structure:
\begin{enumerate}
\item{The innermost loop integrates in the $x-y$ plane. This is a Poison-type (elliptic) equation, thus we employ the operator-split method, integrating the Laplacian operator in the Fourier space and the interaction term (the $EF$ and $EB$ terms) in real space, in the second-order leapfrog scheme. Thus, at every time instant $t_i=i\Delta t$, we start from $z=0$ where we set the condition $F(i\Delta t;z=0;x,y)=F_0(x,y)$, divide the $z$-axis into $N$ steps of size $\Delta z=L/N$, and at every slice $z=j\Delta z$ we perform the frog's leap: we do the Fast Fourier Transform (FFT) to turn the $(x,y)$-dependence into $(q_x,q_y)$-dependence,\footnote{We denote the fields in Fourier space with a tilde, e.g. $\tilde{F}$.} then we advance the Laplacian for $\Delta z/2$ as $F(i\Delta t;j\Delta z;\mathbf{q})\equiv\tilde{F}_{i,j}^{(0)}\mapsto\tilde{F}_{i,j}^{(1)}=\exp\left(-\imath q^2\Delta z/2\right)\tilde{F}_{i,j}^{(0)}$, then we do the inverse FFT and advance the interaction in real space as $F_{i,j}^{(2)}=\exp\left(\imath\Gamma E\left(i\Delta t;j\Delta z;x,y\right)\right)F_{i,j}^{(1)}$. Finally we do the FFT again and advance the Laplacian for the remaining half-step, $\tilde{F}_{i,j+1}=\exp
    \left(-\imath q^2\Delta z/2\right)\tilde{F}_{i,j}^{(2)}$. Once we reach $j=N$ the integration goes backward, along the same lines, updating now the $B$-field (starting from $B_0(x,y)$), where all signs in the exponents of the above formulas are to be reversed. When we reach $z=0$ again, we are done. In this loop we use the field $E_{1,j}$ as already known for all $j$.}
\item{The above loop will in general produce results inconsistent with the charge field $E_{i,j}$ because the equation for $E$ couples $F$ and $B$ and we have ignored that by integrating the two fields one after the other instead of simultaneously. This is, of course, commonplace in two-point boundary value problems: either only one boundary condition can be imposed exactly and the other is shot for or, as in our case, both are imposed exactly but at the cost of the solution being inconsistent with the equations, so we have to iterate the system to arrive at the correct solution everywhere. The second loop thus iterates the first loop $A$ times, at each step updating the charge field as $E_{i,j}^{(a-1)}\mapsto E_{i,j}^{(a)}=E_{i-1,j}-\tau\left(E_{i,j}^{(a-1)}+I_{i,j}^{(a-1)}/\left(1+I_{i,j}^{(a-1)}\right)\right)/(1+I_{i,j}^{(a-1)})$. The number of iterations $A$ is not fixed: we stop iterations when the intensity pattern stabilizes, $\sum_j\sum_{x,y}\left(I_{i,j}^{(a)}-I_{i,j}^{(a-1)}\right)<\epsilon$, for some tolerance $\epsilon$. Here, $I_{i,j}$ refers to total intensity, i.e. $\vert F\vert^2+\vert B\vert^2+I_x$.}
\item{Finally, the outermost loop integrates in time $t$, from $t=0$, with the initial condition $E(t=0)=-I_x/(1+I_x)$ given above. The integration time $t_f$ is divided into $M=t_f/\Delta t$ intervals, and at the end of each step we update $\left(F_{i,j},B_{i,j},E_{i,j}\right)\mapsto\left(F_{i+1,j},B_{i+1,j},E_{i+1,j}\right)$}. Only the charge field is directly integrated (as written above), in the first-order, Euler scheme. The beam envelopes depend on time only parametrically, through $E(t)$, and they evolve by using an updated $E_{i,j}$ in the first two loops at every time step.
\end{enumerate}
This procedure is very close to that in \cite{belicalg0}; the main difference is that we use a second-order (leapfrog) scheme while on the other hand our time integration is of the lowest, linear order instead of second order as in \cite{belicalg0}.

\section{Time-dependent perturbation theory and the existence of equilibrium configurations}
\label{appa}

\subsection{Stability analysis: fixed points and limit cycles}

In this Appendix we consider the time evolution of the CP beams and
show the existence of a stable equilibrium point with nonzero
intensity. This means that the system reaches stationary state for
long times, justifying the basic assumption of the paper that one
can study the vortex configurations within equilibrium statistical
mechanics. Not all patterns are stable: depending on the boundary
conditions and parameter values, the system may or may not have a
stable equilibrium, and non-equilibrium solutions in photorefractive
optics are well-known \cite{fisher,gaussrotpet}. For our purposes,
however, it is enough to identify the region of parameter space
where the equilibrium exists; other cases are not the topic of this
paper.

The time evolution of the beams $\Psi_\alpha$ and the charge field
$E$ in $(k,q)$-space is obtained by differentiating the equations
(\ref{prpsieq}) with respect to time and plugging in $\partial
E/\partial t$ from the relaxation equation (\ref{preeq}): \bea
\label{teom1}\frac{\partial\Psi_\alpha^\pm}{\partial
t}=-\frac{\Gamma}{\tau}\frac{\left(\left(1+I\right)E+I\right)}{\alpha
k-q^2-\Gamma E}\Psi_\alpha^\pm,~~\frac{\partial E}{\partial
t}=-\frac{1}{\tau}\left(\left(1+I\right)E+I\right). \eea This system
has three equilibrium points. One is the "$0$" point, \be
\nonumber\left(\Psi^\pm_+,\Psi^\pm_-,E\right)=\left(0,0,-\frac{I_x}{1+I_x}\right),
\ee and the remaining two are related by a discrete symmetry
$\Psi_\pm\mapsto\Psi_\mp$, so we denote them as "$\pm$" points, the
"$+$" point being \be
\nonumber\left(\Psi^\pm_+,\Psi^\pm_-,E\right)=\left(\sqrt{\frac{E(1+I_x)+I_x}{1+E}}e^{\imath\phi_+},0,E\right),
\ee and the "$-$" point has instead $\Psi_+=0$ and
$\Psi_-=\sqrt{(E(1+I_x)+I_x)/(1+E)}\exp(\imath\phi_-)$. Notice that
the phase $\phi_\pm$ remains free to vary so this solution supports
vortices. The "$0$" point is the trivial vacuum, i.e. the
zero-intensity configuration with only background lattice. The
fluctuation equations about this point to quadratic order read \be
\label{point0eq}\partial_tX=-\left(-f_+X_1X_5,-f_+X_2X_5,-f_-X_3X_5,-f_-X_4X_5,-\frac{1}{1+I_x}\left(X_1^2+X_2^2+X_3^2+X_4^2\right)-\left(1+I_x\right)X_5\right)
\ee where we have introduced the real variables
$X_{1,3}=\Re\delta\Psi_\pm,X_{2,4}=\Im\delta\Psi_\pm,X_5=\delta E$
and \be
\label{point0eqq}f_\pm=\frac{\Gamma\left(1+I_x\right)^2}{\Gamma
I_x\mp\left(1+I_x\right)\left(k\pm q^2\right)} \ee The system
(\ref{point0eq}) is degenerate at linear order thus we need a
quadratic order expansion to analyze stability. The simplest
approach is to construct a Lyapunov function for the equation
(\ref{point0eq}). The function $V(X)=X^2$ is positive for and only
for $X\neq 0$, and its derivative is \be
\label{point0lyap}\frac{dV}{dt}=-2f_+\left(X_1^2+X_2^2\right)X_5-2f_-\left(X_3^2+X_4^2\right)X_5-\frac{1}{1+I_x}\left(X_1^2+X_2^2+X_3^2+X_4^2\right)X_5-\left(1+I_5\right)X_5^2,
\ee which is strictly negative for $X$ nonzero if $f_\pm>0$ and
$X_5>0$. However, we always have $X_5>0$ because $dX_5/dt$ in the
full relaxation equations (\ref{teom1}) has strictly negative
right-hand side and $E$ grows monotonically from zero to
$-I_x/(1+I_x)$, and at any finite $t$ we have
$E(t)-E(t=\infty)=X_5>0$. Thus the trivial equilibrium point is
\emph{locally} stable for $f_+>0,f_->0$, i.e. $k>q^2$. It is much
harder to construct the Lyapunov function for the global equations
(\ref{teom1}): in this case there are no additional symmetries and
the stability of higher-dimensional systems is in general an
extremely difficult topic. Thus there may well be regions far away
from the "$0$" point which do not flow toward it.

The "$\pm$" pair is quite hard to study. All hope of expanding the
system to second order and understanding the resulting complicated
five-variable system is lost. This time, however, we can do a
non-trivial first-order analysis as the system is non-degenerate and
nicely reduces to the $(X_1,X_5)$ subsystem. Rescaling $X_1\mapsto
(1+E_0)^{-3/4}\left(I_x+E_0\left(1+I_x\right)\right)^{1/2}$ and
$t\mapsto
t\left(\left(1+E_0\right)/\left(I_x+E_0\left(1+I_x\right)\right)\right)^{1/4}$,
the equation of motion for the "$\pm$" point reads \be
\label{pointpmeq}\partial_t\left(\begin{matrix}X_1 \\
X_5\end{matrix}\right)=\left(\begin{matrix}-\frac{a_\pm}{\Gamma
E_0+k+q^2} & -1\\ 1 & -\frac{a_\pm}{\Gamma
E_0+k+q^2}\end{matrix}\right)\left(\begin{matrix}X_1 \\
X_5\end{matrix}\right)+O(X_1^2+X_5^2;X_2,X_3,X_4), \ee with $a_\pm$
being some (known) \emph{positive} functions of $\Gamma, E_0, I_x$
(independent of $k,q$). This is precisely the normal form for the
Andronov-Hopf bifurcation \cite{arnold}, and the bifurcation point
lies at $k=-\Gamma E_0-q^2$. To remind, the bifurcation happens when
the off-diagonal element in the linear term changes sign: the fixed
point is stable when $a_\pm/(\Gamma E_0+k+q^2)$ is positive. The
sign of the nonlinear term determines the supercritical/subcritical
nature of the bifurcation. Negative sign means the fixed point is
stable everywhere before the bifurcation and is replaced by a stable
limit cycle after the bifurcation (supercritical). Positive sign
means the fixed point coexists with the stable limit cycle before
the bifurcation and the $(X_1,X_5)$ plane is divided among their
attraction regions; after the bifurcation there is no stable
solution at all (subcritical).\footnote{One should not take the
stability in the whole $(X_1,X_5)$ plane in the supercritical case
to seriously. We have expand the equations of motion in the vicinity
of the fixed points and the expansion ceases to be valid far away
from the origin.}

In conclusion, stable "$+$" equilibrium exists for $k>-\Gamma
E_0-q^2$ where $E_0$ is best found numerically. Exactly the same
condition holds for the "$-$" point. For $k<-\Gamma E_0+q^2$,
dynamics depends on the sign of the nonlinear term in
(\ref{pointpmeq}): for the positive sign we expect periodically
changing patterns. If the term is negative and the bifurcation is
subcritical, various possibilities arise: the system may wander
chaotically between the "$+$" and the "$-$" point, or it may end up
in the attraction region of the "$0$" point and fall onto the
trivial solution with zero intensity. Naively, the attraction
regions of the two fixed points ("$\pm$" and "$0$") are separated by
the condition $-\Gamma E_0-q^2=q^2$, i.e. $q_c=\sqrt{-\Gamma
E_0(\Gamma,\tau)/2}$, where we have emphasized that $E_0$ is in
general non-universal. However, the actual boundary may be more complex, as our analysis is based on finite-order expansion around
the fixed points, which is not valid far away from them.

\begin{figure}\centering
\includegraphics[width=70mm]{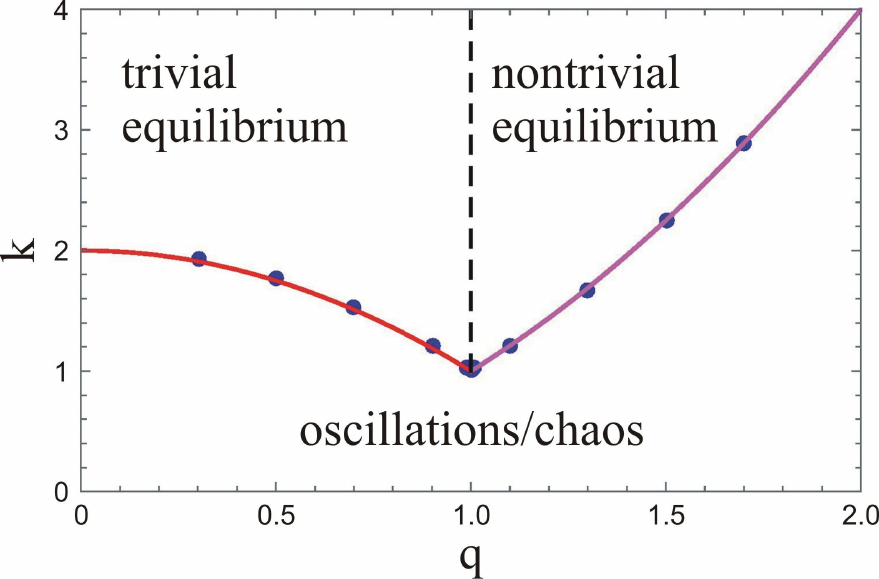}
\caption{\label{figbif} Stability diagram in the $q$-$k$ plane. The
onset of instability for $k<k_c(q)$ is found numerically for a range
of $q$ values. The solid lines are the analytical prediction for the
stability of the "$0$"-point ($k_c=q^2$, magenta) and of the
"$+$"-point ($k_c=\Gamma E_0-q^2\approx\Gamma-q^2$, red). The black
dashed line at $q=q_c\approx 1$ separates the stability regions of
the two points. The domain of applicability of our main results is
the top left corner (nontrivial equilibrium), above
$k>k_\mathrm{min}\sim 1/L$ and for not too large $q$ values.
Parameter values: $\Gamma=2, I_x=0$.}
\end{figure}

The outcome is that the system generically has stable trivial and
nontrivial (nonzero intensity) equilibria, in addition to
time-dependent, periodic or aperiodic solutions. Numerical
integration gives a similar picture of the stability diagram in
Fig.~\ref{figbif}. Numerically we find that the stability limit is
$k>\Gamma-q^2$, i.e. $E_0\approx -1$. The region of applicability of
our formalism lies in the top right corner of the diagram
(nontrivial equilibrium), above $k\approx 1/L$. Formally, both $k$
and $q$ can be any real numbers. In practice, however, $k$ is
discrete and its minimal value is of the order $1/L$. The spatial
momentum $q$ lies between the inverse of the transverse length of
the crystal (which is typically an order of magnitude smaller than
$L$, i.e. minimal $q$ can be assumed equal to zero) and some typical
small-scale cutoff which in our case is the vortex core size. We
made no attempt to study the non-equilibrium behavior in detail or
to delineate the boundary between the oscillatory and the chaotic
regime since it is irrelevant for the main story of the paper.

From a practical viewpoint, the $\Gamma$-$I_x$ plane can be divided
into two regions. One of them has a single stable "$+$" or "$-$"
equilibrium \emph{or} a $+\mapsto -$ limit cycle whose amplitude
vanishes in the thermodynamic limit at all scales, i.e. for all
$(k,\mathbf{q})$. This region can be legitimately described within
the formalism of partition functions and equilibrium field theory.
The second region flows toward the trivial fixed point and does not
support vortices -- this can also (trivially) be described by our
formalism, as it always corresponds to the insulator regime, with no
stable vortices. Thus the consistency check is that our method
predicts no other phases in this region but insulator. In the third
regime where long-term dynamics is either a limit cycle with
amplitude of order unity, or chaos. This regime was studied in
detail in some earlier publications (e.g.~\cite{korn} and references
therein), and it cannot be reached within our present formalism.

\subsection{Numerical checks}

\begin{figure}\centering
\includegraphics[width=150mm]{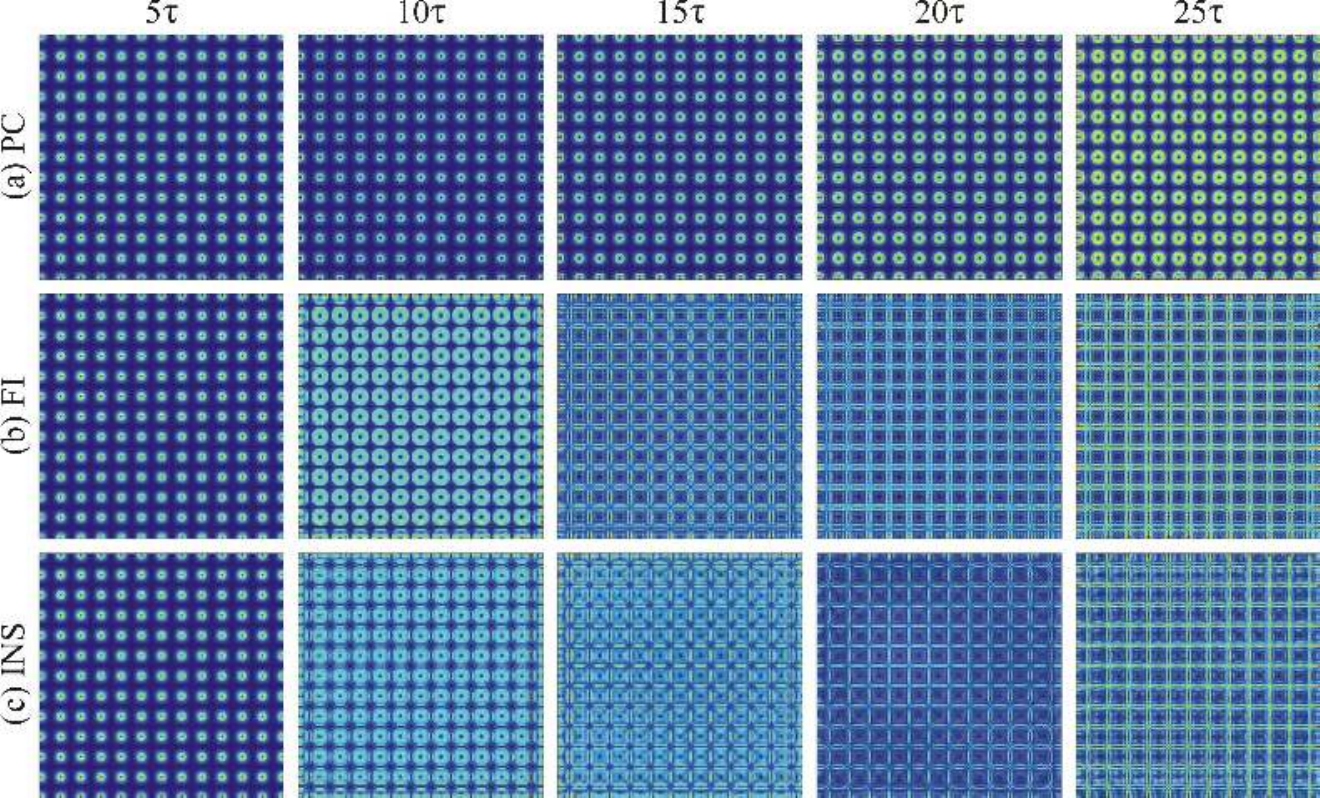}
\caption{\label{figequil} Time evolution of patterns at five
different times: (a) perfect conductor phase, (b) frustrated
insulator phase and (c) insulator phase. In all cases the approach
to equilibrium is obvious, and we expect that for long times a
thermodynamic description is justified. The parameters are the same
as in Fig.~\ref{figcleanlatt}, for the corresponding phases.}
\end{figure}

\begin{figure}\centering
\includegraphics[width=150mm]{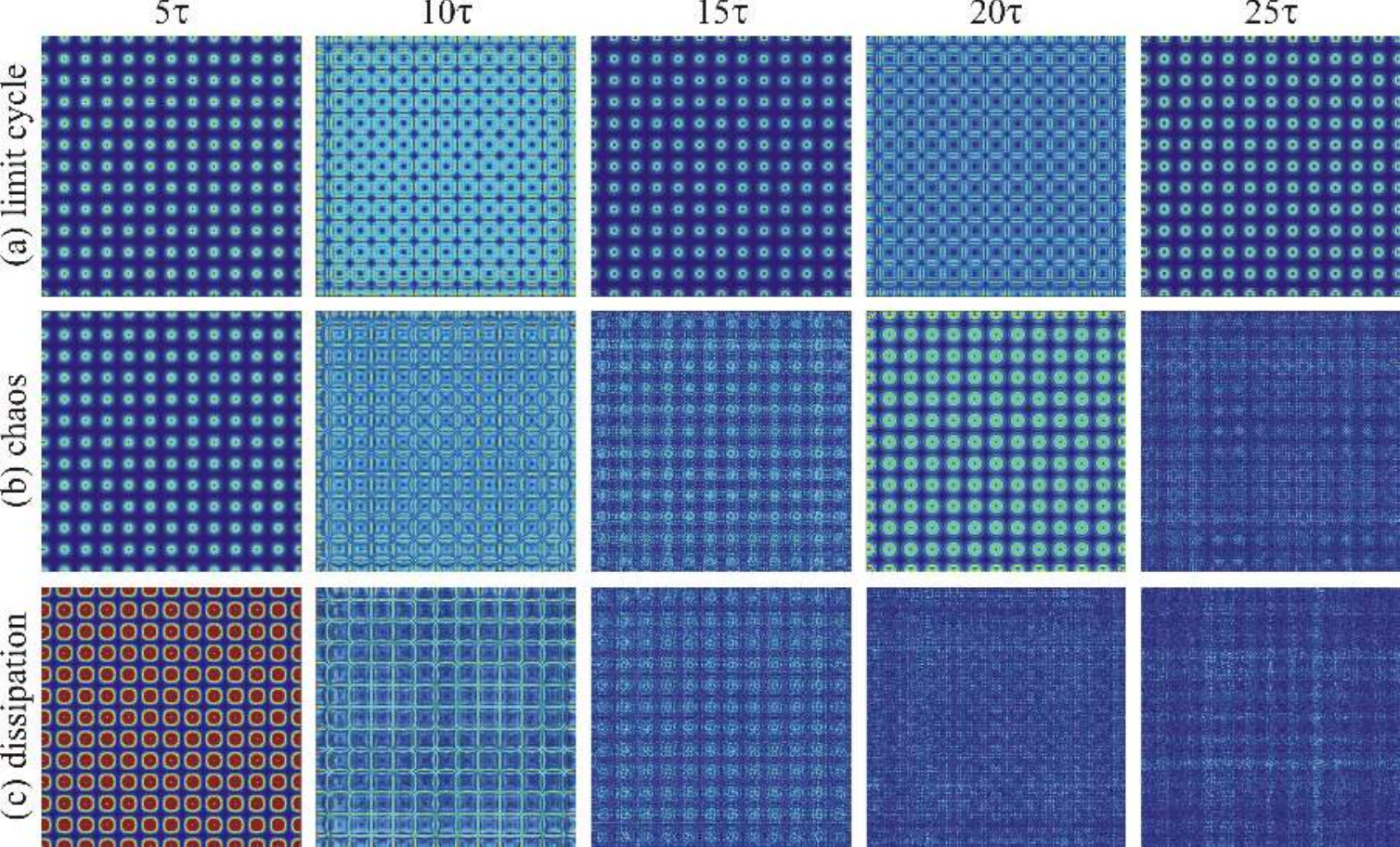}
\caption{\label{fignonequil} Time evolution of non-equilibrium
patterns. In (a) the limit cycle leads to permanent oscillatory
behavior, in (b) wandering along the unstable manifold between the
equilibrium points gives rise to chaos and in (c) dissipation wins
and dynamics dies out. The parameters are the same as in the
previous figure, except that the length $L$ is increased thrice.}
\end{figure}

Now we complement the analytical considerations with numerical
evidence that the phases described in the main text exist as
long-term stable configurations. In Fig.~\ref{figequil} we show the
time evolution of a vortex lattice in three different phases, where
a visual inspection clearly suggests the system approaches
equilibrium. In contrast, in Fig.~\ref{fignonequil} we see first a
pattern that oscillates forever, i.e. follows a limit cycle (A),
becomes incoherent (wandering chaotically over the unstable
manifold, B), or dissipates away (reaching the $0$-fixed point), in
the (C) panel. The loss of stability corresponds to an Andronov-Hopf
bifurcation, as found earlier for non-vortex patterns in
\cite{stabanal}.

\begin{figure}\centering
\includegraphics[width=70mm]{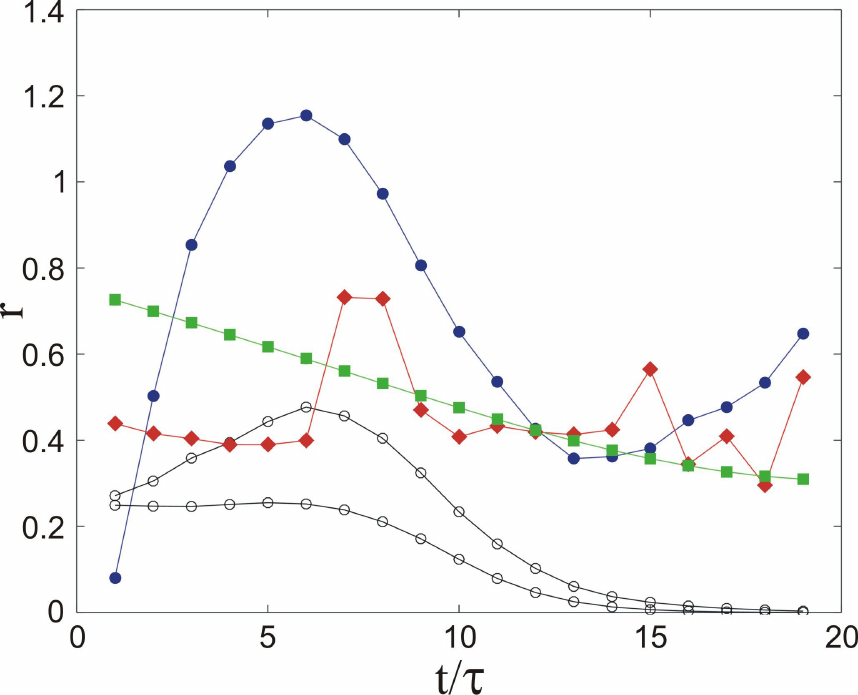}
\caption{\label{fignonequilr} Time evolution of the relaxation rate
$r$ for the various situations from Figs.~\ref{figequil} and
\ref{fignonequil}, illustrating the relaxation to non-trivial
(non-zero-intensity) equilibrium, i.e. "$\pm$"-fixed points
(Figs.~\ref{figequil}(a), \ref{figequil}(c), hollow black circles), limit cycle
(\ref{fignonequil}(a), full blue circles), chaos (\ref{fignonequil}(b), full red romboids) and
the relaxation to trivial (zero-intensity) equilibrium, i.e.
"$0$" fixed point (\ref{fignonequil}c, full green squares). In the main text we
study the cases like the black curves, where time-independent stable
configurations are seen. The symbols are data points from numerics
and the lines are just to guide the eye.}
\end{figure}

Dynamics can be most easily traced by looking at the numerically
computed relaxation rate \be
\label{rrate}\frac{1}{X}\frac{dX}{dt}=\frac{\sum_{x,y}\vert
X\left(t_{j+1}x,y\right)-X\left(t_j;x,y\right)\vert^2}{\sum_{x,y}\vert
X\left(t_j;x,y\right)\vert^2}, \ee which is expected to reach zero
for a generical relaxation process, where in the vicinity of an
asymptotically stable fixed point $X\sim X_\mathrm{eq}+xe^{-rt}$,
will be generically nonzero for a limit cycle or chaos, and will
asymptote to a constant for the $0$-point, where $X_\mathrm{eq}=0$,
so we get $(1/X)dX/dt\sim r$. Fig.~\ref{fignonequilr} summarizes
these possibilities. The black curves, corresponding to
Fig.~\ref{figequil}(a) and Fig.~\ref{figequil}(c), show the
situation which is in the focus of this work -- the approach toward
static equilibrium. The blue curve shows the limit cycle leading to
periodic oscillations. The green curve corresponds to the chaotic
regime with aperiodic dynamics and no relaxation, as in
Fig.~\ref{fignonequil}(b). Finally, the red curve corresponding to
the pattern which radiates away in Fig.~\ref{fignonequil}(b) reaches
a constant value of $r$. In conclusion, the system shows roughly
four classes of dynamics: fixed point, limit cycle, chaos and
incoherence. Our work only covers the first of the four, but the
bifurcation diagrams in the previous subsection give a good hint of
the part of the parameter space which contains them, facilitating
experimental or numerical verification.

\section{Perturbation theory and stability analysis}
\label{appb}

In this Appendix we develop the perturbation theory of the photorefractive beam system starting from the Lagrangian (\ref{psilag}). The perturbation theory yields the criterion for the stability of the intensity patterns as they propagate along the $z$-axis. Formally, it is just the perturbative diagrammatic calculation of the propagator. This calculation explicitly excludes topologically nontrivial patterns and thus is somewhat peripheral for our main goal, understanding the vortex dynamics. But the general ways by which an envelope $\Psi_\pm$ can evolve along the $z$-axis and become unstable remain valid also for vortices. In particular, we will end up with a classification of geometrical symmetries of the intensity pattern $\Psi^\dagger\Psi$; the same symmetries are seen in vortex patterns and are an important guide for numerical and experimental work -- how to recognize instabilities and also phases of the system.

Our system is strongly nonlinear, thus a naive perturbation theory about the "trivial vacuum", i.e. constant beam intensity is out of question. The right way is to perturb about a nontrivial solution, which approximates a stable pattern. This means we treat the light intensity as constant in "time" $z$ but non-constant in space $(x,y)$. This is the hallmark of spatial dynamical solitons: they propagate with a constant profile along the $z$-axis and to a good approximation do not interact with each other and do not radiate \cite{korn}. We thus write $\Psi=\Psi_0+\delta\Psi$, giving $\Psi^\dagger\Psi=I_0+F_0(\delta\Psi_+^\dagger+\delta\Psi_+)+B_0(\delta\Psi_-^\dagger+\delta\Psi_-)+\delta\Psi^\dagger\delta\Psi+O(\vert\delta\Psi\vert^2)$ with $F_0^2+B_0^2=I_0$. The lowest-order Lagrangian for $\Psi_0$ now reads
\be
\label{psimf}\mathcal{L}_0=\Psi_0^\dagger\Delta\Psi_0+\Gamma I_0-\Gamma(1+\tau u+\tau E_0)\log(1+\tau u+I_x+I_0),
\ee
which determines the shape of the solution $\Psi_0(x,y)$ in the first approximation. The dynamical term with $\partial_z\Psi$ drops out (it is proportional to the equation of motion for $\Psi$). Nontrivial propagation in "time" $z$ is obtained from second-order expansion of the potential which is given in the next Appendix in (\ref{psivfluc}) and we will not copy it here. Varying the quadratic expansion with respect to the fluctuation $\delta\Psi$ gives the linearized equation of motion for $\delta\Psi$:
\be
\label{psifluc}\left(\pm\imath\sigma_3\partial_z-q^2+\Gamma-\left(1+\tau u+\tau E_0\right)\right)\delta\Psi^\mp\mp\Gamma\frac{1+\tau u+\tau E_0}{(1+\tau u+I_x+I_0)^2}\delta\Psi^\pm=0,
\ee
where $\delta\Psi^+\equiv\delta\Psi^\dagger,\delta\Psi^-\equiv\delta\Psi$. In homogenous "spacetime" $(z,x,y)$ we can transform to momentum space in both transverse and longitudinal direction. In the transverse plane we get $(x,y)\mapsto (q_x,q_y)$ and $\Delta\mapsto -q^2$. The longitudinal coordinate or "time" $z$ transforms as $z\mapsto k_n$ where $k_n=\pi n/L$, so the "time" maps to discrete frequencies. The reason is of course that its domain is finite, corresponding to the crystal length $L$.

Now we can derive the bare propagator (Green's function) of the fluctuating dynamical field $\delta\Psi$ by inserting the appropriate source $S(z)$ on the right-hand side of the equation (\ref{psifluc}). Normally, the source in the equation for the Green's function is just the Dirac delta function but the counterpropagating nature of our beams imposes a "two-sided" source:
\be
\label{bndcond}S(z)=\left(\begin{matrix}\delta(z) & 0\\ 0 & \delta(z-L)\end{matrix}\right).
\ee
With this source (also Fourier-transformed in $z$), Eq.~(\ref{psifluc}) gives the bare propagator $G^{(0)}_{\alpha\beta}$ for the fields $\delta\Psi^\pm_{\alpha\beta}$:
\be
\label{g0prop}G^{(0)}_{\alpha\beta}(k_n,q)=\left[-\imath k_nS_{\alpha\gamma}(k_n)+A_{\alpha\delta}^* S_{\delta\gamma}(k_n)-B_{\alpha\delta}S_{\delta\gamma}(k_n)\right]\times\left[-k_n^2+A_{\gamma\delta}^*A_{\delta\beta}-B_{\gamma\delta}B_{\delta\beta}^*+\left[A^*,B\right]_{\gamma\beta}\right]^{-1}.
\ee
The auxiliary matrices $A,B$ are defined as follows:
\be
\label{g0propmat}A_{\alpha\beta}=\imath\left(\begin{matrix}P_0+P_1-q^2 & P_0\\ -P_0 & -P_0-P_1+q^2\end{matrix}\right),~~B_{\alpha\beta}=\imath\left(\begin{matrix}P_0 & P_0\\ -P_0 & -P_0\end{matrix}\right),
\ee
where $P_1=(1/4)I_0\Gamma(1+\tau u+\tau E_0)/(1+\tau u+I_x+I_0)^2, P_0=\Gamma-\Gamma(1+\tau u+\tau E_0)/(1+\tau u+I_x+I_0)$, and $S(k_n)=\mathrm{diag}(1,e^{\imath k_nL})$.

Now we have the basic ingredient of the perturbation theory -- the bare propagator. The self-energy correction $\Sigma$ of the propagator from the potential $V_\mathrm{eff}$ can be expanded in a power series over $\delta\Psi$, which gives an infinite tower of vertices. Simple combinatorial considerations give the expansion
\be
\label{veffexp}\Sigma=\sum_{j_1,j_2,j_3\in\mathbb{N}}\frac{(-1)^{j_1+j_2+j_3}(j_1+j_2+j_3-1)!}{j_1!j_2!j_3!}\frac{\Gamma(1+\tau u+\tau E_0)}{(1+\tau u+I_0+I_x)^{j_1+j_2+j_3+1}}(\Psi_0^\dagger\delta\Psi)^{j_1}(\Psi_0\delta\Psi^\dagger)^{j_2}(\delta\Psi_\dagger\delta\Psi)^{j_3},
\ee
and the contraction over the internal indices of $\Psi^\pm,\delta\Psi^\pm$ is understood. Now we can formulate the diagrammatic rules. We have two kinds of propagators, $G^{(0)}$ and its Hermitian conjugate. The mean-field values $\Psi_0^\pm$ are external sources. The term of order $(j_1,j_2,j_3)$ contains $j_1+j_3$ propagator lines $G^{(0)}$ ($j_1$ of them ending with the source $\Psi_0$), and $j_2+j_3$ lines $\left(G^{(0)}\right)^\dagger$ ($j_2$ of them ending with a source $\Psi_0^\dagger$); altogether there are $j\equiv j_1+j_2+2j_3$ lines. The expansion has to be truncated at some $j$. Since the mass dimension of $\Psi$ is $1$, the $(j_1,j_2,j_3)$-diagram has the scaling dimension $2-2(j_1+j_2+2j_3)<0$, so \emph{all} diagrams are irrelevant in the IR. This means we can make a truncation at small $j$.\footnote{We do not worry about the UV divergences: we have an effective field theory and the UV cutoff is physical and finite.} The leading terms are those where the order of the perturbation in $\delta\Psi^\pm$, which equals $j_1+j_2+2j_3$, is the smallest. This gives two classes of diagrams, one with $j_1=1,j_2=j_3=0$ and another with $j_2=0,j_1=j_3=0$. They contain a single external source and introduce the wavefunction renormalization, $G^{(0)}\mapsto ZG^{(0)}$, which does not influence the stability analysis. The four quadratic terms (with $(j_1,j_2,j_3)=(2,0,0),(0,2,0),(1,1,0),(0,0,1)$) introduce a mass operator. Only the terms $(1,1,0)$ and $(0,0,2)$ are trivial (non-interacting); the other two are interacting as they contain $\left(\delta\Psi^\pm\right)^2$, and require the calculation of an internal loop, giving the dressed propagator:
\be
\label{gpropmass}G_{\alpha\beta}^{-1}(k_n,q)=\left(G^{(0)}(k_n,q)\right)_{\alpha\beta}^{-1}+\left(m^2\right)_{\alpha\beta},
\ee
where the mass squared is a \emph{positive} matrix, because the corresponding coefficients in (\ref{veffexp}) have positive sign (from the term $(-1)^{j_1+j_2+j_3}$ with $j_1+j_2+j_3=2$) and the integral of the bare propagator is also positive. Explicitly, it reads
\be
\label{gpropmass2}\left(m^2\right)_{\alpha\beta}=\frac{\Gamma(1+\tau u+\tau E_0)}{(1+\tau u+I_0+I_x)^2}\sum_{k_n}\int_0^\infty dqqG_{\alpha\beta}^{(0)}(k_n,q),
\ee
where the discrete "frequency" $k_n$ is summed in steps of $\pi/L$. Other than the mass renormalization, the dressed propagator has the same structure as the bare one. Now we will consider what this means for the stability of the patterns.

\subsection{The pole structure, stability and dispersion relations}

Consider the poles of the propagator defined by the zeros of the eigenvalues of the matrix $G^{-1}_{\alpha\beta}(k_n,q)$. The stable solution corresponds to the situation where the perturbation $\delta\Psi_\pm$ dies out along $z$, so the stability of the solution is determined by the condition that the pole in $q$ should have a non-positive imaginary part, i.e. that a small perturbation decays. The denominator depends on $k_n,q$ solely through $k_n^2,q^2$; it is linear in $k_n^2$ and quadratic in $q^2$. Therefore, each of the two eigenvalues $\lambda_\pm$ defines two pairs of opposite poles, $\pm q_{*+},\pm q_{*-},\pm q_{**+},\pm q_{**-}$. Out of these, two pairs are positive for all parameter values, so no imaginary part can arise, and we have either two pairs of centrally-symmetric imaginary poles, or one such pair, or none at all. We thus expect the sequence of symmetry-breaking transitions:
\be
\label{symmseq}O(2)\longrightarrow\mathbb{C}_4\longrightarrow\mathbb{C}_2
\ee
Full circular symmetry is expected when there is no instability. With a single pair of unstable eigenvalues, we expect a square-like pattern with $\mathbb{C}_4$ symmetry, and with two pairs only a single reflection symmetry axis remains, yielding the group $\mathbb{C}_2$. Only in the presence of disorder in the background lattice intensity pattern $I_x$ can we expect the full breaking of the symmetry group down to unity, but this is an \emph{explicit} breaking and is not captured by this analysis.

\begin{figure}\centering
\includegraphics[width=70mm]{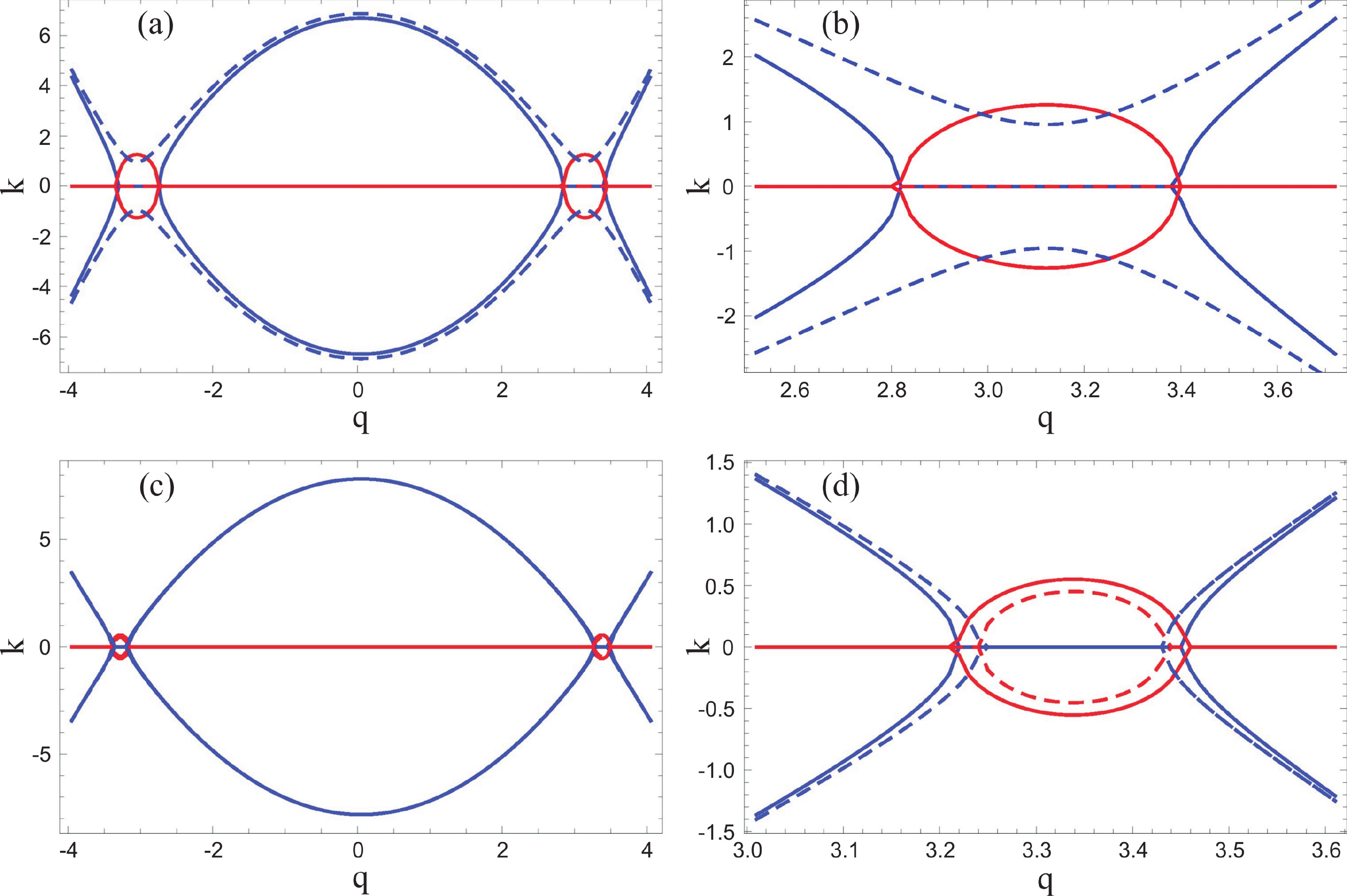}
\caption{\label{figprsolstab} Dispersion relation (position of the
poles of the propagator) $k(q)$, where $k$ is the continuous
approximation of the discrete effective momentum $k_n=\pi n L$, for
$I_x=0$ (a, b) and $I_x=1$ (c, d). The plots (b, d) are zoom-ins
from the plots (a ,c). Blue lines denote $\Re k$ and red lines $\Im
k$. Notice that the propagator contains only $k_n^2$ and $q^2$, so
the pole has two copies with opposite sign and is either real or
pure imaginary. Dashed lines are the corrected relations, with
dressed propagator instead of the bare one. In the top panels, the
region of instability, with $\Im k_n\neq 0$, is cured by the
nonlinear corrections, whereas in the bottom panels the instability
remains. This generically happens at finite $q$ and corresponds to
the edge instability. Parameter values: $I_0=1$, $\Gamma=15$,
$L=10\mathrm{mm}$.}
\end{figure}

The dispersion relation for a typical choice of parameter values is represented in Fig.~\ref{figprsolstab}, where we plot the location of the pole $k(q)$ in the continuous approximation (interpolating between the $k_n$ values), with real parts of the pole in blue and imaginary in red. Since we have two pairs of opposite eigenvalues, the dispersion is $P$-symmetric in both $x,y$ and $z$ (remember that "time" is really another spatial dimension), and any dispersion relation with a nonzero imaginary part will have a branch in the upper half-plane, i.e. an unstable branch. The only way out of instability is that the pole is purely real, i.e. infinitely sharp -- this quasiparticle-like excitation signifies a solitonic solution. In Fig.~\ref{figprsolstab}, the dashed lines are drawn with the bare propagator $G^{(0)}$ and the full lines with the dressed propagator $G$, for the sets of parameter values. The perturbation always reduces the instability, i.e. the magnitude of the imaginary part of the poles -- in (a, b) completely, resulting in zero imaginary part, and in (c, d) only partially. This reduction of instability likely explains the fact that linear stability analysis works extremely well for hypergaussian beams (which have most power at small values of $q$), as found in \cite{stabanal}.

\begin{figure}\centering
\includegraphics[width=70mm]{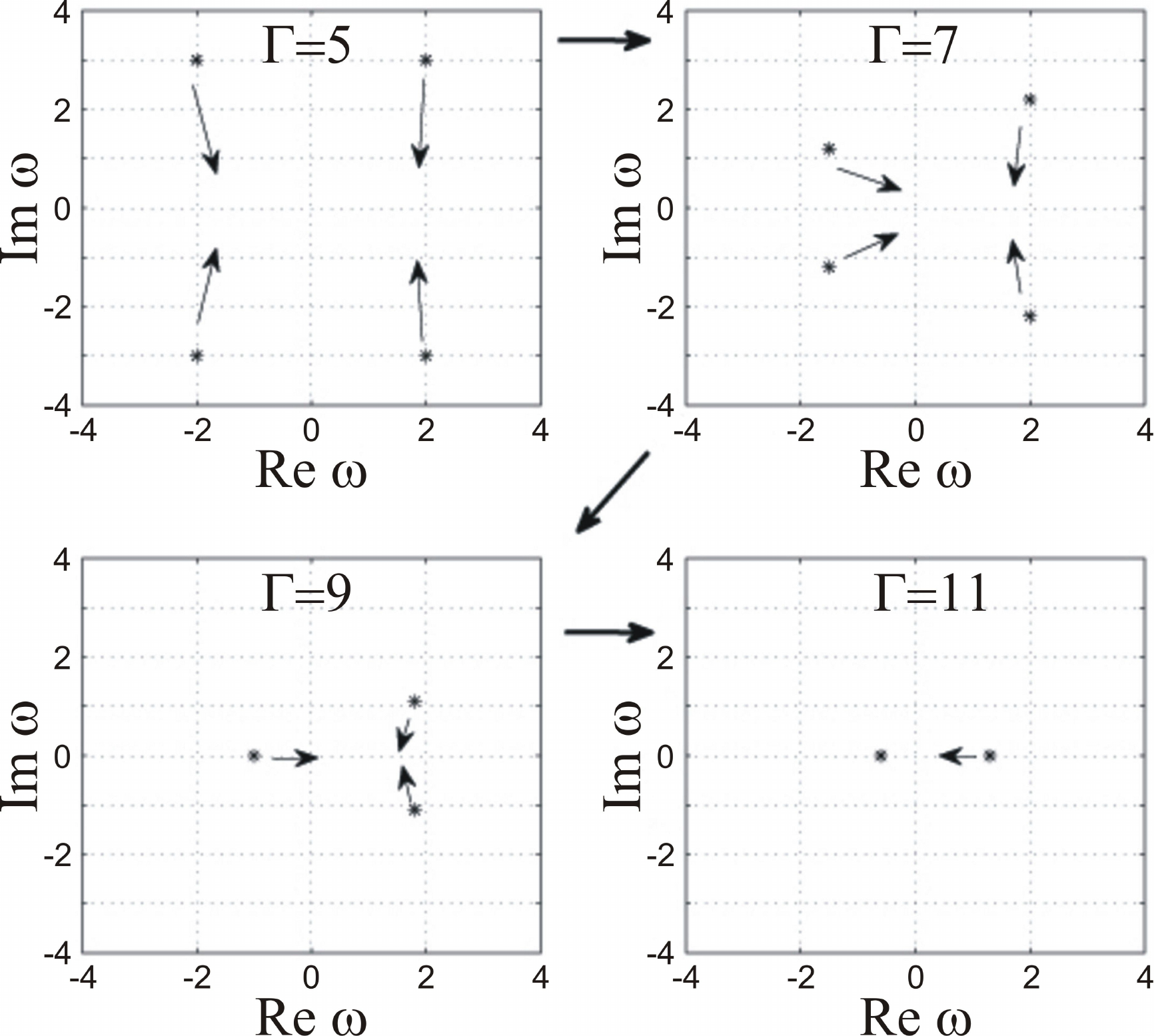}
\caption{\label{figpoles} The movement of the poles in the complex
momentum plane in the case of central instability, for four
different values of the PR coupling constant $\Gamma=5,7,9,11$. The
complex momentum $k$ is denoted by $\omega$. Starting from the
$\mathbb{C}_4$-symmetric situation with two pairs of
complex-conjugate poles, we first break the symmetry down to
$\mathbb{C}_2$ and eventually loose all geometric symmetry as the
two pairs merge into two real poles. Parameter values
$I_0=I_x=1,L=10\mathrm{mm}$.}
\end{figure}

The fact that the imaginary region always lies at finite $q$ implies that the instability always starts at a finite scale, which corresponds to the behavior seen in the edge instability, which is shown, e.g., in Fig.~\ref{figcleangauss1}. In order to understand the central instability, which starts from a single point, corresponding to $q\to\infty$, one needs to take into account also the higher order corrections from the potential (\ref{veffexp}) which, as we discussed, diverge at $q\to\infty$. While we always have a natural UV cutoff, it may happen that the corrections become large (though finite) before that UV scale is reached. We postpone a detailed account for the subsequent publication, and content ourselves to give only the diagram of the movement of the poles in the complex plane. Higher-order terms bring $q$-dependent corrections and break the inversion symmetry, resulting in the evolution of poles as in Fig.~\ref{figpoles}. The instability corresponds to the situations where at least one pole has a positive imaginary part, i.e. the first three situations in the figure. The last pattern, with no symmetry at all and two real poles, is stable (but not asymptotically stable, as there is no pole with non-zero negative imaginary part).

The analysis performed here is obviously incomplete, and we have contented ourselves merely to give a sketch of how the instabilities considered in the main text arise, as well as to formulate a perturbation scheme which allows one to study such phenomena. Further work along the lines of \cite{stabanal,kivshar3,kivshar4} is possible by making use of our formalism, and we plan to address this topic in the future.

\section{Derivation of the vortex Hamiltonian from the microscopic Lagrangian}
\label{app00}

Starting from the vortex solution (\ref{psivort}) we want to obtain an effective Hamiltonian for the vortex-vortex interaction. The task is to separate the kinetic term of the winding phase (with $\nabla\theta_{0\pm}=\sum_iQ_\pm\log\vert\mathbf{r}-\mathbf{r}_i\vert$) from (i) the intensity fluctuations $\delta\psi_\pm$ about some background value $\psi_{0\pm}$ and (ii) the non-vortex phase fluctuations ($\delta\theta_\pm$) in (\ref{psivort}). The first task requires us to integrate out the amplitude fluctuations in the quadratic approximation. We first write $\Psi=\Psi_0+\delta\Psi$ and expand the Lagrangian to quadratic order:
\bea
\nonumber\mathcal{L}=\mathcal{L}_0+\mathcal{L}_2,\\
\nonumber\mathcal{L}_0=\frac{1}{2}\partial_r\psi_0^\dagger\partial_r\psi_0+\frac{I_0}{2r^2}\vert\nabla\theta_{0\alpha}\vert^2,\\
\nonumber\mathcal{L}_2=\frac{1}{2}\partial_r\delta\Psi^\dagger_\alpha\partial_r\delta\Psi_\alpha+\frac{1}{2r^2}\delta\Psi^\dagger_\alpha\vert\nabla\theta_{0\alpha}\vert^2\delta\Psi_\alpha+V_\mathrm{eff}\left(\delta\Psi_\pm\right),\\
\nonumber V_\mathrm{eff}(\delta\Psi^\pm)=-\Gamma\delta\Psi^\dagger\delta\Psi-\Gamma\frac{1+\tau u+\tau E_0}{2(1+\tau u+I_x+I_0)^2}\times\\
\label{psivfluc}\times\left(\left(\Psi_0\delta\Psi^\dagger\right)^2+\left(\Psi_0^\dagger\delta\Psi\right)^2-2\left(1+\tau u+I_x+I_0\right)\delta\Psi^\dagger\delta\Psi-\left(\Psi_0\sigma_2\delta\Psi^\dagger\right)\left(\Psi_0^\dagger\sigma_2\delta\Psi\right)\right).
\eea
The zeroth-order (non-fluctuating) term $\mathcal{L}_0$ determines the intensity $I_0=\psi_0^\dagger\psi_0$, and produces the kinetic term for the vortex phase $\theta_\alpha$, which gives just two decoupled copies of the conventional XY vortex gas. The quadratic part $\mathcal{L}_2$ becomes quite involved when we separate the amplitude $\delta\psi$ and the phase $\delta\theta$. Inserting (\ref{psivort}) into (\ref{psivfluc}), one gets a quadratic action for $\delta\psi_\alpha$ and $\delta\theta_\alpha$. The rest gives a coupled quadratic action for the amplitude and phase fluctuations. Altogether, the Lagrangian is
\be
\label{psivort2}\mathcal{L}=\frac{1}{2}\left(\delta\psi_+'^2+\delta\psi_-'^2\right)+\frac{1}{2r^2}\left(\delta\psi_+^2+\delta\psi_-^2\right)\vert\nabla\theta_\alpha\vert^2+\delta\psi_\alpha\hat{K}_{\alpha\beta}\delta\psi_\beta+\left(\delta\psi_\alpha^\dagger\psi_{\alpha}\nabla\theta_{\alpha}\nabla\delta\theta_\alpha+\mathrm{h.c.}\right)+\ldots,
\ee
where $(\ldots)$ denote all terms of cubic or higher order in amplitude or phase fluctuations $\delta\psi_\alpha,\delta\theta_\alpha$, and we have left out the constant terms independent of all field values. Primes denote the derivatives with respect to $r$. The first term defines the intensity fluctuations through $\psi_\alpha(r)$, and the second term (transformed through partial integration) yields the aforementioned conventional Coulomb gas of vortices after inserting the vortex solution from (\ref{psivort}) for $\theta_\alpha$. The third term has the meaning of stiffness or mass matrix for intensity fluctuations and the last term gives rise to the coupling between the flavors, upon integrating out $\delta\psi$. The matrix $\hat{K}$ is
\be
\label{kmat}\hat{K}=\frac{1}{\left(b+\frac{3}{2}I_0\right)\left(b-\frac{1}{2}I_0\right)}\left(\begin{matrix}b+\frac{I_0}{2} & I_0\\ I_0 & b+\frac{I_0}{2}\end{matrix}\right),~~b=\Gamma\frac{1+\tau E_0}{2(1+\tau E_0)^2}\left(2+2I_x+3I_0\right).
\ee
The action is quadratic in $\delta\psi$; therefore, we know how to integrate it out and obtain an effective action depending only on phase fluctuations. To do that, we need to solve the eigenvalue equation for $\delta\psi$ obtained from (\ref{psivort2}), which reads
\be
\label{eompsi}\partial_{rr}\delta\psi_\alpha-K_{\alpha\beta}\delta\psi_\beta=\left(\frac{\vert\nabla\theta\vert^2}{2r^2}+\lambda_\pm\right)\delta\psi_\alpha,
\ee
and is solved by diagonalizing the system $\delta\psi_\alpha\mapsto\delta\chi_\alpha=U_{\alpha\beta}\delta\psi_\beta$ and reducing it to the Bessel equation:
\be
\label{eompsisol}\delta\chi_\pm(r)=\sqrt{r}\left(C_\pm J_\nu\left(\sqrt{2\left(K_{11}\mp K_{12}-\lambda_\pm\right)}r\right)+D_\pm Y_\nu\left(\sqrt{2\left(K_{11}\mp K_{12}-\lambda_\pm\right)}r\right)\right),
\ee
with $\nu=\sqrt{1/4+\vert\nabla\theta\vert^2}$, where $J_\nu,Y_\nu$ are Bessel functions of the first and second kind, respectively. For well-defined behavior close to the vortex core (for $r\sim a$) we have $D_\pm=0$, and $C_\pm$ are arbitrary as the equation is linear. The eigenvalues $\lambda_{n\pm},n=0,1,ldots$ are obtained by requiring that the fluctuation decays to zero at the crystal edge $r=\Lambda$:
\be
\label{eompsilam}\sqrt{2K_{11}\mp K_{12}-\lambda_{n\pm}}\Lambda=j_n(\nu),
\ee
where $j_n$ is the $n$-th zero of the Bessel function $J_\nu$. The values $\lambda_{n\pm}$ are impossible to express analytically in closed form, however it is not necessary for our purposes as $\Lambda\gg a$. The functional determinant obtained after integrating out $\chi_\pm$ is now expressed in terms of the eigenvalues:
\be
\mathcal{K}_{\alpha\beta}=\log\left(\prod_n\lambda_{n\alpha}\lambda_{n\beta}\right)^{-1/2}=-\frac{1}{2}\sum_n\log\left(K'_\alpha+K'_\beta+\frac{2j_n\left(\nu\right)^2}{\Lambda^2}\right)\sim -\frac{\Lambda}{2}\left(K'_\alpha+K'_\beta\right)+O\left(\frac{1}{\Lambda}\right),
\ee
where $K'_\pm=2K_{11}\mp K_{12}$. Now we are left with a solely phase-dependent quadratic Lagrangian:
\be
\label{psivort3}\mathcal{L}=\psi_{0\alpha}\nabla\theta_\alpha\nabla\delta\theta_\alpha\hat{\mathcal{K}}_{\alpha\beta}\psi_{0\beta}\nabla\theta_\beta\nabla\delta\theta_\beta+\frac{I_0}{2r^2}\vert\nabla\theta_\alpha\vert^2
\ee
The final task is to integrate out the phase fluctuations, which is a trivial Gaussian integration, yielding:
\be
\label{psivort4}\mathcal{L}=\frac{I_0}{2r^2}\vert\nabla\theta_\alpha\vert^2+I_0\nabla\theta_\alpha\mathcal{K}^{-1}_{\alpha\beta}\nabla\theta_\beta
\ee
The resulting Lagrangian now depends only on the vortexing phases $\theta_\alpha$. The first term is carried from the original Lagrangian, and it does not mix the flavors. But the second term, stemming from the fluctuations, has nonzero mixed $\pm$ cross-terms. The quadratic derivative terms can be transformed by partial integration to the familiar Coulomb gas form of the XY model, with the same-flavor coupling which is already present in absence of fluctuations, \emph{and} the coupling between the vortices of different flavors. Thus the existence of two beams together with the fact that amplitude and phase fluctuations do not decouple give us a richer system, with interaction between two vortex flavors. For future use it is more convenient to look at the vortex Hamiltonian $\mathcal{H}_\mathrm{vort}$ -- the difference from the Lagrangian lies just in the sign of the term $V_\mathrm{eff}$. This finally yields the Hamiltonian (for Eq.~(\ref{hamvort}), repeated here for convenience):
\be
\label{hamvortapp}\mathcal{H}_\mathrm{vort}=\sum_{i<j}\left(g\vec{Q}_i\cdot\vec{Q}_j+g'\vec{Q}_i\times\vec{Q}_j\right)\log r_{ij}+\sum_i\left(g_0\vec{Q}_i\cdot\vec{Q}_i+g_1\vec{Q}_i\times\vec{Q}_i\right),
\ee
with $r_{ij}\equiv\vert\mathbf{r}_i-\mathbf{r}_j\vert$, and the indices $1\leq i,j\leq N$ sum over all the vortices. The coupling constants $g,g',g_0,g_1$ are the result of integrating out the intensity fluctuations and in general are given by rather cumbersome (and not very illustrative) functions of $\Gamma,I_0,\tau$. We give the expressions at leading order just for comparison with numerics:
\bea
\nonumber g&=&I_0+\frac{4b+2I_0}{(2b+3I_0)(2b-I_0)}\\
\nonumber g'&=&\frac{4I_0}{(2b+3I_0)(2b-I_0)}\\
\label{gmicro}b&=&\Gamma\frac{1+\frac{\tau}{L}-\tau\frac{I_0+I_x}{1+I_0+I_x}}{2\left(1+\frac{\tau}{L}-\tau\frac{I_0+I_x}{1+I_0+I_x}\right)^2}\left(2+2\frac{\tau}{L}+2I_x+3I_0\right).
\eea
These expressions are used later to redraw the phase diagram in the space of physical parameters $\Gamma,I_0,I_x,L$.

\section{Multi-vortex mean-field theory}
\label{appmf}

For a mean-field treatment of a system with multiple vortices, we start from the Hamiltonian (\ref{hamvort}) and introduce the order parameter fields in the following way. Denote the number of vortices with charge $(1,1)$ by $\rho_{2+}$ and the number of vortices $(1,-1)$ by $\rho_{2-}$; due to charge conservation, this means we also have $\rho_{2+}$ vortices of type $(-1,-1)$ and $\rho_{2-}$ vortices with charge $(-1,1)$. The number of single-charge vortices of type $(1,0)$ and $(0,1)$ is denoted by $\rho_{1+}$ and $\rho_{1-}$, respectively. Denote also $\rho_2\equiv\rho_{2+}+\rho_{2-}$ and $\delta\rho_2\equiv\rho_{2+}-\rho_{2-}$ (notice that $-\rho_2\leq\delta\rho_2\leq\rho_2$), and finally $\rho_1\equiv\rho_{1+}+\rho_{1-}$. We insert this into the vortex Hamiltonian $\mathcal{H}_\mathrm{vort}$, and assume that the long-ranged logarithmic interaction $\log r_{ij}$ justifies the mean-field approximation: for $i\neq j$ we can approximate $\log r_{ij}\sim\log\Lambda$, assuming that average inter-vortex distance is of the same order of magnitude as the system size. For the core energy, we know that $g_0,g_1\sim\log(a/\epsilon)\sim -\log\epsilon\sim\log\Lambda$, where in the last equality we have assumed that the UV cutoff $\epsilon$ is of similar order of magnitude as the inverse of the IR cutoff $1/\Lambda$, which is natural.\footnote{Nevertheless, this is clearly not a rigorous argument. Our mean-field calculation is somewhat sketchy and merely assumes that the long-range interactions can safely be modeled as a uniform vortex charge field.} Thus all terms are proportional to $L\log\Lambda$ and we can write
\be
\label{frenmf}\mathcal{F}_\mathrm{mf}=\beta\log\frac{\Lambda}{a}\left[2\left(g-1\right)\rho_2+2g'\delta\rho_2+\left(g-1\right)\rho_1\right]\equiv A\rho_2+B\delta\rho_2+\frac{B}{2}\rho_1.
\ee
We use the notation $\beta\equiv L$, to emphasize the analogy with the free energy of spin vortices, where $\beta$ is the inverse temperature. The analogy is purely formal as our system is not subject to thermal noise. Now the ground state is determined by minimizing the free energy, i.e. the effective action of the system. Notice that $\mathcal{F}_\mathrm{mf}$ is linear in the fields $\rho_2,\delta\rho_2,\rho_1$ so the optimal configurations have either $\mathcal{F}_\mathrm{mf}=0$ or $\mathcal{F}_\mathrm{mf}\to-\infty$, and the mean-field densities $\rho_{1,2}$ are either zero, or arbitrary (formally infinite). This is a well-known property of the 2D Coulomb gas and has to do with the fact that (assuming the cutoff dependence has been eliminated) this system is conformal invariant in the insulator phase, so all finite densities $\rho$ are equivalent: there is no other scale to compare $\rho$ to. Likewise, the prefactor $\beta$ can be absorbed into the definition of the coupling constants $g,g'$ and thus is not an independent parameter (this is well-known also from the single-flavor case). Minimizing (\ref{frenmf}) is an elementary exercise and we find again four regimes, corresponding to the four phases we guessed based on the single-vortex free energy $\mathcal{F}^{(1)}$:
\begin{enumerate}
\item{For $A>0,A>\vert B\vert$ the minimum is reached for $\rho_2=\delta\rho_2=\rho_1=0$. In the ground state there are no vortices at all -- the system is a vortex insulator.}
\item{For $B>0,A+B<0$, giving $g+g'<1,g'>0$, the free energy has its minimum for $\rho_2>0$ and $\delta\rho_2=-\rho_2$ (notice that $-\rho_2\leq\delta\rho_2\leq\rho_2$). This means $\rho_{2+}=0,\rho_{2-}>0$, so opposite-charged vortices $(Q,-Q)$ proliferate, and the system is dominated by the interactions between the charges. This is the frustrated vortex insulator regime. Since $g'<0$, the single-charge vortices (density $\rho_1$) are suppressed.}
\item{For $B<0,A+B<0$, i.e. $g+g'<1,g'<0$ the minimum is reached for $\rho_2=\delta\rho_2>0$, i.e. $\rho_{2-}=0$ so the vortices $(Q,Q)$ can proliferate. However, since $g'<0$ there is also nonzero single-flavor density $\rho_1$ and the proliferation of vortices $(Q,0)$ and $(0,Q)$ which generically dominate over two-flavor vortices. This is the conductor phase, with mostly single-flavor vortices (as in the standard XY model).}
\item{The point $A=B=0$ is special: naively, from (\ref{frenmf}) arbitrary nonzero $\rho_1,\rho_2,\delta\rho_2$ are allowed. Of course, higher-order corrections will change this but the energy cost of vortex formation will generically be smaller than in previous phases. This is the vortex perfect conductor phase. In the mean-field approach it looks like a single point, but that will turn out to be an artifact of the mean-field approach: for small nonzero $A,B$ the system still remains in this phase.}
\end{enumerate}
In terms of the original parameters $g,g'$, one sees the insulator phase is given by $g+g'>1$, and the conductor and the FI are separated by the line $g'=0$. We can now sketch the phase diagram, which is given in Fig.~\ref{figphdiag}(a), side by side with the more rigorous diagram obtained by the RG flow, in the subsubsection IIIB2.

\section{Counter-propagating boundary conditions}
\label{appc}

In the derivation of the vortex Hamiltonian and its RG analysis, we have pulled under the rug the treatment of the CP boundary conditions: the effective Hamiltonian (and consequently the partition function and the phase diagram) depends solely on the bulk configuration, and nowhere can one see the fact that $\Psi_+(z=0;\mathbf{r},t)$ and $\Psi_-(z=L;\mathbf{r},t)$ are fixed. Now we will explicitly show that these boundary conditions are irrelevant in the RG sense, i.e. they contribute additional, boundary terms to the effective Hamiltonian, but these terms do not change the fixed points to which the solution flows.

The full Hamiltonian with correct CP boundary conditions is obtained by adding the $F$-source at $z=0$ and the $B$-source at $z=L$ to the Lagrangian $\mathcal{L}$ from (\ref{psilag}) or, equivalently, to the equations of motion. The sources impose the conditions $F(z=0;x,y;t)=F_0(x,y)$ and $B(z=L;x,y;t)=B_0(x,y)$ so they equal
\be
\label{cpsource}J_+=F_0(x,y)\delta(z),~~J_-=B_0(x,y)\delta(z-L),
\ee
and the full Lagrangian is
\be
\label{sourceham}\mathcal{L}_\mathrm{CP}=\mathcal{L}+J_+\Psi_++J_-\Psi_-\mapsto\mathcal{L}+F_0(x,y)\Psi_+(z;x,y;t)+e^{\imath kz}B_0(x,y)\Psi_-(z;x,y;t).
\ee
Unlike the Dirac delta source (\ref{bndcond}) for the Green's function, now the source has nontrivial dependence on transverse coordinates. Now we can insert the vortex solution (\ref{psivort}) in both $\Psi_\pm$ and $F_0,B_0$ and repeat the steps from the subsequent derivation. The vortex charges in $F_0$ can be denoted by $\vec{P}_{i'}^{(+)}\equiv (P_{i'+},0)$ and $\vec{P}_{i'}^{(-)}\equiv (0,P_{i'-})$; by definition, the $+$ component of $B_0$ as well as the $-$ component of $F_0$ are zero and thus carry no vorticity. The primed indices refer to the vortices in the input beams, and the non-primed, like before, to the bulk vortices. Notice the source term changes sign upon performing the Legendre transform, appearing as $-J_+\Psi_+-J_-\Psi_-$ in the Hamiltonian.

Now we will check if the RG flow of the Hamiltonian with boundary terms is affected by the sources. In the notation introduced above, the total vortex Hamiltonian is
\be
\label{sourcehamvort}\mathcal{H}_\mathrm{CP}=\sum_{i,j}\left(g\vec{Q}_i\cdot\vec{Q}_j+g'\vec{Q}_i\times\vec{Q}_j\right)\log r_{ij}+\sum_{i',j}\delta(z)P_{i'+}\left(gQ_{j+}+g'Q_{j-}\right)\log r_{i'j}+\sum_{i,j'}\delta(z-L)P_{j'-}\left(gQ_{i-}+g'Q_{i+}\right)\log r_{ij'}.
\ee
Notice there is no source-source interaction: same-flavor interaction cannot exist as $P_{i'-}=P_{j'+}=0$, and cross-flavor interaction does not exist as $J_+$ and $J_-$ exist at different $z$ values, i.e. the cross-term would be proportional to $\delta(z)\delta(z-L)$ and thus vanishes. The presence of sources breaks the spatial homogeneity, complicating the traces (integrals over the positions of virtual vortex-antivortex pairs), but does not change the main line of the calculation. The fluctuation of the partition function due to vortex pair creation is now
\bea
\nonumber\frac{\delta\mathcal{Z}}{\mathcal{Z}}=1+\frac{y^4}{4}\sum_{\vec{q}}\int d^2r\int d^2r_{12}e^{-C\left(\vec{q},\mathbf{r}_1;-\vec{q},\mathbf{r}_2\right)
-\sum_{j'}\left[D_{j'}^+\left(\vec{q},\mathbf{r}_1\right)-D_{j'}^-\left(\vec{q},\mathbf{r}_2\right)\right]}\times\\
\label{sourcezfluc}\times\left[e^{C\left(\vec{Q}_1,\mathbf{R}_1;\vec{q},\mathbf{r}_1\right)+C\left(\vec{Q}_1,\mathbf{R}_1;-\vec{q},\mathbf{r}_2\right)+C\left(\vec{Q}_2,\mathbf{R}_2;\vec{q},\mathbf{r}_1\right)+C\left(\vec{Q}_2,\mathbf{R}_2;-\vec{q},\mathbf{r}_2\right)}-1\right].
\eea
We have denoted $C\left(\vec{Q}_1,\mathbf{R}_1;\vec{Q}_2,\mathbf{R}\right)\equiv\left(g\vec{Q}_1\cdot\vec{Q}_2+g'\vec{Q}_1\times\vec{Q}_2\right)\log R_{12}$, and the coupling to the sources is encapsulated in the function
\be
\label{sourced}D_{j'}^\pm\left(\vec{q},\mathbf{r}\right)\equiv\delta\left(z-z_\pm\right)\left(g_\pm\vec{q}\cdot\vec{P}_{j'}+g'_\pm\vec{q}\times\vec{P}_{j'}\right)\log\vert\mathbf{r}-\mathbf{r}_{j\pm}\vert,
\ee
with $z_+=0,z_-=L$. The coupling constants $g_\pm,g'_\pm$ are obtained from $g,g'$ in (\ref{sourcezfluc}) by replacing
\be
\label{fmicro}I_0\mapsto\sqrt{I_0I_\pm},
\ee
with $I_+=\vert F_0\vert,I_-=\vert B_0\vert$. Now the exponential of the extra term with sources also needs to be expanded in $r_{12}$, as the combinations $2\mathbf{r}=\mathbf{r}_1+\mathbf{r}_2,\mathbf{r}_{12}=\mathbf{r}_1-\mathbf{r}_2$ do not decouple anymore and exact integration is impossible. After writing $D_{j'}^\pm(\vec{q},\mathbf{r}_{1,2})=D_{j'}^\pm\left(\vec{q},\mathbf{r}\right)\pm\mathbf{r}_{12}\cdot\nabla D_{j'}^\pm\left(\vec{q},\mathbf{r}\right)+\ldots$ and similarly for $C$, we notice first that the zeroth-order terms from $D_{j'\pm}$ cancel out: $D_{j'}^\pm\left(\vec{q},\mathbf{r}\right)+D_{j'}^\pm\left(-\vec{q},\mathbf{r}\right)=0$. Then we get (at quadratic order in $r_{12}$ in the integrand):
\bea
\nonumber\frac{\delta\mathcal{Z}}{\mathcal{Z}}&=&1+\frac{y^4}{4}\sum_{j'}\int d^2r\int d^2r_{12}e^{-C\left(\vec{q},\mathbf{r}_{12};-\vec{q},0\right)}\times\\
\nonumber&\times&\left[\mathbf{r}_{12}\cdot\left(\nabla C\left(\vec{Q}_1,\mathbf{R};\vec{q},\mathbf{r}\right)+\nabla C\left(\vec{Q}_2,\mathbf{R};-\vec{q},\mathbf{r}\right)\right)+\frac{r_{12}^2}{2}\left\vert\nabla C\left(\vec{Q}_1,\mathbf{R};\vec{q},\mathbf{r}\right)+\nabla C\left(\vec{Q}_2,\mathbf{R};-\vec{q},\mathbf{r}\right)\right\vert^2\right]\times\\
\nonumber&\times&\left[\mathbf{r}_{12}\cdot\left(1-\nabla D_{j'}^+\left(\vec{q},\mathbf{r}\right)+\nabla D_{j'}^-\left(\vec{q},\mathbf{r}\right)\right)+\left(\mathbf{r}_{12}\cdot\left(\nabla D_{j'}^+\left(\vec{q},\mathbf{r}\right)+\nabla D_{j'}^-\left(\vec{q},\mathbf{r}\right)\right)\right)^2\right]+\ldots\\
\label{sourcezfluc2}&\equiv& 1+\frac{y^4}{4}\left(I_{10}+I_{20}-I_{11}+O\left(r_{12}^3\right)\right).
\eea
The integral $I_{mn}$ is the term with the contribution of order $r_{12}^m$ from the second line in the integrand and with the contribution of order $r_{12}^n$ from the third line. The integrands in $I_{mn}$ are thus of order $m+n$ in $\mathbf{r}_{12}$, $m$ coming from the expansion of $D_{j'}^\pm$ and $n$ from the expansion of $C$. By homogeneity $I_{01}=0$ and $I_{02}$ is the same integral that appears in absence of sources, whose calculation was used in obtaining (\ref{zfluc2}) and which gives the right-hand side of the RG flow (\ref{rgfloweqs}). The remaining integral $I_{11}$ is the new ingredient, and the only one which depends on the sources. Representing it as
\be
I_{11}=\frac{\pi^2}{4}\sum_{j'}\sum_{\alpha=\pm}\sum_{\sigma=1,2}\delta(z-z_\alpha)\left(g_\alpha\vec{Q}_\sigma\cdot\vec{P}_{j'\alpha}+g'_\alpha\vec{Q}_\sigma\times\vec{P}_{j'\alpha}\right)\nabla\frac{1}{\vert\mathbf{R}_{12}\vert}\cdot\nabla\frac{1}{\vert\mathbf{r}_{j\alpha}\vert}\tilde{I}_{j'\alpha},
\ee
we compute the integral $\tilde{I}_{j'\alpha}$ in polar coordinates:
\be
\tilde{I}_{j'\alpha}=\frac{1}{2}\int_0^{2\pi} d\theta_j\log\left(r_{12}^2-2r_{12}r_{j'\alpha}\cos\theta_j+r_{j'\alpha}^2\right)\vert_{\Lambda_1}^{\Lambda_2},
\ee
where $\theta_{j'\alpha}$ is the angle between $\mathbf{r}_{j'\alpha}$ and $\mathbf{r}_{12}$. Assuming the RG scale changes as $\Lambda_1=\Lambda,\Lambda_2=\Lambda(1+\ell)$, for small $\ell$ we can expand the integrand getting:
\be
\tilde{I}_{j'\alpha}=\int_0^{2\pi}d\theta_{j'}\frac{\Lambda^2-\Lambda r_{j'\alpha}\cos\theta_j}{\Lambda^2-2\Lambda r_{j'\alpha}\cos\theta_j+r_{j'\alpha}^2}\ell+O\left(\ell^2\right)=2\pi\ell+O\left(\ell^2\right)
\ee
The complicated dependence on the positions of the sources disappears completely in the first order in $\ell$!\footnote{The additional assumption is that $\Lambda>r_{j'\alpha}$ so the integrand contains no poles; this is clearly justified as $\Lambda$ is the length cutoff.} Altogether, comparing the outcome of (\ref{sourcezfluc2}) to the original Hamiltonian (\ref{sourcehamvort}), we see that the renormalization of the bulk interaction between $\vec{Q}_1$ and $\vec{Q}_2$ is unaffacted by the sources, given as before by the $I_{02}$ term, and the source-bulk coupling renormalizes with a strictly negative shift (as $\tilde{I}_{j'\alpha}=2\pi>0$). The flow equations for $g,g'$ couplings are unchanged, being the same as in (\ref{rgfloweqs}). The bulk-to-boundary couplings $g_\pm,g'_\pm$ have the flow equations
\be
\label{sourcefloweqs}\frac{\partial g_\pm}{\partial\ell}=-\pi^3N\ell,~~\frac{\partial g'_\pm}{\partial\ell}=-\frac{\pi^3}{2}N\ell,
\ee
where $N=\sum_{j'}\sum_\alpha 1$ is the total vorticity of the sources. This obviously flows to $g_\pm,g'_\pm=0$.

Intuitively, one may wonder how come that such an important thing as the CP geometry has no bearing on the vortex dynamics; surely the behavior of a co-propagating system would be expected to differ from a counter-propagating system. The answer is that the CP geometry does enter our calculations -- the $B$-beam has an extra minus sign in the equations of motion (\ref{prpsieq}) (alternatively, in the Lagrangian in Eq.~\ref{psilag}); equivalently, the symmetry group of the effective potential in the Lagrangian is $SU(1,1)$, not $SU(2)$ as it would be for two co-propagating beams. Finally, let us emphasize again that in the numerical simulations we directly solve the propagation equations (\ref{prpsieq}) together with (\ref{preeq}), i.e. we directly take into account the CP boundary conditions -- no analytical approximations whatsoever are used in the numerics, and no use is made of the effective vortex Hamiltonian.

\section{Order parameters and RG analysis of the CP vortices in the presence of disorder}
\label{appd}

\subsection{Saddle-point solutions}

We start by rewriting the replicated partition function $\bar{\mathcal{Z}}^n$ in terms of $p_\alpha,q_{\alpha\beta}$ and inserting the constraints which encapsulate their definition in Eq.~(\ref{glassop}):
\bea
\label{constraint1}1\mapsto\int\mathcal{D}\left[\lambda_{(\mu)}^\alpha\right]\exp\left[\lambda_{(\mu)}^\alpha\left(p_\alpha^{(\mu)}-\frac{1}{N}\sum_{i=1}^NQ_{i\alpha}^{(\mu)}\right)\right]\\
\label{constraint2}1\mapsto\int\mathcal{D}\left[\lambda_{(\mu\nu)}^{\alpha\beta}\right]\exp\left[\lambda_{(\mu\nu)}^{\alpha\beta}\left(q_{\alpha\beta}^{(\mu\nu)}-\frac{1}{N}\sum_{i,j=1}^NQ_{i\alpha}^{(\mu)}Q_{j\beta}^{(\nu)}\right)\right].
\eea
We have five constraints, $\lambda_{(\mu\nu)}^{++},\lambda_{(\mu\nu)}^{--},\lambda_{(\mu\nu)}^{+-}=\lambda_{(\mu\nu)}^{-+},\lambda_{(\mu)}^+,\lambda_{(\mu)}^-$, for the corresponding five order parameters in (\ref{glassop}). We can denote
\be
\hat{K}\equiv\left(\begin{matrix}\lambda_{(\mu\nu)}^{++} & \lambda_{(\mu\nu)}^{+-}\\ \lambda_{(\mu\nu)}^{+-} & \lambda_{(\mu\nu)}^{--}\end{matrix}\right),~~\vec{\lambda}\equiv\left(\begin{matrix}\lambda_{(\mu)}^+\\ \lambda_{(\mu)}^-\end{matrix}\right).
\ee
We will also sometimes leave out the replica indices $\mu,\nu$ to avoid cramming the notation too much. Now we can first integrate out the vortex degrees of freedom $Q_{i\alpha}^{(\mu)}$ from (\ref{rephamvort2}) to get the effective action
\bea
\nonumber S_\mathrm{eff}=-\frac{\beta^2}{4}\sum_{\mu,\nu=1}^n\left[\sigma_{++}^2\left(q_{++}^{(\mu\nu)}\right)^2+2\sigma_{+-}^2\left(q_{+-}^{(\mu\nu)}\right)^2+\sigma_{--}^2\left(q_{--}^{(\mu\nu)}\right)^2\right]-\\
\nonumber-\beta\sum_{\mu=1}^n\left[J_0^+\left(p_+^{(\mu)}\right)^2+2J_0^{+-}p_+^{(\mu)}p_-^{(\mu)}+J_0^-\left(p_-^{(\mu)}\right)^2\right]+\\
\nonumber+\frac{1}{2}\log\det\hat{K}-\frac{1}{4}\vec{\lambda}\hat{K}^{-1}\vec{\lambda}-\\
\label{rephamvort3}-\sum_{\mu,\nu=1}^n\left(\lambda_{(\mu\nu)}^{++}q_{++}^{(\mu\nu)}+\lambda_{(\mu\nu)}^{+-}q_{+-}^{(\mu\nu)}+\lambda_{(\mu\nu)}^{--}q_{--}^{(\mu\nu)}\right)-\sum_{\mu=1}^n\vec{\lambda}_{(\mu)}\cdot\vec{p}^{(\mu)}.
\eea
The saddle-point equations for the constraints give the constraints in terms of the expectation values $q^{(\mu\nu)},p^{(\mu)}$. Luckily, the equation for $\vec{\lambda}$ is easy:
\be
\label{saddlec3}\frac{\partial S_\mathrm{eff}}{\partial\vec{\lambda}}=\hat{K}^{-1}\vec{\lambda}-\vec{p}=0,
\ee
so we immediately solve $\vec{\lambda}=\hat{K}\vec{p}$. Now plugging this into the equations for the three remaining constraints yields
\bea
\label{saddlec1}\frac{\partial S_\mathrm{eff}}{\partial\lambda_{\pm\pm}}=\frac{1}{2}\frac{X\lambda_{\pm\pm}^{-1}}{X^2-Y^2}-q_{\pm\pm}+\frac{1}{4}\left(p_\pm\right)^2=0\\
\label{saddlec2}\frac{\partial S_\mathrm{eff}}{\partial\lambda_{+-}}=\frac{Y\lambda_{+-}^{-1}}{X^2-Y^2}-q_{+-}+\frac{1}{2}p_+p_-=0.
\eea
We have denoted $X=\det\lambda_{++}=\det\lambda_{--},Y=\det\lambda_{+-}$ (these have a well-defined limit for $n\to 0$). It is trivial to write $\lambda_{\pm\pm},\lambda_{+-}$ from the above expressions, and we can feed the solutions for all the constraints into the effective action and then solve the saddle-point equations for the order parameters $p_\pm,q_{++},q_{--},q_{+-}$. Full equations are too complex to be solved, even approximately. We will simplify the problem with the following reasoning. The sums over single-replica order parameters generically scale as $\sum_\mu p^{(\mu)}_\pm\sim\sum_\mu\left(p^{(\mu)}_\pm\right)^2\sim n$, whereas the double-replica parameters have $\sum_{\mu,\nu}q^{(\mu\nu)}_{\alpha\beta}\sim n^2$. This means that in the limit $n\to 0$ the $p_\pm$-terms dominate over $q_{\alpha\beta}$-terms. Therefore, if $p_\pm\neq 0$ we can disregard the quantities $q_{\alpha\beta}$ or expand in a series over them, simplifying the equations significantly. Only if the replica symmetry breaking imposes $p_\pm=0$ (not every replica-symmetry-breaking configuration does so), the $q_{\alpha\beta}$ order parameters are significant, and the saddle-point equations with $p_\pm=0$ are again approachable.

Consider first the case $p_\pm=0$. After some algebra, the effective action is now
\be \label{repseff1}S_\mathrm{eff}=-\frac{\beta^2}{4}\sum_{\mu,\nu=1}^n\left(\sigma_{++}^2\left(q_{(\mu\nu)}^{++}\right)^2+2\sigma_{+-}^2\left(q_{(\mu\nu)}^{+-}\right)^2+\sigma_{--}^2\left(q_{(\mu\nu)}^{++}\right)^2\right)+
\frac{1}{2}\log\left(X^2\vert q^{++}\vert^{-1}\cdot\vert q^{--}\vert^{-1}-4Y^2\vert q^{+-}\vert^{-2}\right).
\ee
Consider first the ansatz when the $q^{\pm\pm}$ fields are nonzero, whereas the mixed-flavor field $q^{+-}$ is zero. In this case, the second term in (\ref{repseff1}), coming from the determinant $\hat{K}$, simplifies further and we get the saddle-point equation
\be
\label{saddleq11}-\frac{\beta^2}{2}\sigma_{\pm\pm}^2q^{\pm\pm}-\frac{1}{2}\left(q^{\pm\pm}\right)^{-1}=0,
\ee
which is the same as for the infinite-range spin-glass Ising model \cite{cardy,spinglass}. One obvious solution is $q^{\pm\pm}=q^{+-}=0$, the completely disordered system with no vortex proliferation -- the familiar insulator phase. It is easy to check that this is indeed a minimum of the effective action $S_\mathrm{eff}$. There is also a replica-symmetric but nontrivial solution
\be
\label{ansatzrsymm}q^{\pm\pm}_{(\mu\nu)}=\mathcal{Q}_0^{\pm\pm}+\left(1-\mathcal{Q}_0^{\pm\pm}\right)\delta_{\mu\nu},
\ee
which yields the solution $\mathcal{Q}_0^{\pm\pm}=1-1/(\beta\sigma^{\pm\pm})$. However, this solution is unstable and is not observable. A stable non-trivial solution is obtained if the replica symmetry is broken. The ansatz is well-known from the spin-glass literature (e.~g.,~\cite{spinglass}) and has a $\rho\times\rho$ matrix $\hat{Q}^{\pm\pm}$ on the block-diagonal and the constant zero elsewhere, with
\be
\label{ansatzrbr}\hat{Q}^{\pm\pm}=\mathcal{Q}_1^{\pm\pm}+\left(1-\mathcal{Q}_1^{\pm\pm}\right)\delta_{\mu\nu},~~\mu,\nu=1,\ldots,\rho.
\ee
Eq.~(\ref{saddleq11}) suggests that $\mathcal{Q}_1^{\pm\pm}>0$ for sufficiently large $\beta$, i.e. small $L$. However, no analytical solution for the elements $\mathcal{Q}_1^{\pm\pm}$ exists and they have to be solved for numerically, by plugging in the solution into the effective action and minimizing it. This is an easy task (for chosen values of the parameters and disorder statistics) but we will not do it here as we do not aim at quantitative accuracy anyway; we merely want to sketch the phase diagram. Now if the third field $q^{+-}$ is nonzero, it satisfies the same equation as (\ref{saddleq11}) just with $\sigma_{++}^2\mapsto 2\sigma_{+-}^2$. The three combinations of nonzero order parameters correspond to the three familiar phases -- $q^{\pm\pm}\neq 0$ is the conductor, $q^{+-}\neq 0$ is the frustrated insulator and $q^{\pm\pm},q^{+-}\neq 0$ is the perfect conductor.

\begin{figure}\centering
\includegraphics[width=80mm]{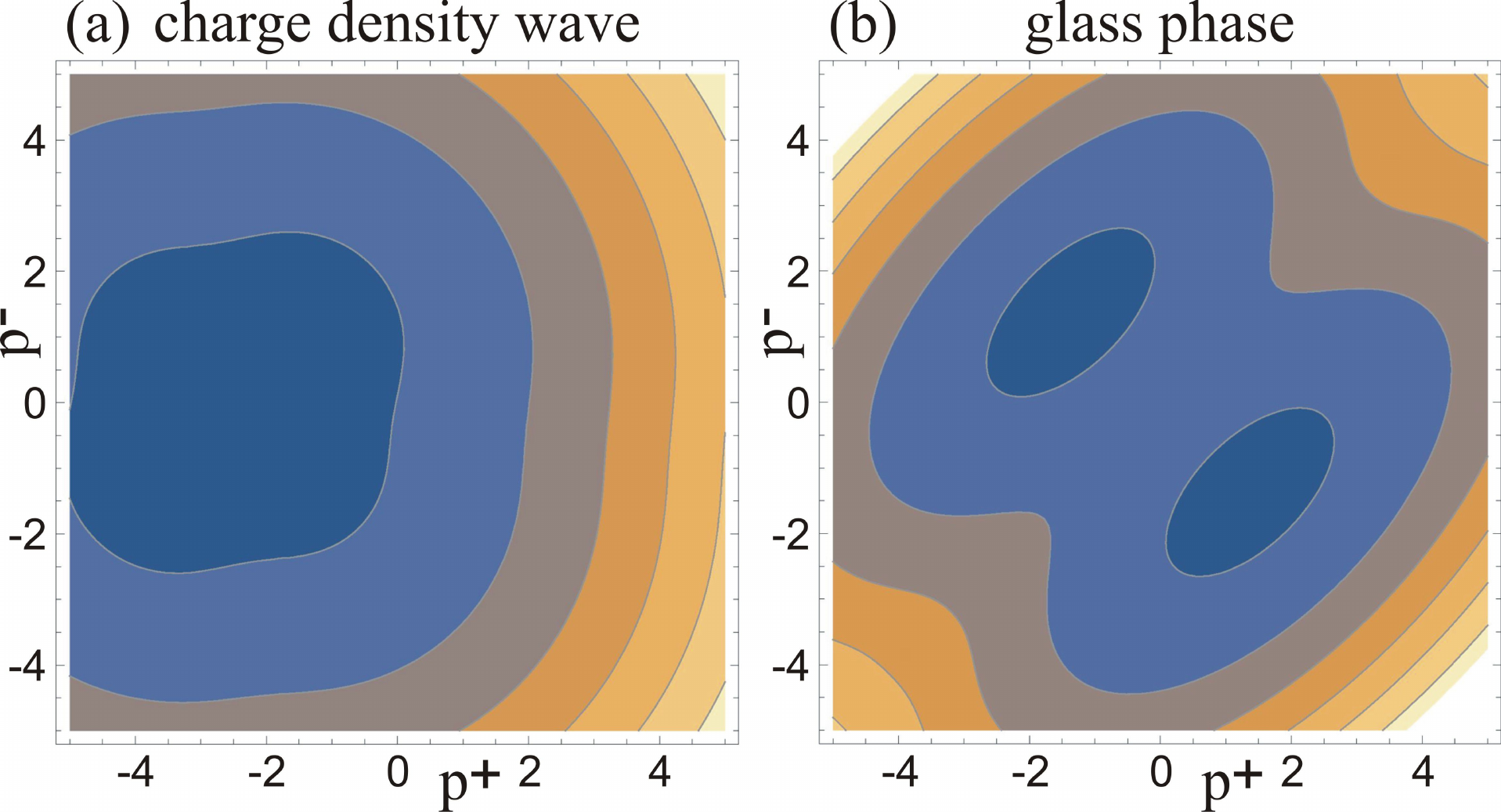}
\caption{\label{figdirtyphd} Free energy (effective action
$S_\mathrm{eff}^{(\mu)}$) in a given replica subsystem in a photonic
lattice with quenched disorder, for the case when the order
parameter $p^\pm=\sum_iQ_{i\pm}$ has a nonzero saddle-point solution
for the action in a given subsystem (replica). Darker (blue) tones
are lower values. The ground states of the system are the local
minima. In the panel (a) for $J^+=-J^-=1$, there is a single local
minimum. In the case (b), for $J^+=-J^-=1$, we see two distinct
minima of equal height, for two different nonzero values of $p^\pm$.
Such potential energy landscape fits the description of glassy
systems.}
\end{figure}

The solutions with $\vec{p}\neq 0$ yield new physics. In this case, we have at leading order $\lambda_{\pm\pm}=-2X/(X^2-Y^2)(p^{\pm})^{-2},\lambda_{+-}=-2Y/(X^2-Y^2)(p^+p^-)^{-1}$, so the effective action is
\bea
\nonumber S_\mathrm{eff}=-\beta\sum_{\mu=1}^n\left(J_0^+\left(p^+_{(\mu)}\right)^2+2J_0^{+-}p^+_{(\mu)}p^-_{(\mu)}+J_0^-\left(p^-_{(\mu)}\right)^2\right)-\\
\label{repseff2}-\log p^+p^-+O\left(\vert q^{\alpha\beta}\vert^2\right)+O\left(\frac{\vert q^{\alpha\beta}\vert}{\vert\vec{p}\vert^2}\right),
\eea
giving the saddle-point equations
\be
\label{saddlep}J_0^\pm p^\pm_{(\mu)}+\frac{J_0^{+-}}{\beta}p^\mp_{(\mu)}+\left(p^\pm_{(\mu)}\right)^{-1}=0,
\ee
which easily gives
\be
\label{psol}p^\pm=s_1\sqrt{\frac{1}{J_0^\pm+s_2\frac{J_0^{+-}}{\beta}\sqrt{\frac{J_0^+}{J_0^-}}}},
\ee
with $s_{1,2}\in\lbrace\pm 1\rbrace$. The solution is the same for every $\mu$ and $p^\pm_{(\mu)}=\left(p^\pm,p^\pm,\ldots, p^\pm\right)$. Now depending on the sign of the determinant $J_0^+J_0^--\left(J_0^{+-}\right)^2$ the solutions for different $s_{1,2}$ may be minima or saddle points. In either case, we have a phase with nonzero local charge density, which is the meaning of $\vec{p}$. If there are multiple minima we call this phase vortex glass. The reader may argue that true glass should satisfy more stringent conditions and that our phase is not a true glass. Depending on the viewpoint this may well be accepted and we use the term "glass phase" merely as shorter and more convenient than "phase with power-law correlation decay, no long-range order and frustrated free energy landscape". The phase with a single minimum will be called charge density wave, as it has a unique ground state configuration yielding macroscopically nonzero charge density, i.e. it has a true long-range order. On the other had, with multiple minima the replica-averaged charge density sums to zero. The landscape, i.e. the effective action of the system for given replica ($\mu$) as a function of $p^\pm_{(\mu)}$, is given in Fig.~\ref{figdirtyphd} as the density map of the function $S_\mathrm{eff}(p^+,p^-)$ dependence for $J_0^+=-J_0^-=1$ (glass phase, A) and $J_0^+=J_0^-=1$ (charge density wave, B). We see that the glassy phase shows two inequivalent minima in each replica, with $s_1=-s_2=\pm 1$ in Eq.~(\ref{psol}), so the total action, the sum of actions of all replica subsystems, can have one and the same value for many configurations, the definition of a highly frustrated system, one of the reasons we dub this phase glass. The charge density wave only has a single minimum for $s_1=s_2=1$.

\subsection{RG flow equations}

The starting Hamiltonian is the same as in (\ref{rephamvort2}). Now we will write it out more explicitly, keeping the distance-dependent parts:
\be
\label{rephamrg}\beta\mathcal{H}_\mathrm{eff}=\beta\sum_{\mu=1}^n\sum_{i,j}\left(\bar{g}_c\vec{Q}^{(\mu)}_i\cdot\vec{Q}^{(\mu)}_j+\bar{g}'_c
\vec{Q}^{(\mu)}_i\times\vec{Q}^{(\mu)}_j\right)\log r_{ij}-\frac{\beta^2}{2}\sum_{\mu,\nu=1}^n\sum_{i,j}Q^{(\mu)}_{i\alpha}Q^{(\nu)}_{i\beta}\sigma_{\alpha\beta}^2Q^{(\mu)}_{j\alpha}Q^{(\nu)}_{j\beta}.
\ee
We have denoted the elements of $J_0$ by $J_0^{++}=J_0^{--}=\bar{g}_c,J_0^{+-}=J_0^{-+}=\bar{g}'_c$ (the bars over the letter remind us that these are disorder-averaged values). The fluctuation of the partition function is completely analogous to the clean case, only it has the additional non-local quartic term. It can again be expanded over $r_{12}$ as in (\ref{zfluc}) but the quartic term contains no small parameter for the power series expansion and has to be kept in the exponential form. Starting from the expression for the fluctuation analogous to the clean case (\ref{zfluc}), we get
\bea
\nonumber\frac{\delta\mathcal{Z}}{\mathcal{Z}}=1+\frac{y^4}{4}\sum_{\vec{q}^{(\rho)},\vec{q}^{(\sigma)}}e^{-\frac{\beta^2}{2}(\vec{q}^{(\rho)},-\vec{q}^{(\sigma)},\vec{q}^{(\rho)},-\vec{q}^{(\sigma)})+\frac{\beta^2}{2}\left(\vec{Q}^{(\mu)},\vec{q}^{(\rho)},\vec{Q}^{(\nu)},\vec{q}^{(\sigma)}\right)}\int dr_{12}r_{12}^3e^{g\vec{q}^{(\rho)}\cdot\vec{q}^{(\rho)}+g'\vec{q}^{(\rho)}\times\vec{q}^{(\rho)}}\times\\
\label{zflucdis}\times\left[\int drr^2\left(g\vec{Q}_1^{(\mu)}\cdot\vec{q}^{(\rho)}+g'\vec{Q}_1^{(\mu)}\times\vec{q}^{(\rho)}\right)\nabla\log\vert\mathbf{R}_1-\mathbf{r}\vert+\left(g\vec{Q}_2^{(\mu)}\cdot\vec{q}^{(\rho)}+g'\vec{Q}_2^{(\mu)}\times\vec{q}^{(\rho)}\right)\nabla\log\vert\mathbf{R}_2-\mathbf{r}\vert\right]^2.
\eea
We have used the notation:
\be
(\vec{q}_1,\vec{q}_2,\vec{q}_3,\vec{q}_4)\equiv\sigma_{++}^2q_{1+}q_{3+}q_{2+}q_{4+}+\sigma_{+-}^2\left(q_{1+}q_{3-}q_{2+}q_{4-}+q_{1-}q_{3+}q_{2-}q_{4+}\right)+\sigma_{--}^2q_{1-}q_{3-}q_{2-}q_{4-}.
\ee
Now we trace out the fluctuations first by integrating over $r$ and doing some simple algebra:
\bea
\nonumber\frac{\delta\mathcal{Z}}{\mathcal{Z}}&=&\left[1+16y^4\left(g+g'\right)^2\cosh\left(\beta^2\sigma^2_{++}+\beta^2\sigma_{+-}^2\right)\cosh\left(\beta^2\sigma_{--}^2+\beta^2\sigma^2_{+-}\right)\left(\vec{Q}_1^{(\mu)}\cdot\vec{Q}_2^{(\nu)}+\vec{Q}_1^{(\mu)}\times\vec{Q}_2^{(\nu)}\right)\log R_{12}\right]\times\\
\nonumber &\times&\left[1+16y^4\left(g-g'\right)^2\cosh\left(\beta^2\sigma_{++}^2-\beta^2\sigma_{+-}^2\right)\cosh\left(\beta^2\sigma_{--}^2-\beta^2\sigma_{+-}^2\right)\left(\vec{Q}_1^{(\mu)}\cdot\vec{Q}_2^{(\nu)}-\vec{Q}_1^{(\mu)}\times\vec{Q}_2^{(\nu)}\right)\log R_{12}\right]\times\\
\nonumber&\times&\left[1-2\pi y^4e^{-\frac{\beta^2}{2}\left(\sigma_{++}^2\left(q_+^{(\mu)}q_+^{(\nu)}\right)^2+\sigma_{+-}^2\left(q_+^{(\mu)}q_-^{(\nu)}\right)^2+\sigma_{+-}^2\left(q_-^{(\mu)}q_+^{(\nu)}\right)+\sigma_{--}^2\left(q_-^{(\mu)}q_-^{(\nu)}\right)^2\right)}\int drr^{1-\beta\left(g\vec{q}^{(\mu)}\cdot\vec{q}^{(\mu)}+g'\vec{q}^{(\mu)}\times\vec{q}^{(\mu)}\right)}\right].\\
\label{zflucdis2}\eea
The next step is the summation over all possible $\pm 1$ charges of virtual vortices $\vec{q}^{(\mu)},\vec{q}^{(\nu)}$ (the two replica indices mean two summations from $1$ to $n$), which requires quite some algebra. The renormalized partition function $\bar{\mathcal{Z}}^n$ finally gives the RG flow equations:
\bea
\nonumber\frac{\partial g}{\partial\ell}=-8\pi\left(g+g'\right)^2y^4\cosh\left(\beta^2\sigma_{++}^2+\beta^2\sigma_{+-}^2\right)\cosh\left(\beta^2\sigma_{--}^2+\beta^2\sigma_{+-}^2\right)-\\
\nonumber -8\pi\left(g-g'\right)^2y^4\cosh\left(\beta^2\sigma_{++}^2-\beta^2\sigma_{+-}^2\right)\cosh\left(\beta^2\sigma_{--}^2-\beta^2\sigma_{+-}^2\right)\\
\nonumber\frac{\partial g'}{\partial\ell}=-\pi\left(g+g'\right)^2y^4\cosh\left(\beta^2\sigma_{++}^2+\beta^2\sigma_{+-}^2\right)\cosh\left(\beta^2\sigma_{--}^2+\beta^2\sigma_{+-}^2\right)-\\
\nonumber -\pi\left(g-g'\right)^2y^4\cosh\left(\beta^2\sigma_{++}^2-\beta^2\sigma_{+-}^2\right)\cosh\left(\beta^2\sigma_{--}^2-\beta^2\sigma_{+-}^2\right)\\
\nonumber\frac{\partial y}{\partial\ell}=2\pi\left(1-g-g'-\frac{\beta^2}{4}\left(\sigma_{++}^2+2\sigma_{+-}^2+\sigma_{--}^2\right)\right)y\\
\label{rgflowdirty}\frac{\partial\sigma^2_{\alpha\beta}}{\partial\ell}=-2\pi\beta^4\sigma_{\alpha\beta}^4y^4.
\eea
As discussed in the main text, the fixed point must lie either at $y=0$ or $y\to\infty$, depending on the magnitude of $g+g'+\beta^2\sigma^2$. For $y\to 0$, three clean fixed points remain, which flow to zero disorder: these correspond to PC, FI and conductor. The disordered fixed point also has $y\to 0$ but the disorder is nonzero: this is the CDW phase from the mean-field analysis, the dirty analog of the insulator. Finally, when $y\to\infty$ and nonzero $\sigma^2$ at the fixed point we expect glassy behavior.

\end{document}